\documentclass[12pt,english,aps,manuscript,showpacs]{revtex4-1}
\usepackage[T1]{fontenc}
\usepackage[latin9]{inputenc}
\usepackage{geometry}
\usepackage{color}
\usepackage{rotating}
\usepackage{textcomp}
\usepackage{amsthm}
\usepackage{amsmath}
\usepackage{amssymb}
\usepackage{graphicx}

\makeatletter

\providecommand{\tabularnewline}{\\}

\@ifundefined{textcolor}{}
{%
 \definecolor{BLACK}{gray}{0}
 \definecolor{WHITE}{gray}{1}
 \definecolor{RED}{rgb}{1,0,0}
 \definecolor{GREEN}{rgb}{0,1,0}
 \definecolor{BLUE}{rgb}{0,0,1}
 \definecolor{CYAN}{cmyk}{1,0,0,0}
 \definecolor{MAGENTA}{cmyk}{0,1,0,0}
 \definecolor{YELLOW}{cmyk}{0,0,1,0}
}

\makeatother

\usepackage{babel}
\begin{document}

\title{Chen's derivative rule revisited: Role of tip-orbital interference in STM}

\author{Gábor Mándi}

\author{Krisztián Palotás}

\address{Department of Theoretical Physics, Budapest University of Technology and Economics,
Budafoki út 8., H-1111 Budapest, Hungary}

\begin{abstract}
On the occasion of its 25th anniversary, we revise Chen's derivative rule for electron tunneling
[C.\ J.\ Chen, Phys.\ Rev.\ B 42, 8841 (1990)] for the purpose of
computationally efficient simulations of scanning tunneling microscopy (STM) based on first principles electronic structure data.
The revised model allows the weighting of tunneling matrix elements of different tip orbital characters by an arbitrary
energy independent choice or based on energy dependent weighting coefficients obtained by an expansion of the tip single electron
wavefunctions/density of states projected onto the tip apex atom. Tip-orbital interference in the STM junction
is included in the model by construction and can be analyzed quantitatively. As a further advantage, arbitrary
tip geometrical orientations are included in the revised model by rotating the coordinate system of the tip apex
using Euler angles and redefining the weighting coefficients of the tunneling matrix elements. We demonstrate the reliability of
the model by applying it to two functionalized surfaces of recent interest where quantum interference effects play an important
role in the STM imaging process: N-doped graphene and a magnetic Mn$_2$H complex on the Ag(111) surface.
We find that the proposed tunneling model is 25 times faster than the Bardeen method concerning computational time, while
maintaining good agreement. Our results show that the electronic structure of the tip has a considerable effect
on STM images, and the Tersoff-Hamann model does not always provide sufficient results in view of quantum
interference effects. For both studied surfaces we highlight the importance of interference between $s$ and $p_z$
tip orbitals that can cause a significant contrast change in the STM images.
Our method, thus, provides a fast and reliable tool for calculating STM images based on
Chen's derivative rule taking into account the electronic structure and local geometry of the tip apex.
\end{abstract}


\maketitle

\section{Introduction}

The role of the tip in scanning tunneling microscopy (STM) on the imaging contrast mechanisms has been extensively studied in
recent years. A reliable interpretation of experimental findings can be obtained by STM simulations that need electron tunneling
models capable of dealing with a diversity of tip parameters in a consistent manner. Furthermore, it is required that STM
simulations are computationally inexpensive and user-friendly in order to provide a useful tool not only for theoretician experts
of electronic structure methods but also for experimental STM groups.

The pioneering electron tunneling model was proposed by Bardeen who derived the tunneling current formula based on
first order perturbation theory \citep{Bardeen1961}. While this method contains the effect of the electronic structure of the
STM tip explicitly, it is only applicable in the pure tunneling regime and does not account for multiple scattering effects.
An extension of the Bardeen method including multiple scattering has been proposed by Palotás and Hofer \citep{Palotas05}
that has been implemented in the BSKAN code \citep{Hofer03}. Using the Keldysh nonequilibrium Green's function formalism,
Mingo {\it{et al.}} developed an electron transport model that is valid in both tunneling and close-contact regimes
\citep{Mingo19962225}. These methods require the single electron wavefunctions or Green's functions of the sample surface and
the tip or of the coupled system that can be obtained from first principles electronic structure calculations. Although these
single electron quantities can routinely be calculated, e.g., by using density functional theory (DFT), STM image simulations are
still considered as computationally demanding if performed at high levels of electron transport theory. The reason is the numerous
parameters affecting the electron tunneling that have to be considered if aiming at an accurate interpretation of STM experiments, 
e.g., the bias voltage, the tunneling current or tip-sample distance and the tip characteristics:
material, geometry, functionalization, termination, orientation, and the corresponding electronic structure. Note that the
modeling of the tip needs special effort as the above listed tip parameters can be chosen in practically infinite combinations.

To overcome the computational drawback of high level electron transport theories, several simplifications have been introduced.
Tersoff and Hamann assumed an $s$-wave tip \citep{Tersoff19831998,Tersoff1985805}. This way the Bardeen current formula
has been recast to contain the electronic structure of the sample surface only. In the Tersoff-Hamann model the tunneling current
is proportional to the integrated local density of states of the sample at the position of the tip. The energy integral
corresponds to the bias voltage. This approach has become the most widely used method for simulating STM and scanning tunneling
spectroscopy (STS) due to its relative simplicity, although it completely neglects the effects of the STM tip.

Allowing more generality for the tip orbitals, Chen derived the so-called derivative rule from the Bardeen current formula,
where the resulting tunneling matrix element is proportional to the linear combination of spatial derivatives of single electron
wavefunctions of the sample surface in the vacuum \citep{Chen19908841}. The spatial derivatives are determined by the orbital
characters of the STM tip. However, practical calculations of the coefficients for the linear combination were not reported.
The determination of the energy dependent coefficients in Chen's derivative rule from first principles calculations is one of the
subjects of the present work.
Palotás {\it{et al.}} developed a tunneling model using the three-dimensional Wentzel-Kramers-Brillouin (3D-WKB) theory
\citep{Palotas14-3DWKB} that accounts for the energy dependent combination of tip orbital characters determined from the orbitally
decomposed DOS projected to the tip apex atom \citep{Palotas-OrbDepWKB}. We will show that the orbital-decomposed projected tip
DOS can be related to an approximation of the tunneling matrix weighting coefficients in the revised Chen's derivative rule.

The above mentioned tunneling models have been successfully used in theoretical investigations of a multitude of STM junctions.
To select a few examples, we focus on studies concerned with the role of the tip geometry, orbital character and functionalization
on the STM imaging. Teobaldi {\it{et al.}} investigated the effect of bias voltage and tip electronic structure on the
STM contrast formation of the highly oriented pyrolytic graphite (HOPG) surface using tungsten tips with different terminations
and sharpnesses \citep{Teobaldi2012}. Chaika {\it{et al.}} demonstrated that by using oriented single crystalline
tungsten tips it is possible to select a particular tip electron orbital for high-resolution imaging of HOPG
\citep{Chaika2010,Chaika2013}. Channel selective tunneling was also examined by Wong {\it{et al.}} using tip functionalization
with hexa-peri-hexabenzocoronene (HBC) molecules \citep{Wong-SelectiveTunneling2012}. Employing Chen's derivative rule,
Gross {\it{et al.}} simulated STM images of pentacene and naphthalocyanine molecules using CO-functionalized tips
\citep{Gross-p_tip_PRL}. The increased lateral resolution achieved by these tips demonstrated the significant contribution of
$p$-type tip states. Siegert {\it{et al.}} studied the influence of $s$- and $p$-wave tip symmetries on the STM maps of
$\pi$-conjugated molecules using the reduced density matrix formalism combined with Chen's derivative rule
\citep{Siegert2013-TipSym}. The orbital-dependent 3D-WKB model was extended for spin-polarized STM in Ref.\ \cite{Mandi-Fe110}.
The effect of tip orbital symmetries on the scanning tunneling spectra was also investigated by probing
the cuprate high-temperature superconductor $\mathrm{Bi_{2}Sr_{2}CaCu_{2}O_{8+\delta}}$. Suominen {\it{et al.}} found that the
symmetry of the tip can radically change the topographic image due to the overlap of sample and tip orbitals \citep{Suominen2011},
while da Silva Neto {\it{et al.}} stated that the apparent nematic behavior of the lattice is likely related to a realistic
STM tip probing the band structure of the material \citep{daSilva2013}. They also pointed out the importance of tunneling
interference effects in the STM junction.

The effect of inter- and intra-atomic interference of electron orbitals has also been in the focus of several studies.
Telychko {\it{et al.}} investigated the nitrogen-doped graphene surface with tungsten and diamond tips and found significantly
smaller current (dip) above the nitrogen atom than above the neighboring carbon atoms at constant height
\citep{Telychko-Graphene-N}. This finding has been explained by a destructive quantum interference essentially resulting
from the C-N $\pi$-bond.
Using the Keldysh Green's function formalism, Jurczyszyn and Stankiewicz and Mingo {\it{et al.}} extensively investigated
inter-orbital interference effects in various tip-sample combinations and found that the interference has a considerable
influence on STM images and STS spectra \citep{Jurczyszyn2003185,Jurczyszyn200570,Mingo19962225}.
Sachse {\it{et al.}} showed that an antiferromagnetic alignment of Mn spin moments in a $\mathrm{Mn_{2}H}$ complex on the Ag(111)
surface explains the experimental STM observation of a dip above the middle of the Mn dimer \citep{Sachse-AgMn2H}.
In the present work we point out that a destructive quantum interference between $s$ and $p_z$ tip orbitals contributes to the
emergence of such a dip in the STM image.

Very recently, the role of the spatial orientation of the tip in the tunneling process has been the subject of a few studies.
Hagelaar {\it{et al.}} performed STM simulations of NO adsorbed on Rh(111) surface over a wide range of tip apex terminations and
orientations and compared them with experimental STM images \citep{Hagelaar2008}. They found that asymmetric tip orientations
provide good qualitative agreement with the experiments under certain tunneling conditions. Lakin {\it{et al.}} developed a
technique to recover the relative orientation of a $\mathrm{C}_{60}$-functionalized tip and a $\mathrm{C}_{60}$ molecule
adsorbed on the Si(111)-(7\texttimes{}7) surface based on Chen's derivative rule \citep{Lakin-MoleculeOriChen}.
Using the 3D-WKB method, Mándi {\it{et al.}} studied tip rotational effects on the imaging of HOPG(0001) \citep{Mandi-HOPGstipe}
and W(110) surfaces \citep{Mandi-ArbTipOriW110} and found that the STM images can considerably be distorted due to different
spatial orientations of the tip. Although the 3D-WKB method lacks the inclusion of electron interference, it proved to be useful
for statistically determining the geometry and likely orientations of the tip in bias voltage dependent STM experiments of HOPG
based on previously inaccessible large scale simulations of tip models in a computationally efficient way \citep{Mandi-Corr}.
The conclusion of all these listed studies is that the relative orientation of the sample surface and the local tip apex
geometry is far from being highly symmetric that is usually considered in standard STM simulations nowadays.
Therefore, it is clear that spatial orientations of the tip have to be integrated into STM simulation models, thus,
the second subject of the present work is to include arbitrary tip geometrical orientations into Chen's derivative rule.

The paper is organized as follows. The revised Chen's derivative rule is presented in section \ref{sec:Simulation-method}.
Computational details of the electronic structure calculations of surface and tip models are reported in section \ref{sec:Comp}.
Results of simulated STM images and comparisons with other electron tunneling methods are found in section \ref{sub:Graphene-N}
and \ref{sub:MnH-on-Ag(111)} for the N-doped graphene and for the antiferromagnetic $\mathrm{Mn_{2}H}$ complex on Ag(111) surface,
respectively. We summarize our findings in section \ref{sec:Conclusions}.

\section{Revised Chen's derivative rule\label{sec:Simulation-method}}

In the tunneling regime the current $I$ depending on the bias voltage $V$ between the sample surface and the tip can be calculated
using Bardeen's tunneling formula \citep{Bardeen1961},
\begin{equation}
I(V)=\frac{2\pi e}{\hbar}{\displaystyle \sum_{\mu\nu}f\left(E_{\mu}\right)\left[1-f\left(E_{\nu}+eV\right)\right]\left|M_{\mu\nu}\right|^{2}\delta\left(E_{\mu}-E_{\nu}-eV\right)},\label{eq:Bardeen tunneling}
\end{equation}
where $e$ is the elementary charge, $\hbar$ is the reduced Planck constant, $f$ is the Fermi distribution function and
$M_{\mu\nu}$ is the tunneling matrix element between two single electron states of the sample ($\mu$) and the tip ($\nu$)
involved in the tunneling. $E_{\mu}$ and $E_{\nu}$ denote corresponding Kohn-Sham eigenenergies that can be obtained from
first principles calculations. Note that $\mu$ and $\nu$ denote composite indices of the band ($n$), wave vector
($\mathbf{k}_{\parallel}$) and spin ($\sigma$) in the separate sample and tip subsystems, respectively.
The tunneling matrix element can be calculated as an integral over the $\mathcal{S}$ separation surface in the vacuum between the
sample and the tip,
\begin{equation}
M_{\mu\nu}=-\frac{\hbar^2}{2m}{\displaystyle \int_{\mathcal{S}}\left(\chi_{\nu}^{\star}\boldsymbol{\nabla}\psi_{\mu}-\psi_{\mu}\boldsymbol{\nabla}\chi_{\nu}^{\star}\right)\mathbf{dS}.}\label{eq: Matrix element}
\end{equation}
In the model the tunneling is assumed to be elastic and energy conservation is ensured by the Dirac-delta in
Eq.\ (\ref{eq:Bardeen tunneling}). At finite temperature the thermal broadening of the electron states has to be taken into account.
This is usually done by approximating the Dirac-delta with a Gaussian function,
\begin{equation}
\delta(E_{\mu}-E_{\nu}-eV)\sim\frac{1}{\sqrt{2\pi\Delta^{2}}}\exp\left[-\frac{\left(E_{\mu}-E_{\nu}-eV\right)^{2}}{2\Delta^{2}}\right].\label{eq:Delta}
\end{equation}
In principle, all $\mu-\nu$ transitions have to be considered with the probability given by this Gaussian factor
(and $|M_{\mu\nu}|^2$) when calculating the tunneling current but practically transitions with significantly low probability
can be neglected, e.g., if $|E_{\mu}-E_{\nu}-eV|\geq 3\Delta$, where $\Delta=k_{B}T$ is the thermal broadening of the states
at $T$ temperature with $k_B$ the Boltzmann constant.

Chen's approach is based on the expansion of the tip wavefunction into spherical harmonic components around the tip apex position
$\mathbf{r}_{0}$ \citep{Chen19908841},
\begin{equation}
\chi_{\nu}\left(\mathbf{r}\right)={\displaystyle \sum_{lm}C_{\nu lm}k_{l}\left(\kappa_{\nu}r\right)Y_{lm}\left(\vartheta,\varphi\right)},
\end{equation}
where $r=\left|\mathbf{r}-\mathbf{r}_{0}\right|$, $k_{l}$ is the spherical modified Bessel function of the second kind, $Y_{lm}$
is the spherical harmonic function depending on the azimuthal $(\vartheta)$ and polar $(\varphi)$ angles and $\kappa_{\nu}$ is
the vacuum decay of the tip wavefunction. The expansion can also be performed according to the real space orbital characters
$\beta\in\{s,p_y,p_z,p_x,d_{xy},d_{yz},d_{3z^2-r^2},d_{xz},d_{x^2-y^2}\}$ introducing the notation of
$\tilde{Y}_{\nu\beta}(r,\vartheta,\varphi)=k_{\beta}(\kappa_{\nu}r)Y_{\beta}(\vartheta,\varphi)$
with $Y_{\beta}$ real spherical harmonics as
\begin{equation}
\chi_{\nu}\left(\mathbf{r}\right)={\displaystyle \sum_{\beta}C_{\nu\beta}\tilde{Y}_{\nu\beta}\left(r,\vartheta,\varphi\right)}.\label{eq:beta_expansion}
\end{equation}
Using this expansion of the tip wavefunction in Eq.\ (\ref{eq: Matrix element}) leads to Chen's derivative rule,
and $|M_{\mu\nu}|^2$ can be written as
\begin{equation}
\left|M_{\mu\nu}\right|^{2}=\frac{4\pi^{2}\hbar^{4}}{\kappa_{\nu}^{2}m^{2}}\left|{\displaystyle \sum_{\beta}}C_{\nu\beta}\hat{\partial}_{\nu\beta}\psi_{\mu}(\mathbf{r}_{0})\right|^{2}.\label{eq:M2_orig}
\end{equation}
We introduce a new notation, $M_{\mu\nu\beta}=C_{\nu\beta}\hat{\partial}_{\nu\beta}\psi_{\mu}(\mathbf{r}_{0})$ that corresponds
to the tunneling matrix element of a given orbital symmetry ($\beta$). Here, the differential operator $\hat{\partial}_{\nu\beta}$
acts on the sample wavefunction at the tip apex position $\mathbf{r}_{0}$. Note that $\hat{\partial}_{\nu\beta}$ is dimensionless
as it contains a factor $\kappa_{\nu}^{-l}$ with $l$ being the angular quantum number. The differential operators for the given
orbital characters are summarized in Table \ref{tab:Derivatives} following Chen \citep{Chen19908841}.

\begin{table}[h]
\begin{centering}
\begin{tabular}{|c|c|c|c|c|c|c|c|c|c|}
\hline
$\beta$ & $s$ & $p_y$ & $p_z$ & $p_x$ & $d_{xy}$ & $d_{yz}$ & $d_{3z^2-r^2}$ & $d_{xz}$ & $d_{x^2-y^2}$ \tabularnewline
\hline
$\hat{\partial}_{\nu\beta}$ & 1 & $\frac{1}{\kappa_{\nu}}\frac{\partial}{\partial y}$ & $\frac{1}{\kappa_{\nu}}\frac{\partial}{\partial z}$ & $\frac{1}{\kappa_{\nu}}\frac{\partial}{\partial x}$ & $\frac{1}{\kappa_{\nu}^2}\frac{\partial^2}{\partial x\partial y}$ & $\frac{1}{\kappa_{\nu}^2}\frac{\partial^2}{\partial y\partial z}$ & $\frac{3}{\kappa_{\nu}^2}\frac{\partial^2}{\partial z^2}-1$ & $\frac{1}{\kappa_{\nu}^2}\frac{\partial^2}{\partial x\partial z}$ & $\frac{1}{\kappa_{\nu}^2}(\frac{\partial^2}{\partial x^2}-\frac{\partial^2}{\partial y^2})$ \tabularnewline
\hline
\end{tabular}
\par\end{centering}
\protect\caption{Differential operators $\hat{\partial}_{\nu\beta}$ for given orbital symmetries ($\beta$) according to Chen
\citep{Chen19908841}. \label{tab:Derivatives} }
\end{table}

Rewriting Eq.\ (\ref{eq:M2_orig}) as
\begin{equation}
\left|M_{\mu\nu}\right|^{2}=\frac{4\pi^{2}\hbar^{4}}{\kappa_{\nu}^{2}m^{2}}\sum_{\beta}\sum_{\beta'}M_{\mu\nu\beta}^{\star}M_{\mu\nu\beta'}=\frac{4\pi^{2}\hbar^{4}}{\kappa_{\nu}^{2}m^{2}}\left[\sum_{\beta}|M_{\mu\nu\beta}|^{2}+\sum_{\beta\ne\beta'}2Re\{M_{\mu\nu\beta}^{\star}M_{\mu\nu\beta'}\}\right]\label{eq:M2_interference}
\end{equation}
allows the investigation of different contributions to the tunneling current.
The first term of the equation on the right hand side is the sum of the absolute value squares of
$M_{\mu\nu\beta}=C_{\nu\beta}\hat{\partial}_{\nu\beta}\psi_{\mu}(\mathbf{r}_{0})$, which is always positive, hence this term
provides a positive contribution to the tunneling current. The second term is an interference term concerning tip orbitals,
which is real and the sign of the individual $\beta\ne\beta'$ components can be positive or negative, respectively contributing
as constructive or destructive interference to the tunneling current. The analysis of the ratios and polarities of the listed
components of $|M_{\mu\nu}|^2$ gives the opportunity to obtain a deeper physical understanding of the electron tunneling process.
Note that in Eq.\ (\ref{eq:M2_orig}) the sample wavefunction $\psi_{\mu}$ can also be expanded into spherical harmonics similarly to
Eq.\ (\ref{eq:beta_expansion}). This way the interference of the sample orbitals and interference between sample and tip orbitals
can be investigated as well. A similar decomposition of tunneling matrix elements has been used by Jurczyszyn and Stankiewicz
\citep{Jurczyszyn2003185,Jurczyszyn200570} and Mingo {\it{et al.}} \citep{Mingo19962225}.

\subsection{Calculation of spatial derivatives\label{sub:Derivatives}}

The spatial derivatives (see Table \ref{tab:Derivatives}) of the sample wavefunction for
$M_{\mu\nu\beta}=C_{\nu\beta}\hat{\partial}_{\nu\beta}\psi_{\mu}(\mathbf{r}_{0})$ can be calculated straightforwardly when using a
plane wave expansion of the wavefunctions. There are many DFT codes, which use a plane wave basis set, e.g., VASP \citep{VASP},
ABINIT \citep{ABINIT} and Quantum-Espresso \citep{QE} to name a few popular ones. Thus, the presented forms of the spatial
derivatives can be potentially useful for future implementations of the revised Chen's derivative rule. In the present work we
use wavefunctions obtained from the VASP code. Let us assume that the single electron wavefunctions of the sample surface are
given in the vacuum at position vector $\mathbf{r}$ in a two-dimensional (2D) Fourier-grid as
\begin{equation}
\psi_{\mu}(\mathbf{r})=\psi_{n^S\mathbf{k}_{\parallel}^S\sigma^S}(\mathbf{r})={\displaystyle \sum_{\mathbf{G_{\parallel}}}A_{n^S\mathbf{k}_{\parallel}^S\sigma^S}(\mathbf{G_{\parallel}},z)\exp\left[i(\mathbf{k}_{\parallel}^S+\mathbf{G_{\parallel}})\mathbf{r}_{\parallel}\right]},\label{eq:plane_wave_expansion}
\end{equation}
where $\mu=(n^S\mathbf{k}_{\parallel}^S\sigma^S)$ is the composite index for single electron states of the sample with
$\mathbf{k}_{\parallel}^S=(k_{x}^S,k_{y}^S)$ the lateral component of the wave vector.
The derivation with respect to $z$ (the direction perpendicular to the sample surface) acts on the expansion coefficients only,
\begin{equation}
\frac{\partial}{\partial z}\psi_{n^S\mathbf{k}_{\parallel}^S\sigma^S}(\mathbf{r})={\displaystyle \sum_{\mathbf{G_{\parallel}}}\left(\frac{\partial}{\partial z}A_{n^S\mathbf{k}_{\parallel}^S\sigma^S}(\mathbf{G_{\parallel}},z)\right)\exp\left[i(\mathbf{k}_{\parallel}^S+\mathbf{G_{\parallel}})\mathbf{r}_{\parallel}\right]},
\end{equation}
while the $x$- and $y$-derivatives act on the phase factor,
\begin{eqnarray}
\frac{\partial}{\partial x}\psi_{n^S\mathbf{k}_{\parallel}^S\sigma^S}(\mathbf{r}) & = & {\displaystyle \sum_{\mathbf{G_{\parallel}}}i(k_{x}^S+G_{x})A_{n^S\mathbf{k}_{\parallel}^S\sigma^S}(\mathbf{G_{\parallel}},z)\exp\left[i(\mathbf{k}_{\parallel}^S+\mathbf{G_{\parallel}})\mathbf{r}_{\parallel}\right]},\\
\frac{\partial}{\partial y}\psi_{n^S\mathbf{k}_{\parallel}^S\sigma^S}(\mathbf{r}) & = & {\displaystyle \sum_{\mathbf{G_{\parallel}}}i(k_{y}^S+G_{y})A_{n^S\mathbf{k}_{\parallel}^S\sigma^S}(\mathbf{G_{\parallel}},z)\exp\left[i(\mathbf{k}_{\parallel}^S+\mathbf{G_{\parallel}})\mathbf{r}_{\parallel}\right].}
\end{eqnarray}
The same procedure can be applied for higher order derivatives listed in Table \ref{tab:Derivatives}.

\subsection{Determination of weighting coefficients\label{sub:Coeff}}

We report on three ways for the choice of the weighting coefficients $C_{\nu\beta}$ for
$M_{\mu\nu\beta}=C_{\nu\beta}\hat{\partial}_{\nu\beta}\psi_{\mu}(\mathbf{r}_{0})$.

(i)
The simplest choice is the assumption of an idealized tip with a given set of energy independent weighting factors $\{C_{\beta}\}$.
Such examples can be found in the literature. In the study of Gross {\it{et al.}} a CO-functionalized tip was modeled as
a combination of $s$ and $p$ orbitals and interference terms were neglected \citep{Gross-p_tip_PRL}. Siegert {\it{et al.}}
employed the reduced density matrix formalism combined with Chen's derivative rule with the inclusion of interference effects and
they considered a similar combination of $s$ and $p$ orbitals \citep{Siegert2013-TipSym}.
Generally, $C_{\beta}$ can be a complex number. We restrict the choice of the set of $\{C_{\beta}\}$ to fulfill the condition:
$\sum_{\beta}|C_{\beta}|^2=1$. Moreover, in this idealized tip model case we choose the vacuum decay of the tip states
$\kappa_{\nu}=$1 \AA$^{-1}$ for all $\nu$.
Examples of the effect of idealized tips as pure $s$ and pure $p_z$ orbitals and a combination of $(s+p_z)/\sqrt{2}$ on the STM
image of N-doped graphene will be shown in section \ref{sub:Graphene-N}. We will also point out that the effect of interference is
remarkable in this case causing a significant contrast change.

(ii)
Based on Eq.\ (\ref{eq:beta_expansion}), $C_{\nu\beta}$ complex numbers can be obtained as
\begin{equation}
C_{\nu\beta}=\left\langle\left.\tilde{Y}_{\nu\beta}(\mathbf{r})\right|\chi_{\nu}(\mathbf{r})\right\rangle =\langle k_{\beta}(\kappa_{\nu}r)Y_{\beta}(\vartheta,\varphi)|\chi_{\nu}(\mathbf{r})\rangle\label{eq:coeff}
\end{equation}
with $\nu=(n^T\mathbf{k}_{\parallel}^T\sigma^T)$ composite index for single electron states of the tip, where
$\mathbf{k}_{\parallel}^T$ is the lateral component of the wave vector.
We calculate these coefficients explicitly in the Wigner-Seitz sphere ($W-S$) of the tip apex atom with the VASP code.
Since symmetry properties of the model tip geometry are taken into account in VASP, we obtain a reduced set of $C_{\nu\beta}$
corresponding to $\mathbf{k}_{\parallel}^T$ being in the irreducible part of the Brillouin zone. We can calculate how these
coefficients change under 2D transformations ($\mathcal{T}$) of the tip's symmetry group in order to obtain
$C_{\nu\beta}$ in the full 2D Brillouin zone. For this, the plane wave expansion of the tip wavefunction is needed,
\begin{equation}
\chi_{\nu}(\mathbf{r})=\chi_{n^T\mathbf{k}_{\parallel}^T\sigma^T}(\mathbf{r})={\displaystyle \sum_{\mathbf{G_{\parallel}}}B_{n^T\mathbf{k}_{\parallel}^T\sigma^T}(\mathbf{G_{\parallel}},z)\exp\left[i(\mathbf{k}_{\parallel}^T+\mathbf{G_{\parallel}})\mathbf{r}_{\parallel}\right]},\label{eq:plane_wave_expansion1}
\end{equation}
similarly to Eq.\ (\ref{eq:plane_wave_expansion}).
Since the $B$ expansion coefficients are invariant under the $\mathcal{T}$ transformation, i.e.,
$B_{n^T\mathbf{k}_{\parallel}^T\sigma^T}=B_{n^T\mathcal{T}\left(\mathbf{k}_{\parallel}^T\right)\sigma^T}$,
the transformation of the tip wavefunction comes from that of the phase factors.
Using Eqs.\ (\ref{eq:coeff}) and (\ref{eq:plane_wave_expansion1}) we obtain the following for the transformed coefficients,
\begin{eqnarray}
C_{n^T\mathcal{T}\left(\mathbf{k}_{\parallel}^T\right)\sigma^T\beta} & = & {\displaystyle \int_{W-S}k_{\beta}(\kappa_{\nu}r)Y_{\beta}(\mathbf{r})\sum_{\mathbf{G_{\parallel}}}B_{\nu}(\mathbf{G_{\parallel}},z)\exp\left[i\mathcal{T}(\mathbf{k}_{\parallel}^T+\mathbf{G_{\parallel}})\mathbf{r}_{\parallel}\right]d^3r}\nonumber \\
& = & \int_{W-S}k_{\beta}(\kappa_{\nu}r)Y_{\beta}(\mathcal{T}\mathbf{r})\chi_{\nu}(\mathbf{r})d^3r.\label{eq:C-transform}
\end{eqnarray}
Note that $\mathcal{T}$ are represented by $2\times 2$ real matrices and the transformation of the coordinates is
$\mathcal{T}\mathbf{r}=(\mathcal{T}_{11}x+\mathcal{T}_{12}y,\mathcal{T}_{21}x+\mathcal{T}_{22}y,z)$. Using the real spherical
harmonics in Cartesian coordinates, we can calculate their transformations by substituting the transformed lateral
coordinates into their normalized form. The results are shown in Table \ref{tab:Transformation}.
Thus, $C_{\nu\beta}$ is determined in the full 2D Brillouin zone, and we can directly apply them in the formula of the
tunneling matrix elements in Eq.\ (\ref{eq:M2_orig}).

\begin{table}[h]
\begin{centering}
\begin{tabular}{|c|c|c|}
\hline
Orbital & $Y(x,y,z)$ & Transformed orbital\tabularnewline
\hline
\hline
$s$ & $\frac{1}{2\sqrt{\pi}}$ & $s$\tabularnewline
\hline
$p_{y}$ & $\frac{1}{2}\sqrt{\frac{3}{\pi}}\frac{y}{r}$ & $\mathcal{T}_{21}p_{x}+\mathcal{T}_{22}p_{y}$\tabularnewline
\hline
$p_{z}$ & $\frac{1}{2}\sqrt{\frac{3}{\pi}}\frac{z}{r}$ & $p_{z}$\tabularnewline
\hline
$p_{x}$ & $\frac{1}{2}\sqrt{\frac{3}{\pi}}\frac{x}{r}$ & $\mathcal{T}_{11}p_{x}+\mathcal{T}_{12}p_{y}$\tabularnewline
\hline
$d_{xy}$ & $\frac{1}{2}\sqrt{\frac{15}{\pi}}\frac{xy}{r^{2}}$ & $\left(\mathcal{T}_{11}\mathcal{T}_{22}+\mathcal{T}_{12}\mathcal{T}_{21}\right)d_{xy}+2\mathcal{T}_{11}\mathcal{T}_{21}d_{x^{2}-y^{2}}$\tabularnewline
\hline
$d_{yz}$ & $\frac{1}{2}\sqrt{\frac{15}{\pi}}\frac{yz}{r^{2}}$ & $\mathcal{T}_{21}d_{xz}+\mathcal{T}_{22}d_{yz}$\tabularnewline
\hline
$d_{3z^{2}-r^{2}}$ & $\frac{1}{4}\sqrt{\frac{5}{\pi}}\frac{3z^{2}-r^{2}}{r^{2}}$ & $d_{3z^{2}-r^{2}}$\tabularnewline
\hline
$d_{xz}$ & $\frac{1}{2}\sqrt{\frac{15}{\pi}}\frac{xz}{r^{2}}$ & $\mathcal{T}_{11}d_{xz}+\mathcal{T}_{12}d_{yz}$\tabularnewline
\hline
$d_{x^{2}-y^{2}}$ & $\frac{1}{4}\sqrt{\frac{15}{\pi}}\frac{x^{2}-y^{2}}{r^{2}}$ & $\left(\mathcal{T}_{11}^{2}-\mathcal{T}_{21}^{2}\right)d_{x^{2}-y^{2}}+2\mathcal{T}_{11}\mathcal{T}_{12}d_{xy}$\tabularnewline
\hline
\end{tabular}
\par\end{centering}
\protect\caption{Transformation of real spherical harmonics under 2D symmetry operations ($\mathcal{T}$) of the tip. \label{tab:Transformation} }
\end{table}

(iii)
The third suggestion for $C_{\nu\beta}$ is based on the orbital-decomposed DOS projected to the tip apex atom,
$n^{TIP}(E)=\sum_{\beta}n^{TIP}_{\beta}(E)=\sum_{\beta}\sum_{\nu}n^{TIP}_{\nu\beta}\delta(E-E_{\nu})$
obtained from first principles calculation.
Using the expansion of the tip wavefunction in Eq.\ (\ref{eq:beta_expansion}) and the approximation of orthonormality for
$\tilde{Y}_{\nu\beta}(r,\vartheta,\varphi)$ within the Wigner-Seitz ($W-S$) sphere of the tip apex atom,
$\left\langle\left.\tilde{Y}_{\nu\beta}(r,\vartheta,\varphi)\right|\tilde{Y}_{\nu\beta'}(r,\vartheta,\varphi)\right\rangle_{W-S}\approx\delta_{\beta\beta'}$, the following is obtained,
\begin{equation}
n^{TIP}(E)={\displaystyle \sum_{\nu}\sum_{\beta}n^{TIP}_{\nu\beta}\delta(E-E_{\nu})=\sum_{\nu}\left\langle \left.\chi_{\nu}\right|\chi_{\nu}\right\rangle_{W-S}\delta(E-E_{\nu})}\approx\sum_{\nu}\sum_{\beta}|C_{\nu\beta}|^2\delta(E-E_{\nu}).\label{eq:PDOS}
\end{equation}
Thus, we can approximate the complex $C_{\nu\beta}$ coefficients with real values, $C_{\nu\beta}\approx\sqrt{n^{TIP}_{\nu\beta}}$.
This way Eq.\ (\ref{eq:M2_orig}) is recast to
\begin{equation}
\left|M_{\mu\nu}\right|^{2}=\frac{4\pi^{2}\hbar^{4}}{\kappa_{\nu}^{2}m^{2}}\left|{\displaystyle \sum_{\beta}}\sqrt{n^{TIP}_{\nu\beta}}\hat{\partial}_{\nu\beta}\psi_{\mu}(\mathbf{r}_{0})\right|^{2}.\label{eq:M2_pdos}
\end{equation}
Since the calculation of the orbital-decomposed atom-projected DOS is routinely available in DFT codes, the presented
approximation applied to the tip apex atom gives a widely accessible choice for the weighting coefficients in the revised
Chen's derivative rule. In section \ref{sub:Graphene-N} we will demonstrate that the STM images obtained by the
$C_{\nu\beta}\approx\sqrt{n^{TIP}_{\nu\beta}}$ approximation provide good agreement with those calculated using the proper complex
$C_{\nu\beta}$ coefficients according to Eq.\ (\ref{eq:coeff}).

\subsection{Inclusion of arbitrary tip orientations\label{sub:Tip-rotation}}

Since the electronic structures of the sample surface and the tip are generally calculated independently to allow more flexibility
with their geometries, arbitrary orientations of the tip can be included into the revised Chen's method. This can be done by
redefining the spatial derivatives of the sample wavefunctions corresponding to the orbital characters in the rotated coordinate
system of the tip with respect to the sample surface. This rotation is described by a coordinate transformation, which is
represented by a $3\times 3$ matrix $\mathcal{R}$ with elements $\mathcal{R}_i^j$. We use the explicit form of $\mathcal{R}$ as in
Refs.\ \citep{Mandi-ArbTipOriW110,Mandi-Corr},
{\footnotesize
\begin{equation}
\mathcal{R}=\left[\begin{array}{ccc}
\cos\varphi_{0}\cos\psi_{0}-\sin\varphi_{0}\sin\psi_{0}\cos\vartheta_{0} & \cos\varphi_{0}\sin\psi_{0}+\sin\varphi_{0}\cos\psi_{0}\cos\vartheta_{0} & \sin\varphi_{0}\sin\vartheta_{0}\\
-\sin\varphi_{0}\cos\psi_{0}-\cos\varphi_{0}\sin\psi_{0}\cos\vartheta_{0} & -\sin\varphi_{0}\sin\psi_{0}+\cos\varphi_{0}\cos\psi_{0}\cos\vartheta_{0} & \cos\varphi_{0}\sin\vartheta_{0}\\
\sin\psi_{0}\sin\vartheta_{0} & -\cos\psi_{0}\sin\vartheta_{0} & \cos\vartheta_{0}
\end{array}\right]
\end{equation}
}
with the Euler angles $(\vartheta_{0},\varphi_{0},\psi_{0})$. Using the Einstein summation convention, the relationship
between the two set of coordinates, the rotated tip coordinates $x^{\prime j}\in\{x',y',z'\}$ and the sample coordinates
$x^{i}\in\{x,y,z\}$, is the following:
\begin{equation}
x^{\prime j}=\frac{\partial x^{\prime j}}{\partial x^{i}}x^{i}=\mathcal{R}_{i}^{j}x^{i};\; x^{i}=\frac{\partial x^{i}}{\partial x^{\prime j}}x^{\prime j}=\left(\mathcal{R}^{-1}\right)_{j}^{i}x^{\prime j}.
\end{equation}
With the help of these, we can relate the derivatives of the sample wavefunction $\psi$ with respect to the
rotated tip coordinates $x^{\prime j}$ to the derivatives with respect to the sample coordinates $x^{i}$ as
\begin{equation}
\frac{\partial\psi}{\partial x^{\prime j}}=\frac{\partial\psi}{\partial x^{i}}\frac{\partial x^{i}}{\partial x^{\prime j}}=\frac{\partial\psi}{\partial x^{i}}\left(\mathcal{R}^{-1}\right)_{j}^{i}.\label{eq:first_der}
\end{equation}
Similarly, the second derivatives are
\begin{equation}
\frac{\partial^{2}\psi}{\partial x^{\prime k}\partial x^{\prime j}}=\left(\frac{\partial^{2}\psi}{\partial x^{l}\partial x^{i}}\right)\left(\mathcal{R}^{-1}\right)_{j}^{i}\left(\mathcal{R}^{-1}\right)_{k}^{l}.\label{eq:second_der}
\end{equation}
Using Eqs.\ (\ref{eq:first_der}) and (\ref{eq:second_der}) the transformed $\hat{\partial}'_{\nu\beta}$ differential operators
corresponding to the rotated tip coordinate system can be constructed and employed in Eq.\ (\ref{eq:M2_orig}) for the
tunneling matrix elements. Since the transformation is linear, this, in turn, results in redefined $C_{\nu\beta}$ weighting
coefficients in Eq.\ (\ref{eq:M2_orig}) for the $\hat{\partial}_{\nu\beta}$ operators given in the coordinate system of the sample
listed in Table \ref{tab:Derivatives}.

\section{Computational details\label{sec:Comp}}

Using the revised Chen's derivative rule implemented in the BSKAN code \citep{Hofer03,Palotas05}, STM imaging of two
functionalized surfaces of current interest is investigated: N-doped graphene and an antiferromagnetic $\mathrm{Mn_{2}H}$ complex
on the Ag(111) surface in combination with several tip models. Geometrical relaxations and electronic structure calculations of
the surface and tip models were performed separately using the VASP code \citep{VASP} employing the projector augmented wave (PAW)
method \citep{PAW}.

N-doped graphene is modeled as a free-standing single-layer graphene sheet in a $7\times 7$ surface unit cell following
Ref.\ \citep{Telychko-Graphene-N} and 16 \AA-wide vacuum perpendicular to the surface to avoid unphysical interactions between
neighboring slabs. One carbon atom is replaced by nitrogen in the given supercell. The generalized gradient approximation (GGA)
and the exchange-correlation (XC) functional parametrized by Perdew and Wang (PW91) \citep{PW} were used together
with a plane wave basis set energy cut-off of 400 eV and an $11\times 11\times 1$ Monkhorst-Pack \citep{MP} k-point sampling of
the Brillouin zone. We found a planar lattice structure after geometrical relaxation following N-doping,
in agreement with Ref.\ \citep{Telychko-Graphene-N}.

For the revised Chen's derivative rule, idealized model tips of pure $s$, pure $p_z$ and a combination of $(s+p_z)/\sqrt{2}$
orbitals are initially considered, see section \ref{sub:Coeff} (i) for details.
Since N-doped graphene has been experimentally probed with tungsten tips \citep{Telychko-Graphene-N}, we consider three tungsten
tip models with different sharpnesses and compositions: $\mathrm{W_{blunt}}$, $\mathrm{W_{sharp}}$ and $\mathrm{W_{C-apex}}$.
The $\mathrm{W_{blunt}}$ tip model is represented by an adatom adsorbed on the hollow site of the W(110) surface,
the $\mathrm{W_{sharp}}$ tip is modeled as a pyramid of three-atoms height on the W(110) surface, and the $\mathrm{W_{C-apex}}$
tip is a sharp tungsten tip with a carbon apex atom accounting for a likely carbon contamination from the sample.
More details on the used tip geometries can be found in Ref.\ \citep{Teobaldi2012}.

Geometrical relaxations, search for the magnetic ground state and electronic structure calculations of an Mn monomer, Mn dimer
and $\mathrm{Mn_{2}H}$ on the Ag(111) surface have been reported in Ref.\ \citep{Sachse-AgMn2H}, and an antiferromagnetic
ground state for the $\mathrm{Mn_{2}H}$/Ag(111) system has been found. We use their electronic structure results in the
present paper, for more details on the modeled geometries and DFT calculations please refer to Ref.\ \citep{Sachse-AgMn2H}.
In their STM experiments silver tips have been used. Therefore, we consider blunt tips as an adatom adsorbed on the hollow site of
the silver surface in two different orientations: Ag(001) and Ag(111), in a $3\times 3$ surface unit cell and at least 15 \AA-wide
vacuum perpendicular to the surface to avoid unphysical interactions due to the slab geometry. GGA and XC functional parametrized
by Perdew, Burke and Ernzerhof (PBE) \citep{PBE} are employed. Moreover, a plane wave basis set energy cut-off of 250 eV
and an $11\times 11\times 1$ Monkhorst-Pack \citep{MP} k-point grid centered on the $\Gamma$ point are used.
The convergence criterion for the forces acting on relaxed atoms (adatom and first full layer) is 0.01 eV\AA$^{-1}$.

\section{Results and discussion\label{sec:Results-and-discussion}}

We demonstrate the reliability of the revised Chen's derivative rule for the mentioned N-doped graphene and antiferromagnetic
$\mathrm{Mn_{2}H}$ complex on the Ag(111) surface, where quantum interference effects play an important role in the STM imaging
process. This demonstration is done by qualitative and quantitative comparisons of simulated STM images with
corresponding results obtained by Tersoff-Hamann and Bardeen tunneling methods. Quantitative comparison is
facilitated by calculating Pearson product-moment correlation coefficients between the STM datasets
\citep{Mandi-HOPGstipe,Mandi-Corr}. Importantly, we find that the revised Chen's model is 25 times faster than the Bardeen
method concerning computational time taking the same tunneling channels, while maintaining good agreement.
The effects of electronic structure, orbital interference and spatial orientation of the tip on the STM images are highlighted.
Since the detailed analysis of quantum interference effects and arbitrary tip orientations in STM junctions are presently
highly demanding and enormously time-consuming using the Bardeen method, the implementation of the revised Chen's derivative rule
in the BSKAN code \citep{Hofer03,Palotas05} is a very promising tool for more efficient STM simulations providing a deeper
understanding of a wide variety of physical phenomena in STM junctions, e.g., quantum interference and tip geometry effects.

\subsection{N-doped graphene\label{sub:Graphene-N}}

Experimental studies have shown that in the constant height STM images of N-doped graphene the tunneling current above the N atom
is significantly lower than above the neighboring C atoms \citep{Telychko-Graphene-N}. At first sight this seems to be in
contradiction with the fact that the density of states of the N atom is larger than that of the neighboring C atoms close to the
Fermi level. The current dip above the N atom has been explained by a destructive interference between the orbitals of the N and
the nearest neighbor C atoms, a pure sample effect \citep{Telychko-Graphene-N}. Such a quantum interference effect is an ideal
candidate to study with the revised Chen's method.

We have calculated constant height STM images of the N-doped graphene surface using four different tunneling models: 3D-WKB,
Tersoff-Hamann, revised Chen and Bardeen. The constant height STM simulations were performed at relatively small tip-sample
distance (4 \AA) at two selected bias voltages ($\pm 0.4$ V) corresponding to the STM experiments by Telychko
{\it{et al.}} \citep{Telychko-Graphene-N}. First, the 3D-WKB method \citep{Palotas14-3DWKB} has been used. This model takes into
account the orbital characteristics and electronic structure of the sample and the tip as well, but uses the atom-projected
density of states (amplitudes) instead of the explicit wavefunctions (amplitudes and phases), thus electron interference effects
are not considered. Using the 3D-WKB method the N atom always shows up as a protrusion in the STM image as it is expected from
the relation of the density of states of the N and C atoms \citep{Telychko-Graphene-N}.

\begin{figure}[h]
\begin{centering}
\begin{tabular}{|c|c|c|c|c|}
\cline{2-5}
\multicolumn{1}{c|}{} & Tersoff-Hamann & \multicolumn{3}{c|}{Revised Chen} \tabularnewline
\cline{3-5}
\multicolumn{1}{c|}{} & & $s$ tip & $p_z$ tip & $(s+p_z)/\sqrt{2}$ tip \tabularnewline
\hline
\begin{turn}{90}
$-0.4$ V
\end{turn} & \includegraphics[width=0.15\textwidth]{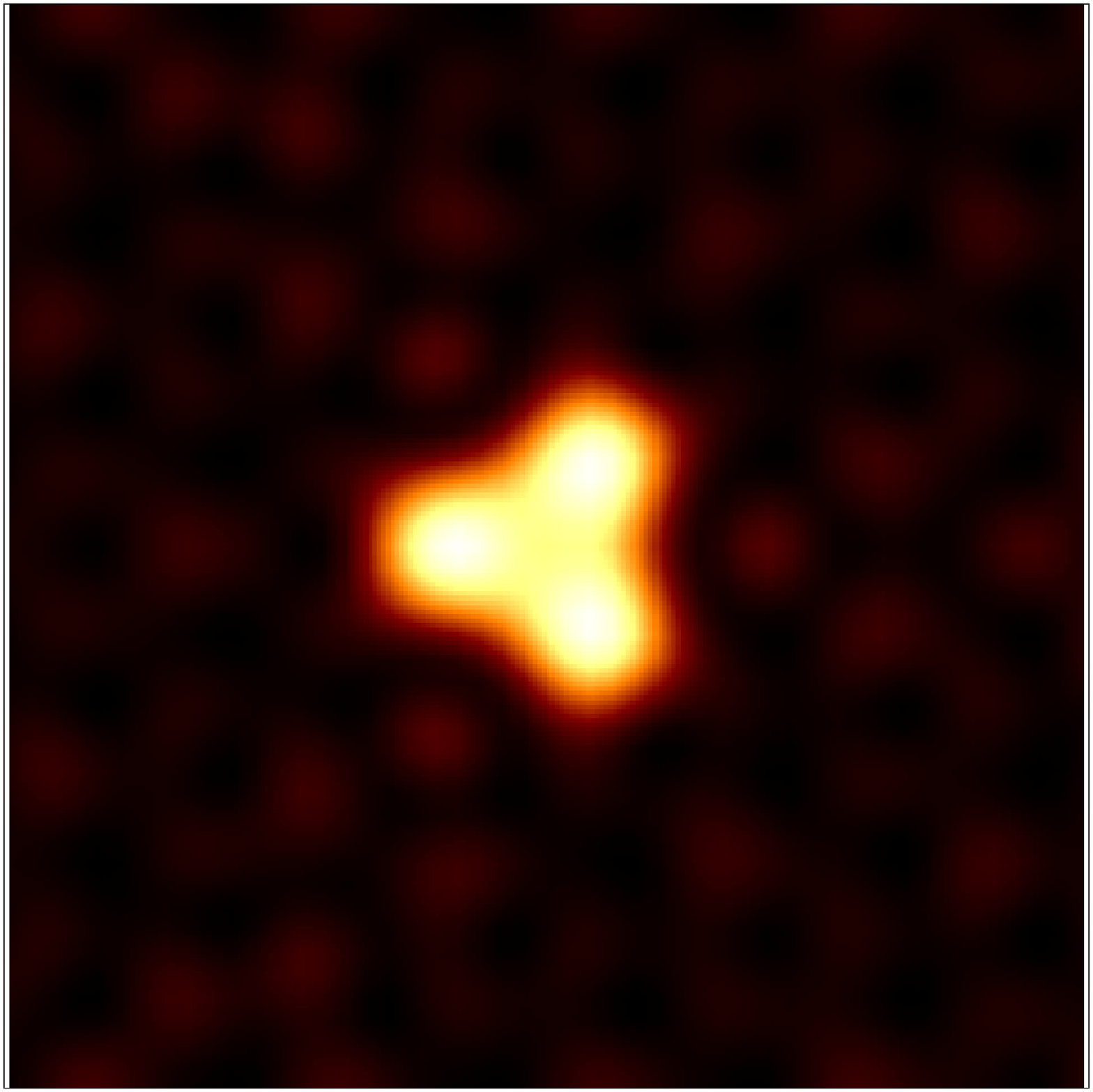} & \includegraphics[width=0.15\textwidth]{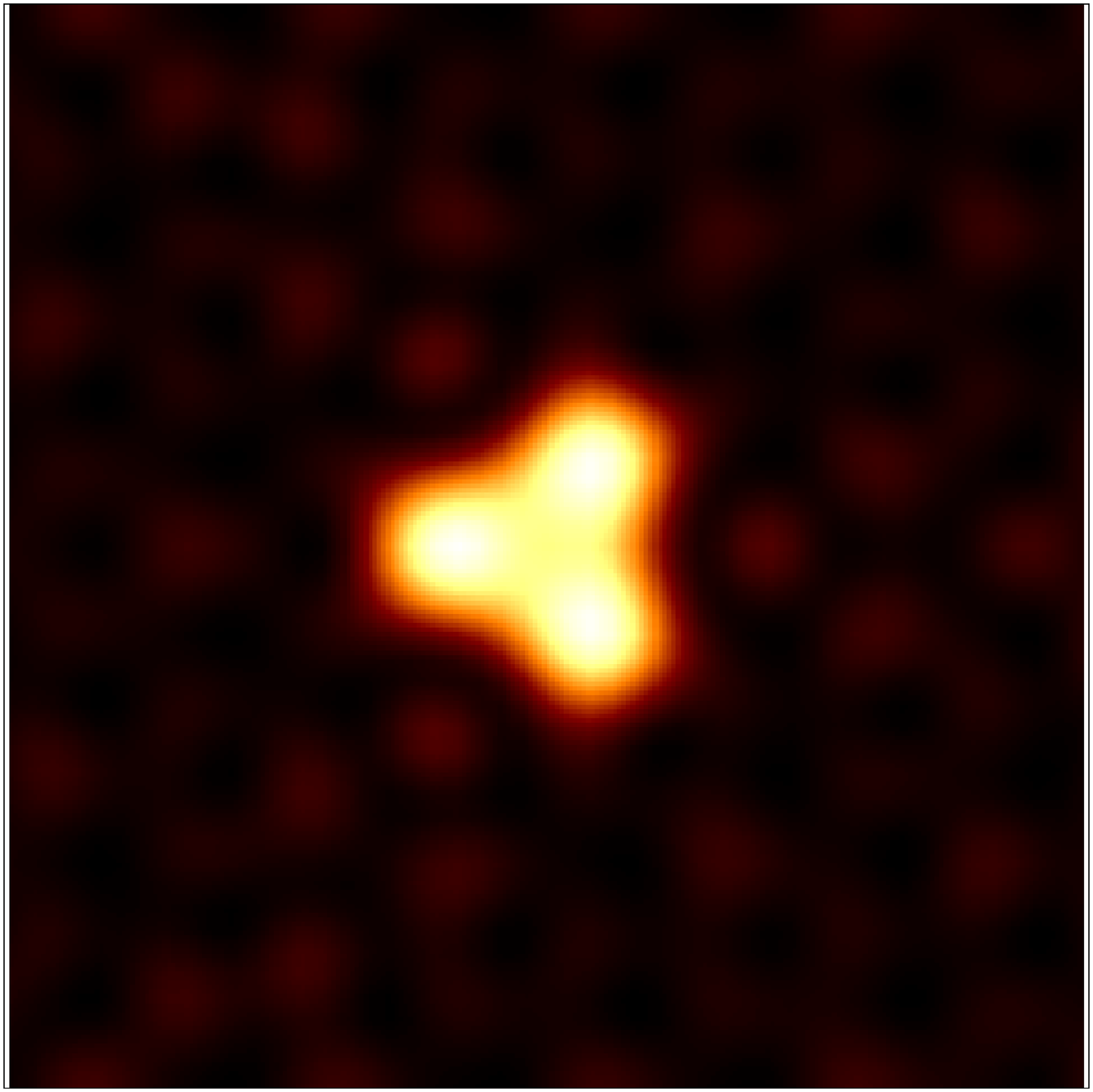} & \includegraphics[width=0.15\textwidth]{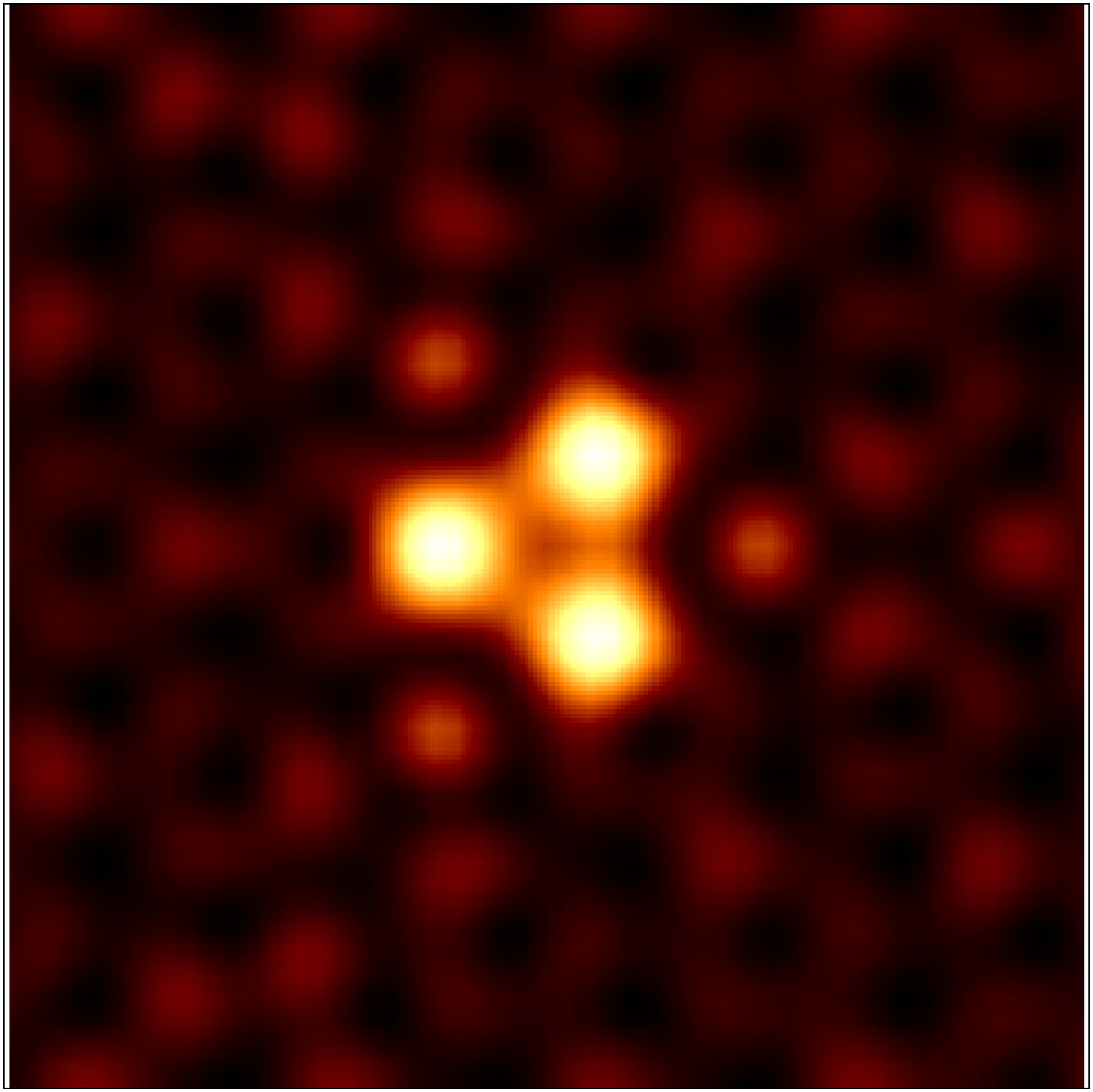} & \includegraphics[width=0.15\textwidth]{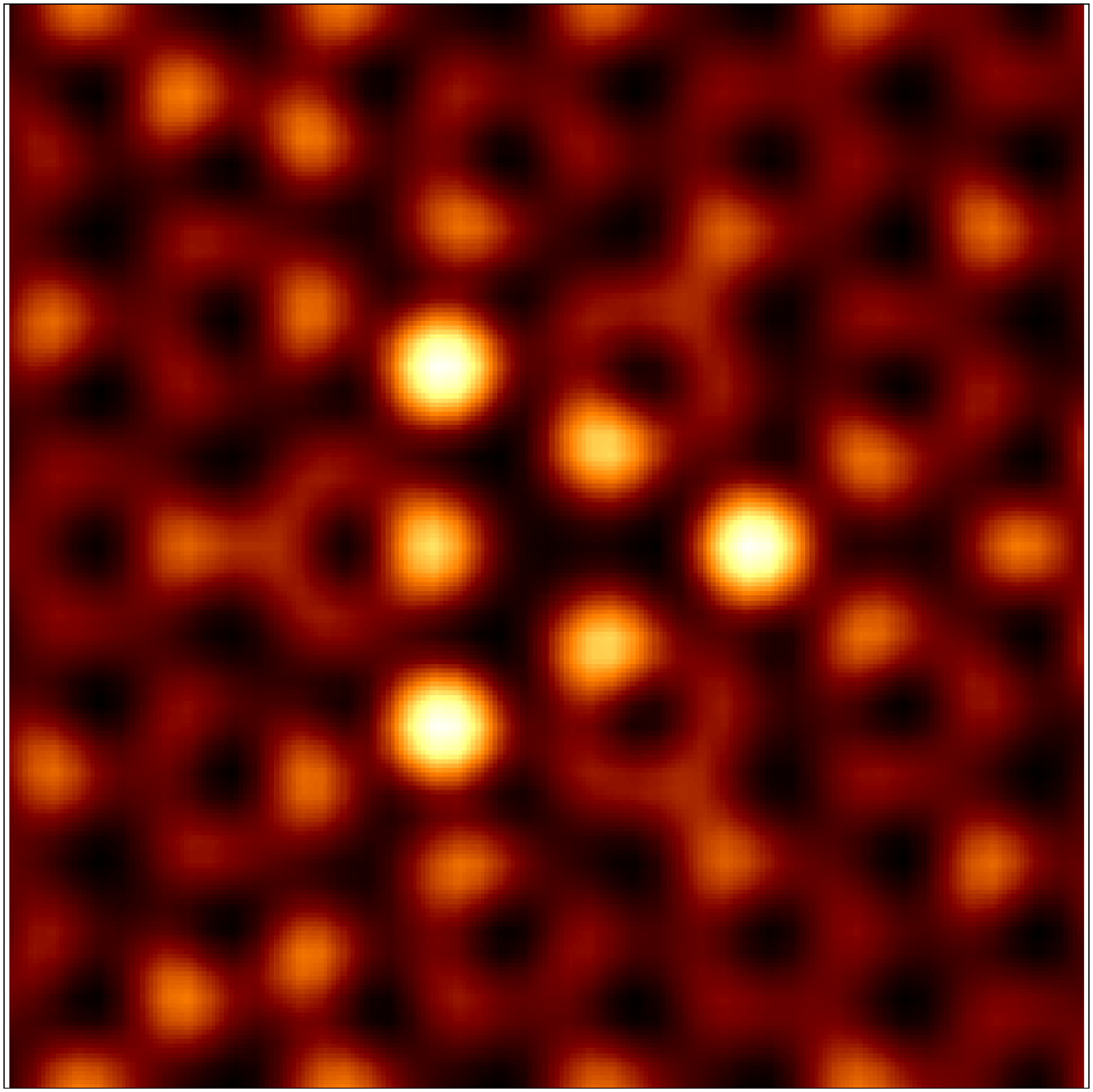} \tabularnewline
\hline
\begin{turn}{90}
$+0.4$ V
\end{turn} & \multicolumn{1}{c}{\includegraphics[width=0.15\textwidth]{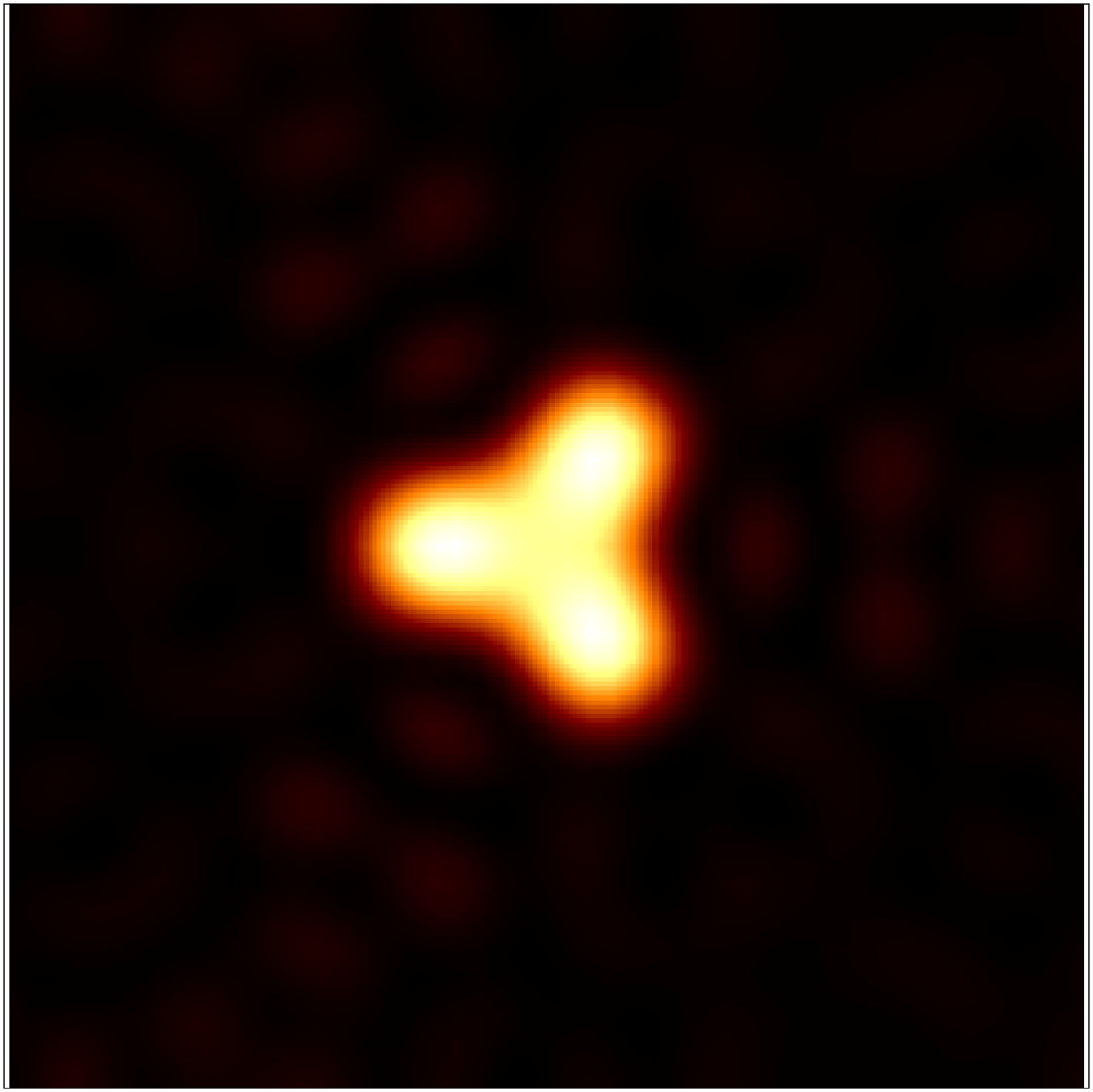}} & \includegraphics[width=0.15\textwidth]{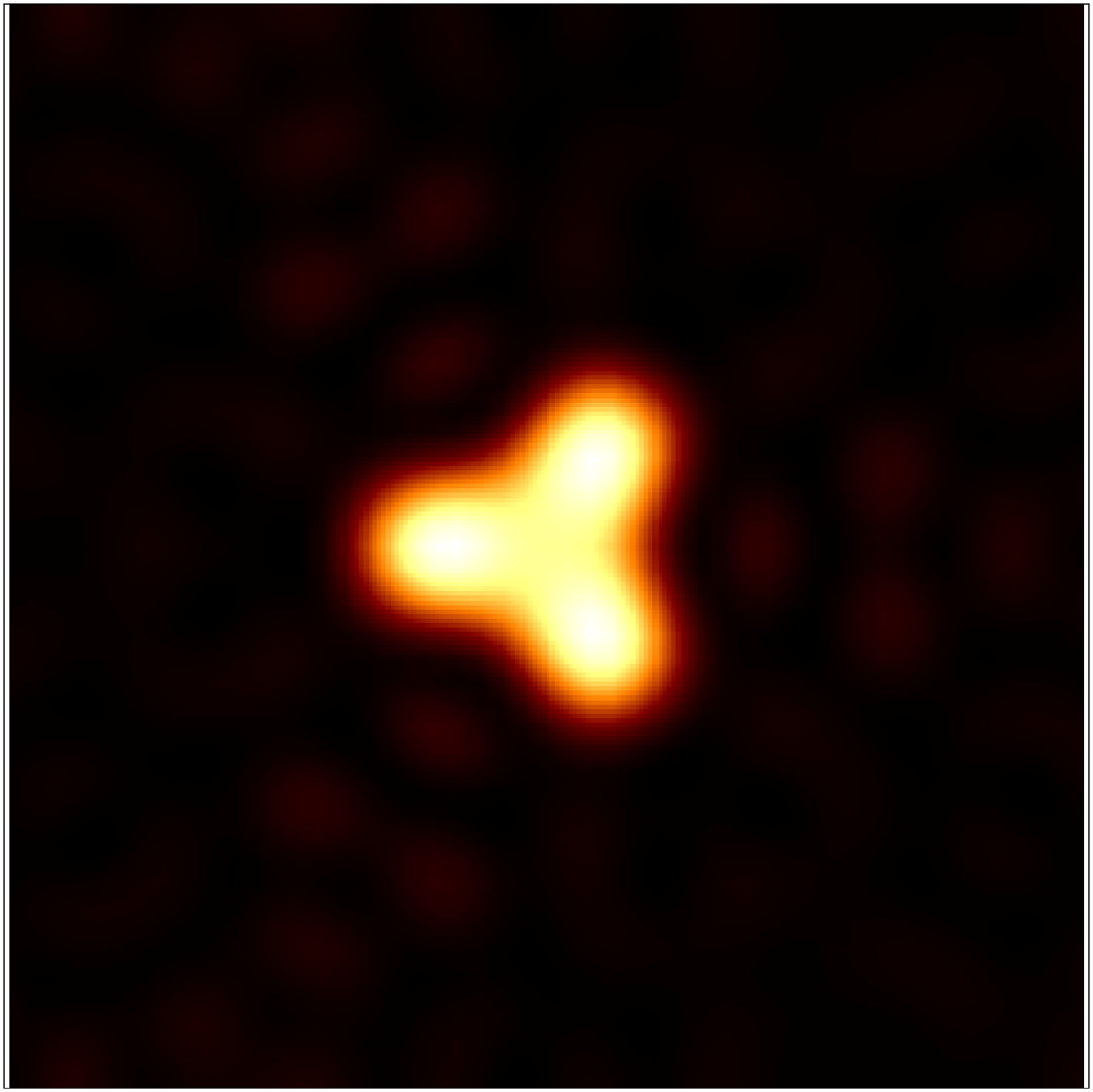} & \includegraphics[width=0.15\textwidth]{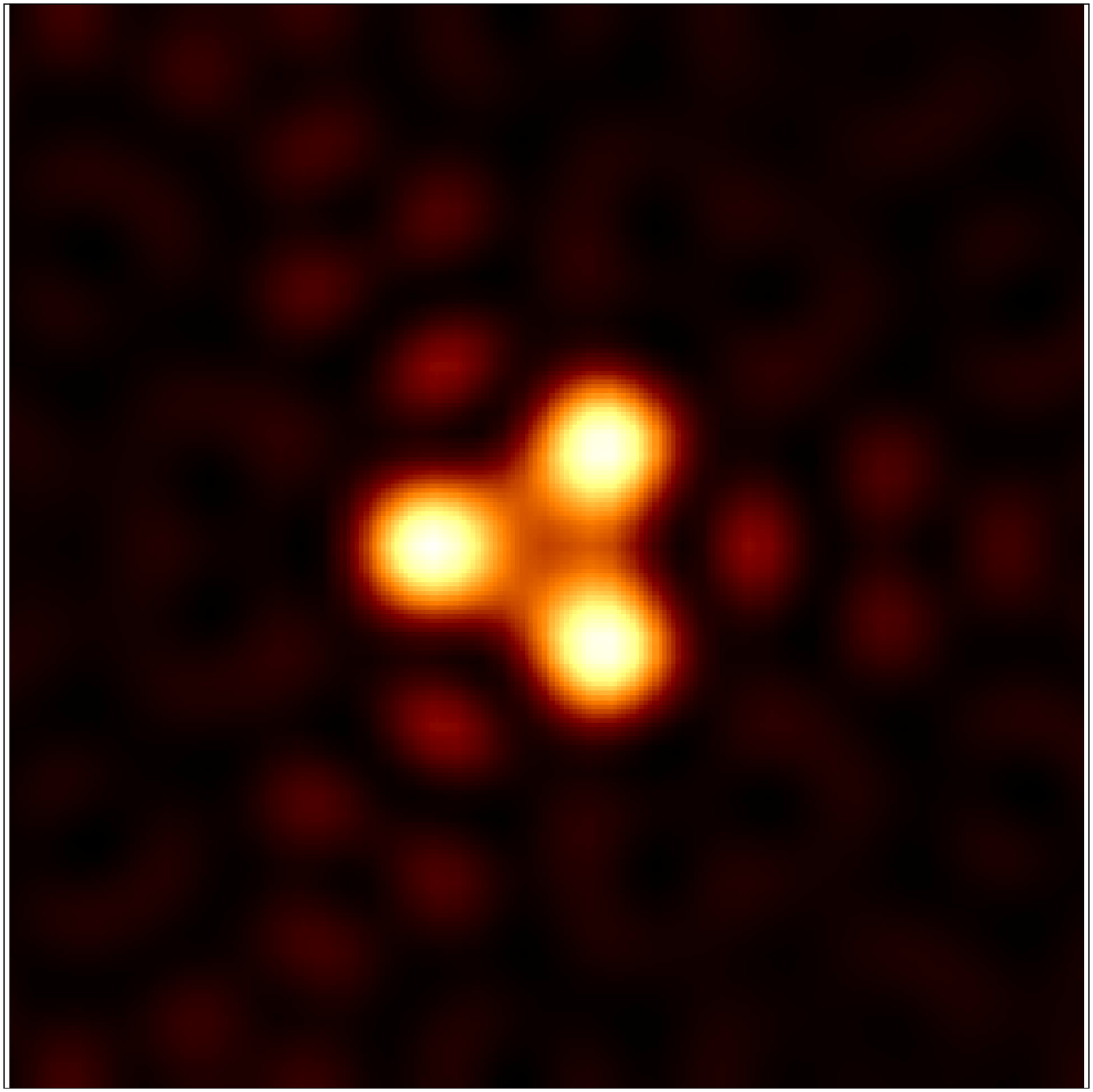} & \includegraphics[width=0.15\textwidth]{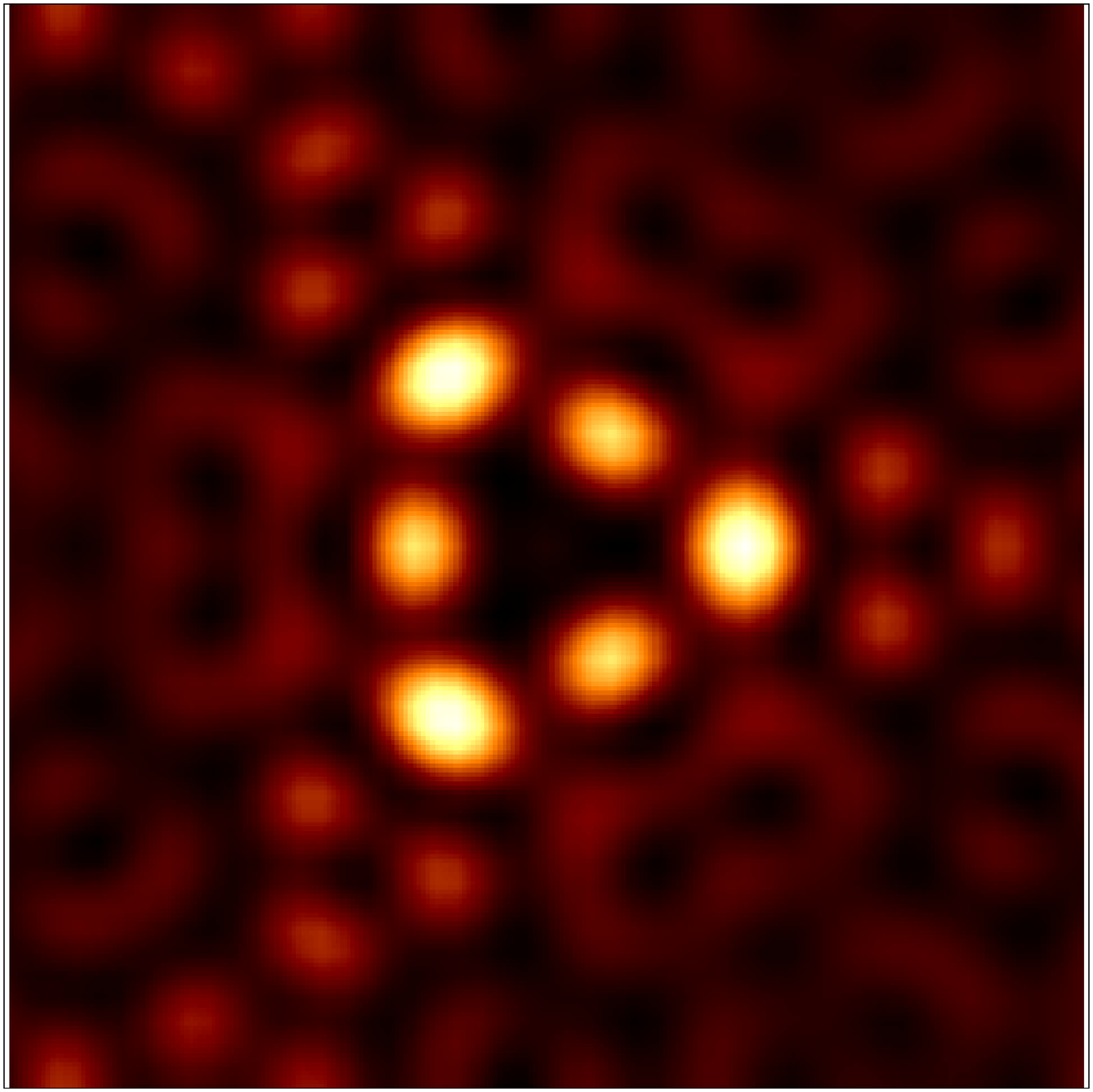} \tabularnewline
\hline
\end{tabular}
\par\end{centering}
\protect\caption{Constant height STM images of N-doped graphene at 4 $\textrm{\AA}$ tip-sample distance and $\pm 0.4$ V bias.
Comparison between Tersoff-Hamann model and revised Chen's method with selected tip orbitals: $s$, $p_z$, $(s+p_z)/\sqrt{2}$.
The N defect is located in the middle of the images.
\label{fig:Comparison-Graphene-TH}}
\end{figure}

Next, we focus on the comparison of the revised Chen's method with two conventional STM simulation models: Tersoff-Hamann and
Bardeen. Fig.\ \ref{fig:Comparison-Graphene-TH} shows that the STM images obtained by the revised Chen's method using a pure $s$
tip quantitatively agree with those calculated by the Tersoff-Hamann model, i.e., a correlation value of 1
between the corresponding STM images is found. We obtain qualitatively similar STM images assuming
an ideal tip of pure $p_z$ orbital, where the current dip above the N atom is slightly more pronounced than with the $s$ tip. For
these tip models the energy independent weighting factors in section \ref{sub:Coeff} (i) were used. Furthermore, we point out the
importance of quantum interference of tip orbitals in Fig.\ \ref{fig:Comparison-Graphene-TH}. Therefore, we consider an ideal tip
of a linear combination of $s$ and $p_z$ orbitals of equal weights. As can be seen, the $(s+p_z)/\sqrt{2}$ tip shows a remarkable
contrast change compared to the STM images of pure $s$ or pure $p_z$ orbitals, where the current dip above the N atom is even
more pronounced and the bright triangle showing C atoms is larger and reversed. This effect is clearly related to the orbital
interference of the tip's $s$ and $p_z$ states, and shows an additional effect to the destructive quantum interference arising
from the sample's C-N bond in the formation of the STM contrast of N-doped graphene. We stress again that the $s-p_z$ tip orbital
interference results in a much more pronounced current dip above the N atom than the destructive quantum interference of the
sample itself, the latter is imaged by the Tersoff-Hamann method. Interestingly, STM images obtained by the $(s+p_z)/\sqrt{2}$ tip
resemble results calculated by a C(111) tip model (see Ref.\ \citep{Telychko-Graphene-N})
having these dominant orbitals in the electronic structure. Note that both types of STM contrast of
N-doped graphene in Fig.\ \ref{fig:Comparison-Graphene-TH} have been experimentally observed in Ref.\ \citep{Telychko-Graphene-N}.

\begin{figure}[h]
\begin{centering}
\begin{tabular}{|cc|cccccc|}
\cline{3-8} 
\multicolumn{2}{c|}{} & \multicolumn{2}{c|}{$\mathrm{W_{blunt}}$ tip} & \multicolumn{2}{c|}{$\mathrm{W_{sharp}}$ tip} & \multicolumn{2}{c|}{$\mathrm{W_{C-apex}}$ tip}\tabularnewline
\cline{3-8} 
\multicolumn{2}{c|}{} & \multicolumn{1}{c|}{$-0.4$ V} & \multicolumn{1}{c|}{$+0.4$ V} & \multicolumn{1}{c|}{$-0.4$ V} & \multicolumn{1}{c|}{$+0.4$ V} & \multicolumn{1}{c|}{$-0.4$ V} & $+0.4$ V\tabularnewline
\hline 
\begin{turn}{90}
Bardeen
\end{turn} & & \includegraphics[width=0.15\columnwidth]{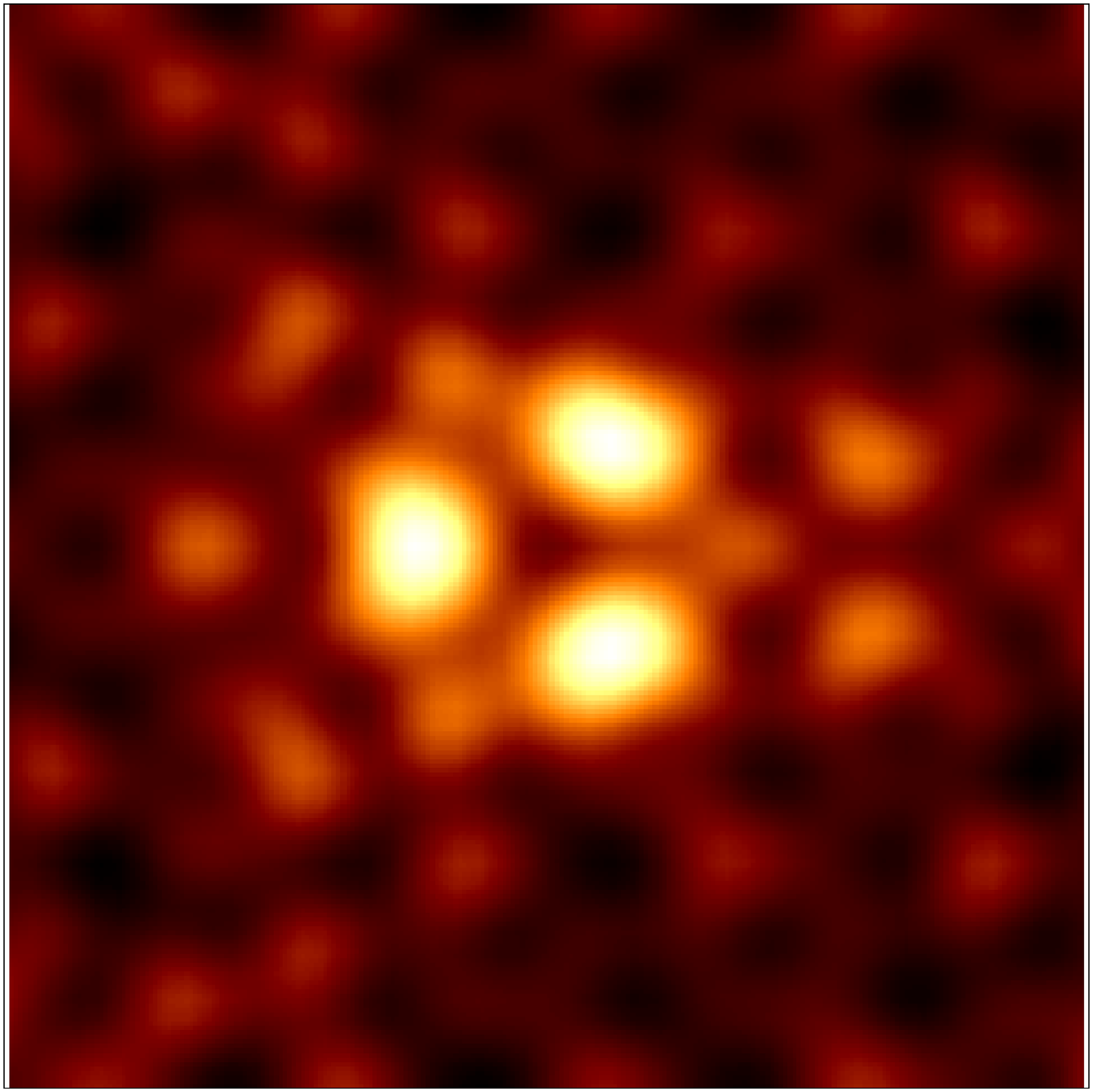} & \includegraphics[width=0.15\textwidth]{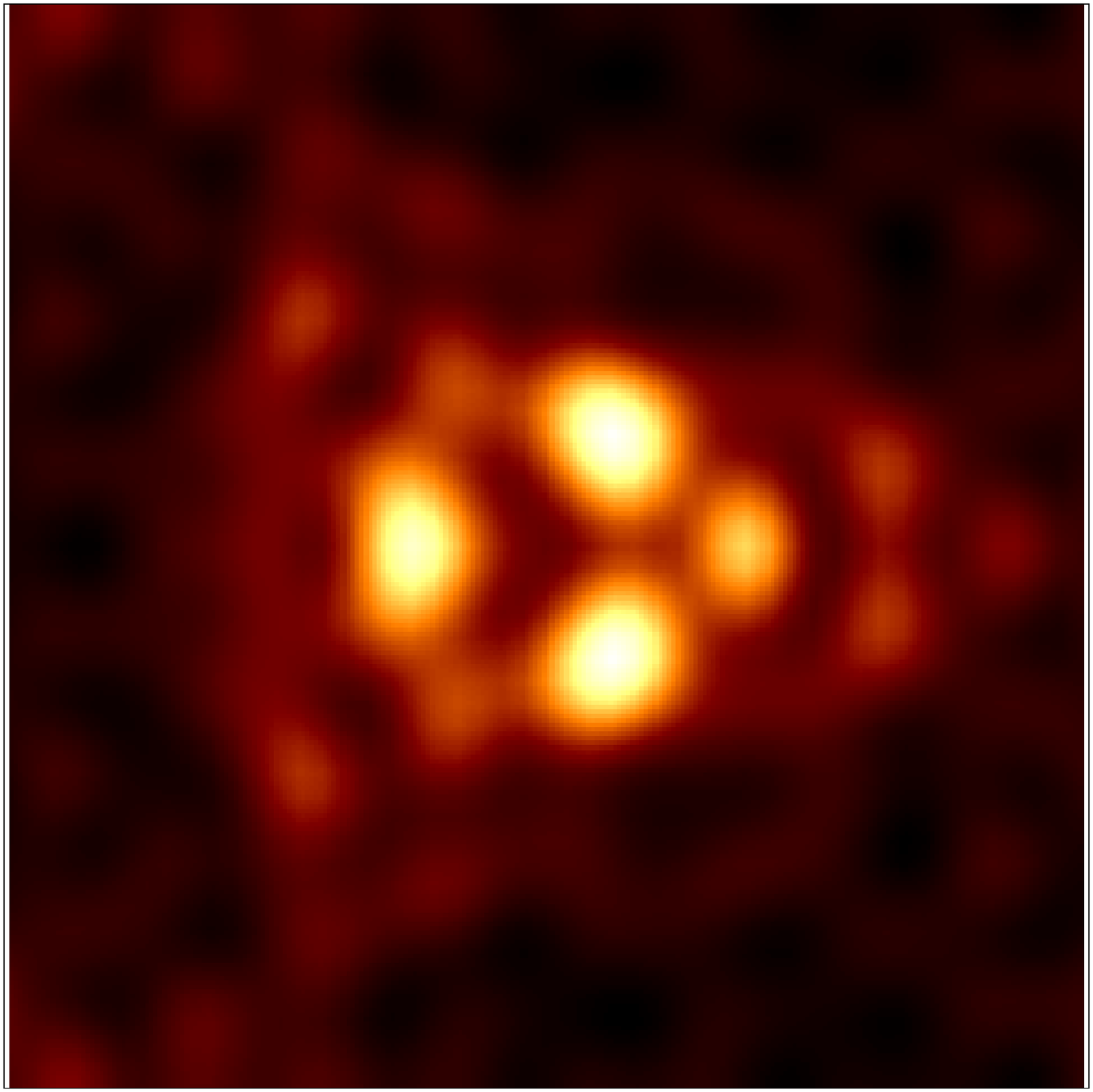} & \includegraphics[width=0.15\textwidth]{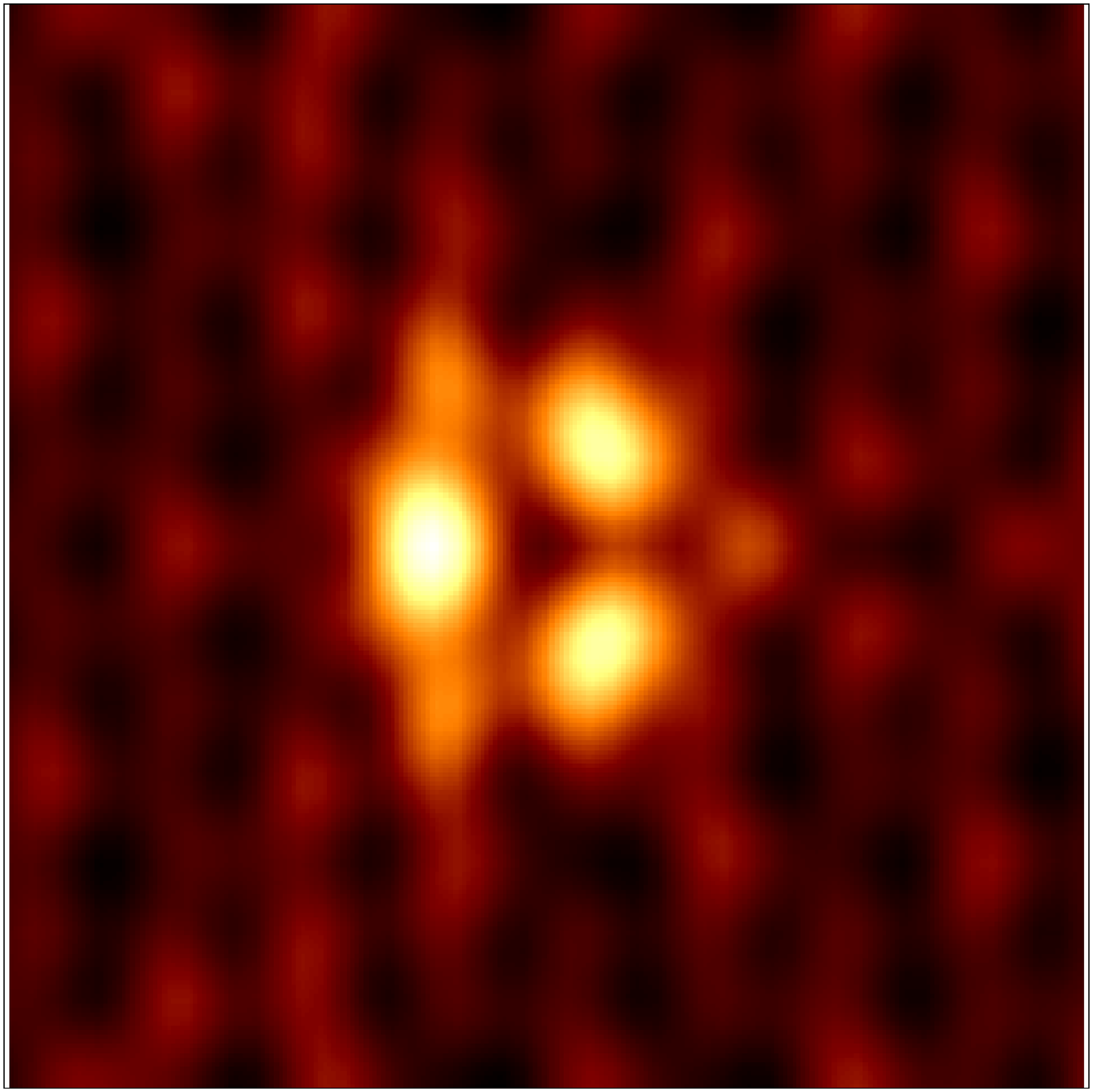} & \includegraphics[width=0.15\textwidth]{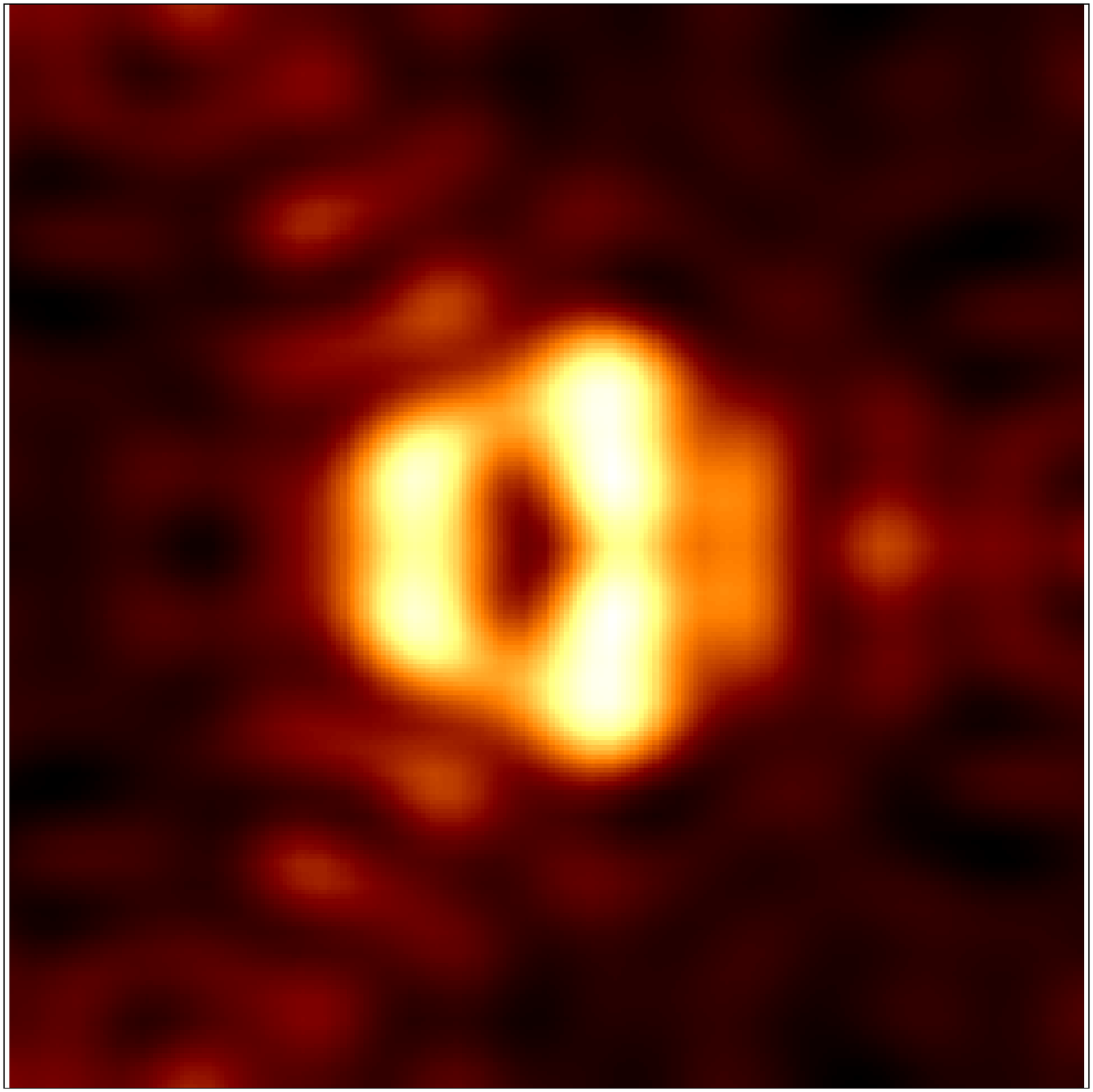} & \includegraphics[width=0.15\textwidth]{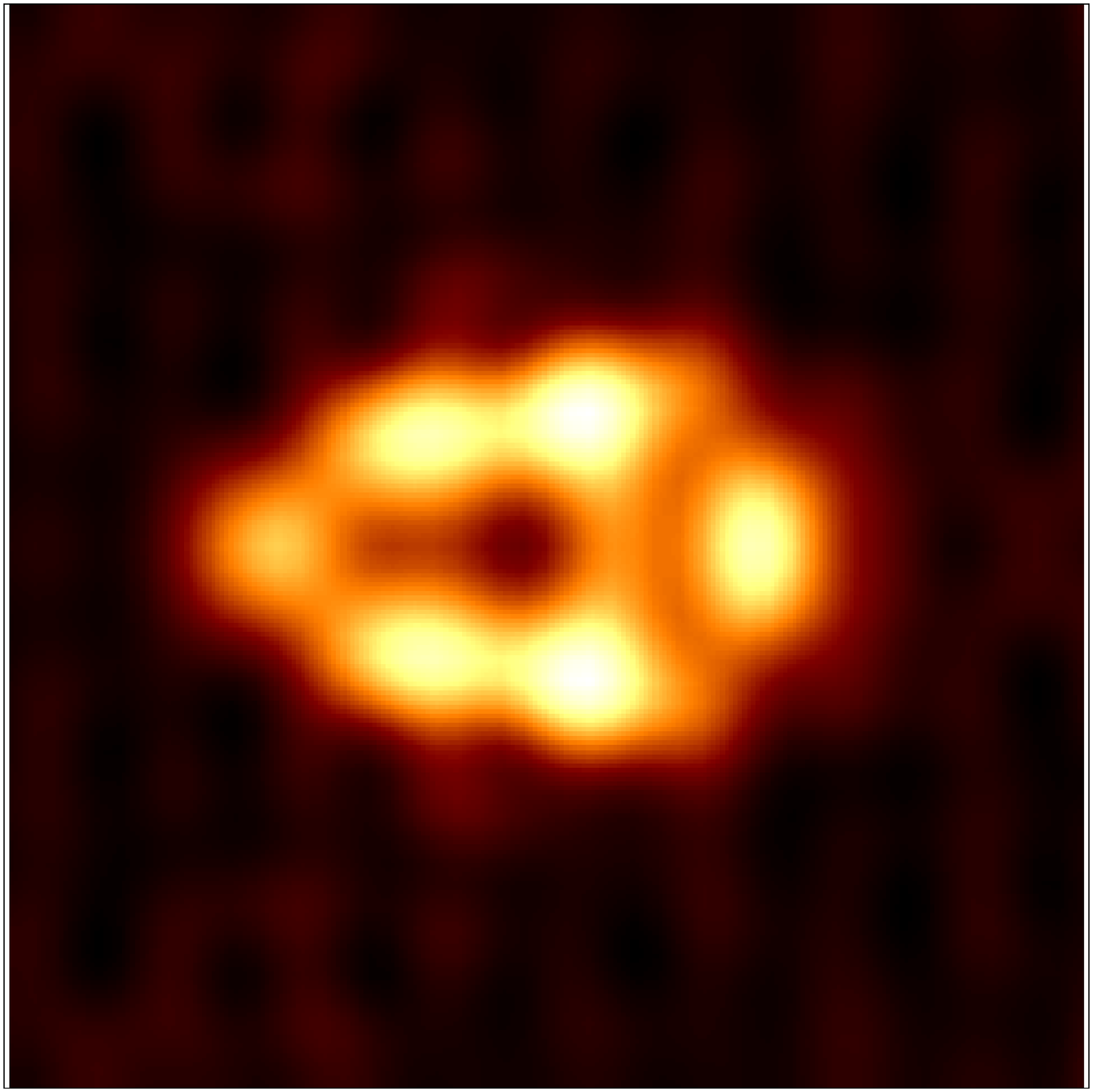} & \includegraphics[width=0.15\textwidth]{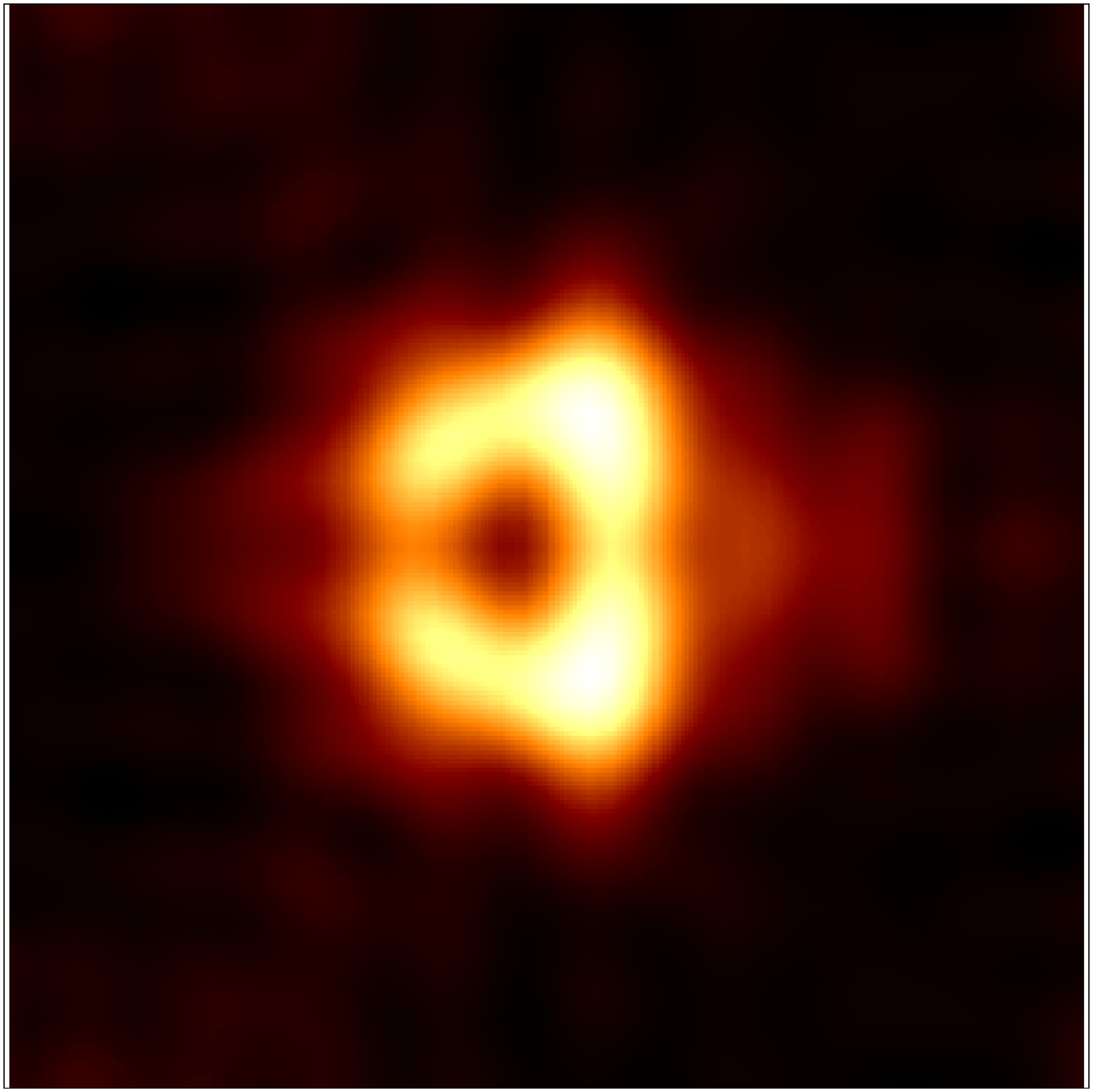}\tabularnewline
\cline{1-2} 
\begin{turn}{90}
Revised Chen
\end{turn} &
\begin{turn}{90}
$C_{\nu\beta}$ in Eq.(\ref{eq:coeff})
\end{turn} & \includegraphics[width=0.15\textwidth]{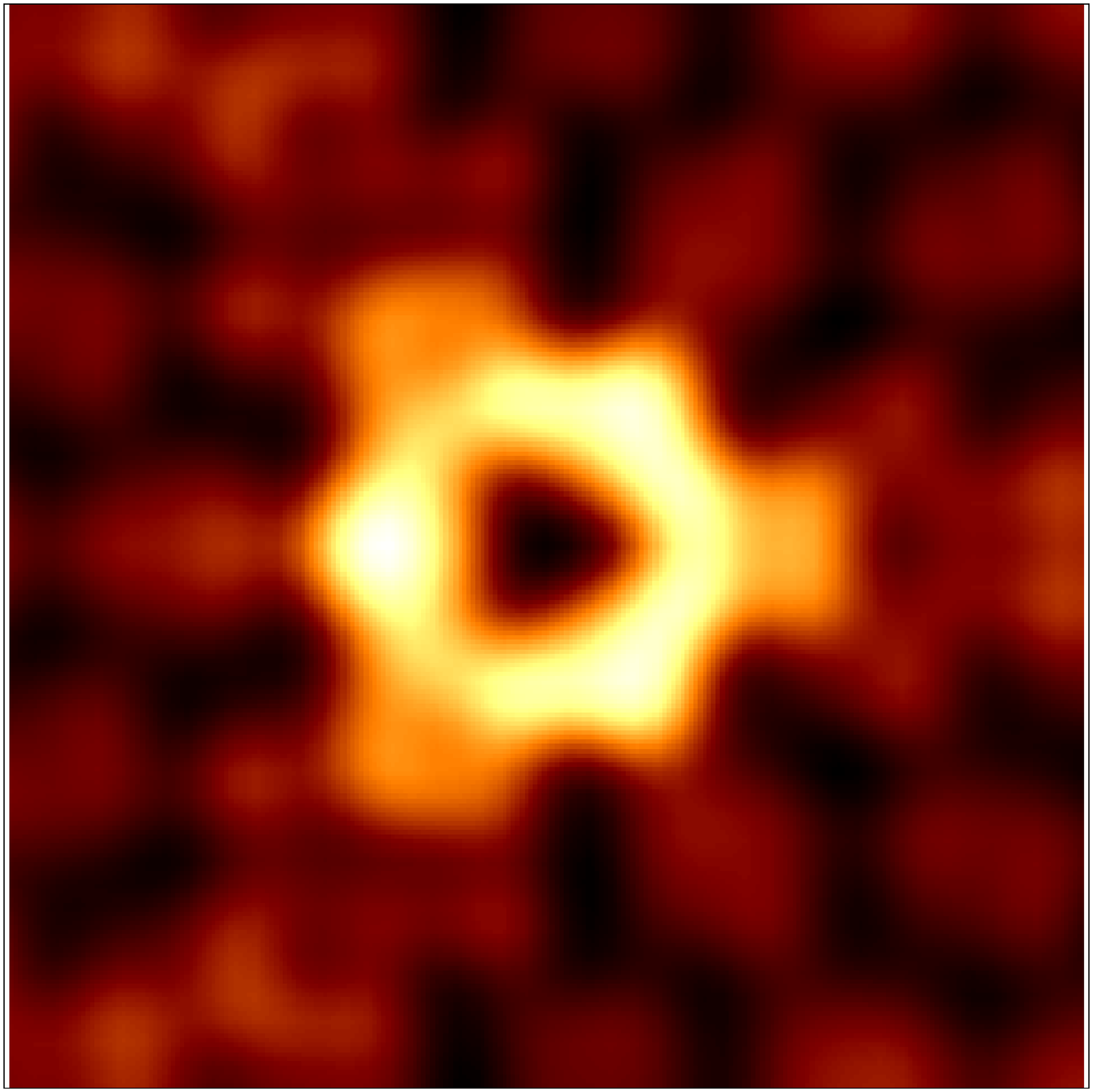} & \includegraphics[width=0.15\textwidth]{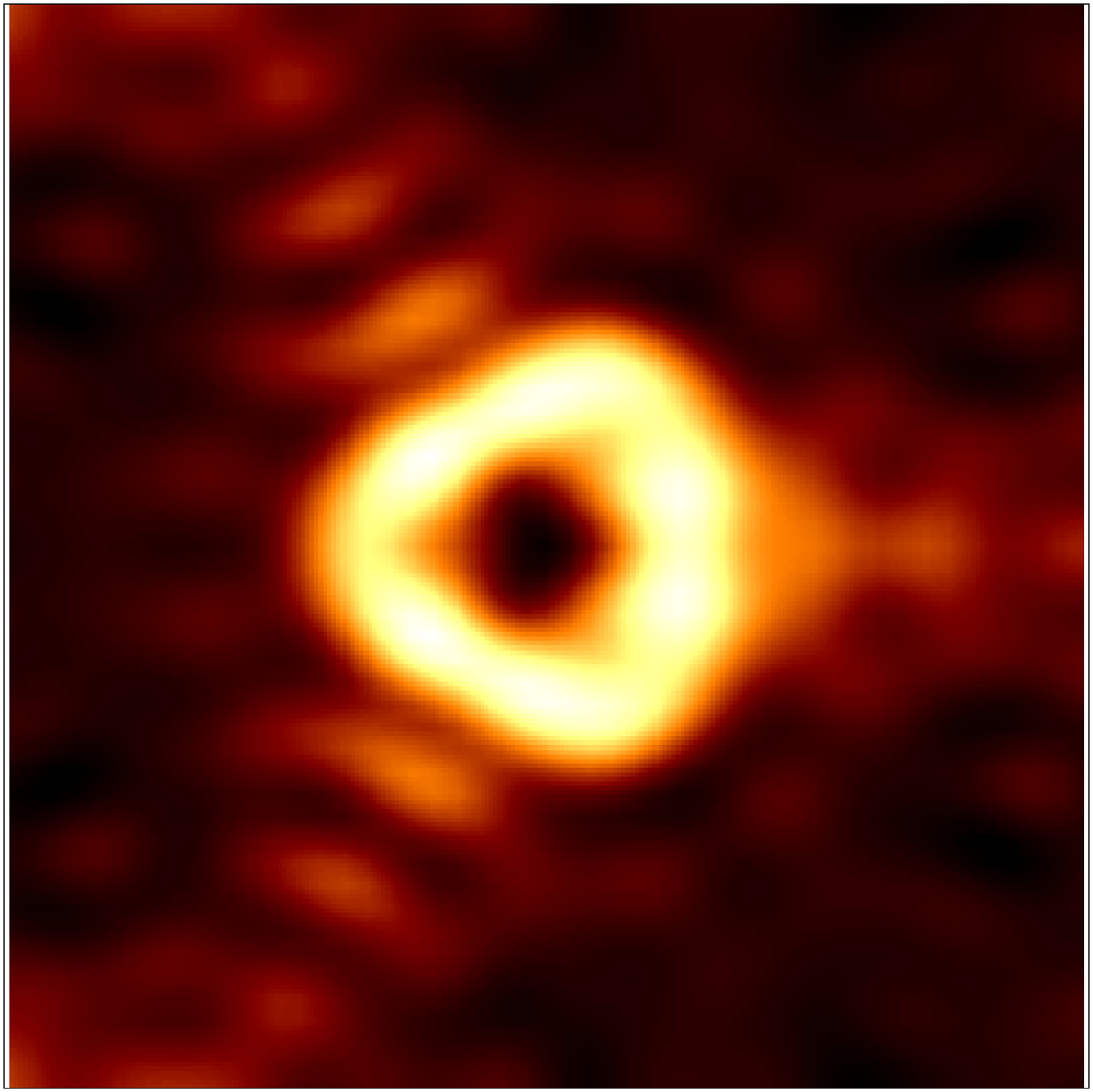} & \includegraphics[width=0.15\textwidth]{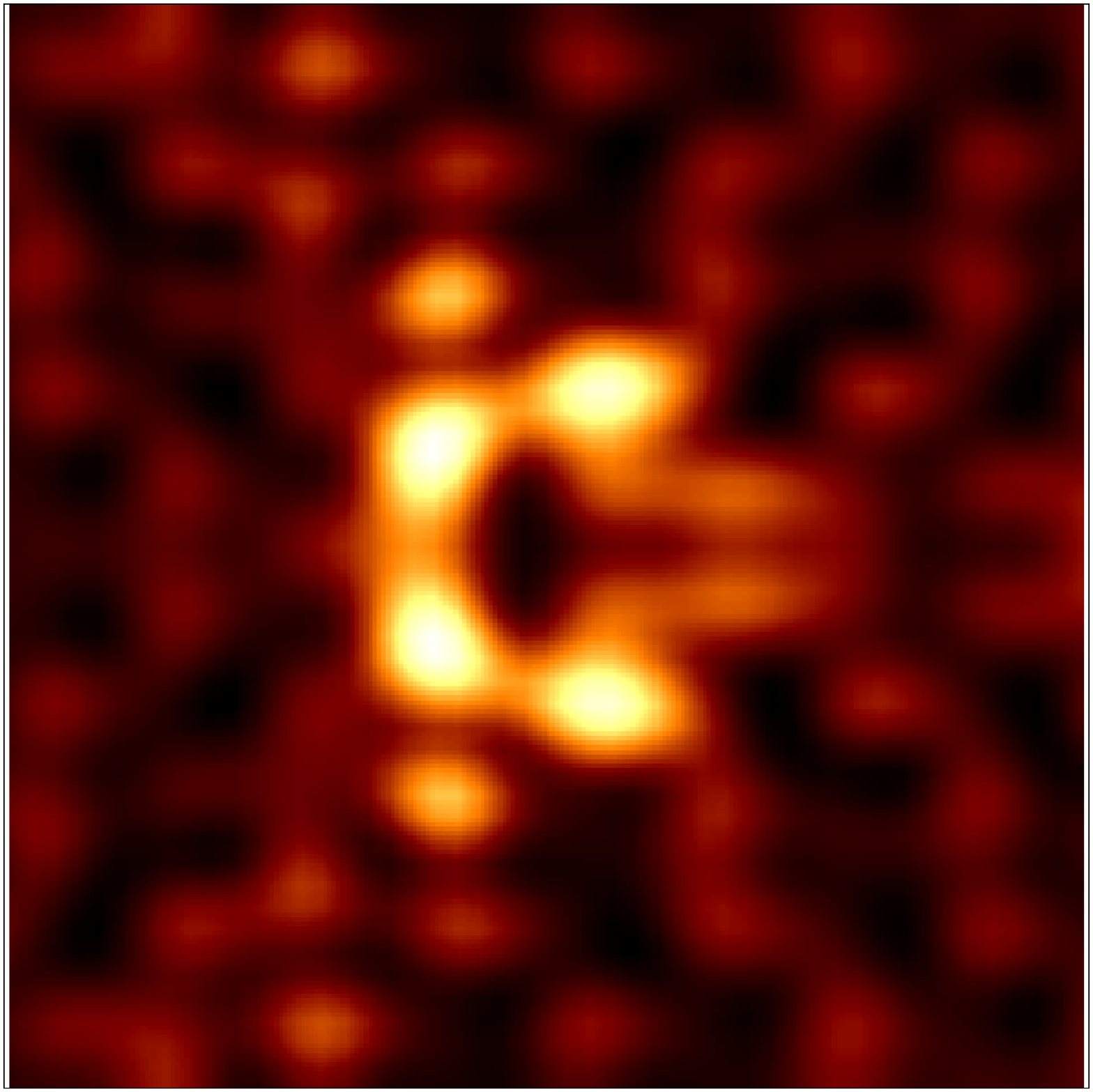} & \includegraphics[width=0.15\textwidth]{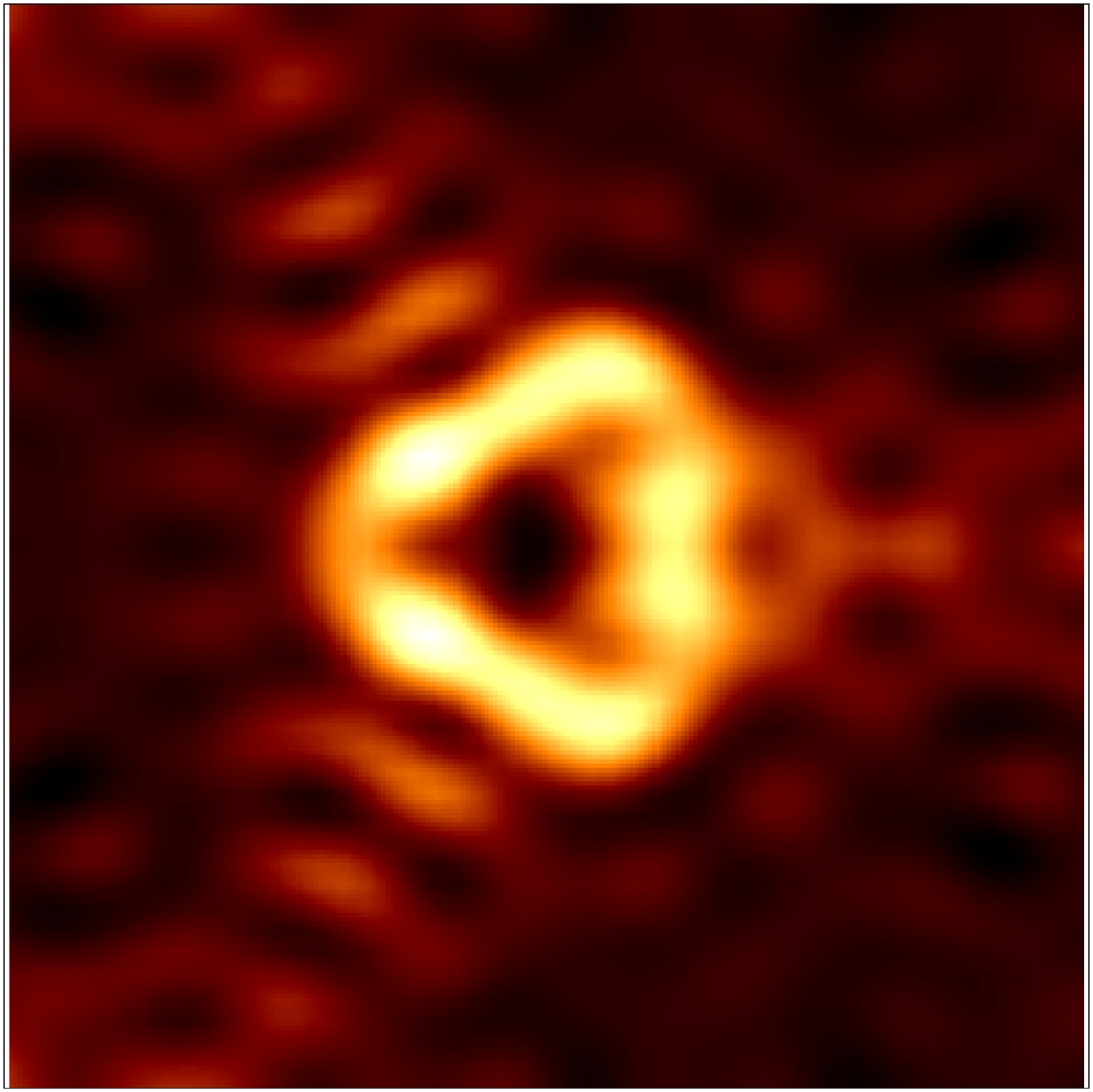} & \includegraphics[width=0.15\textwidth]{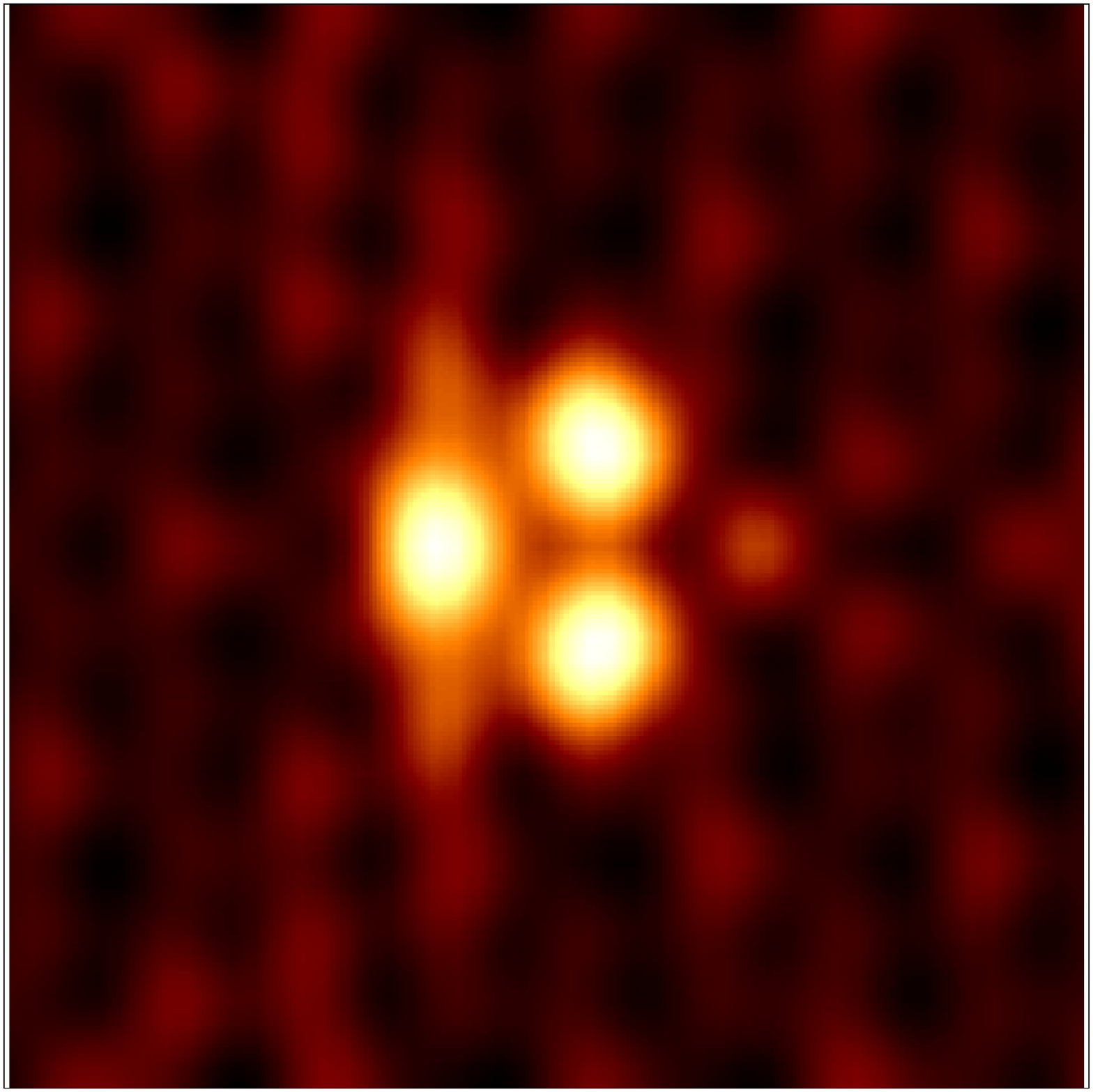} & \includegraphics[width=0.15\textwidth]{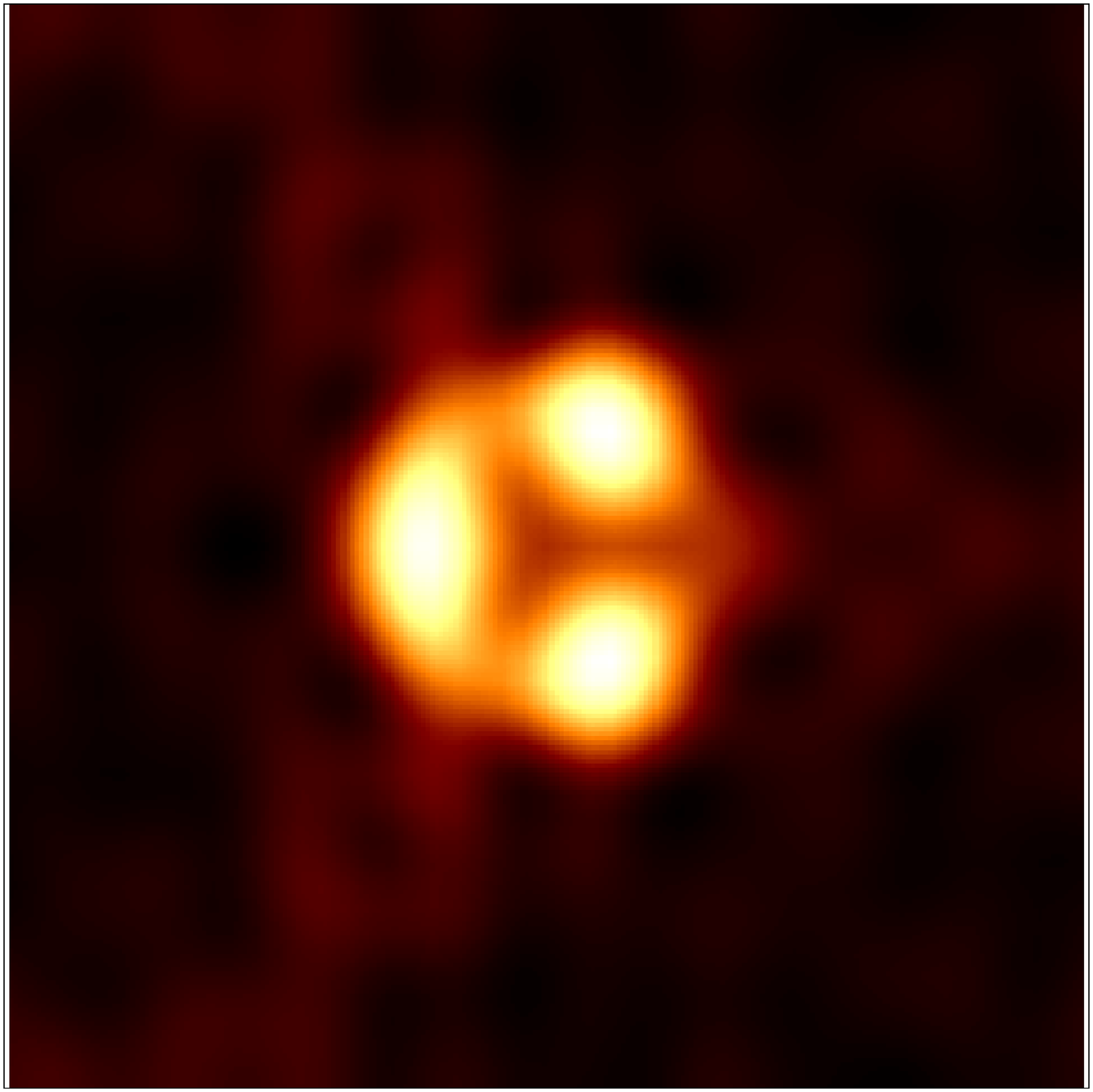}\tabularnewline
\cline{1-2} 
\begin{turn}{90}
Revised Chen
\end{turn} &
\begin{turn}{90}
$C_{\nu\beta}\approx\sqrt{n^{TIP}_{\nu\beta}}$
\end{turn} & \includegraphics[width=0.15\textwidth]{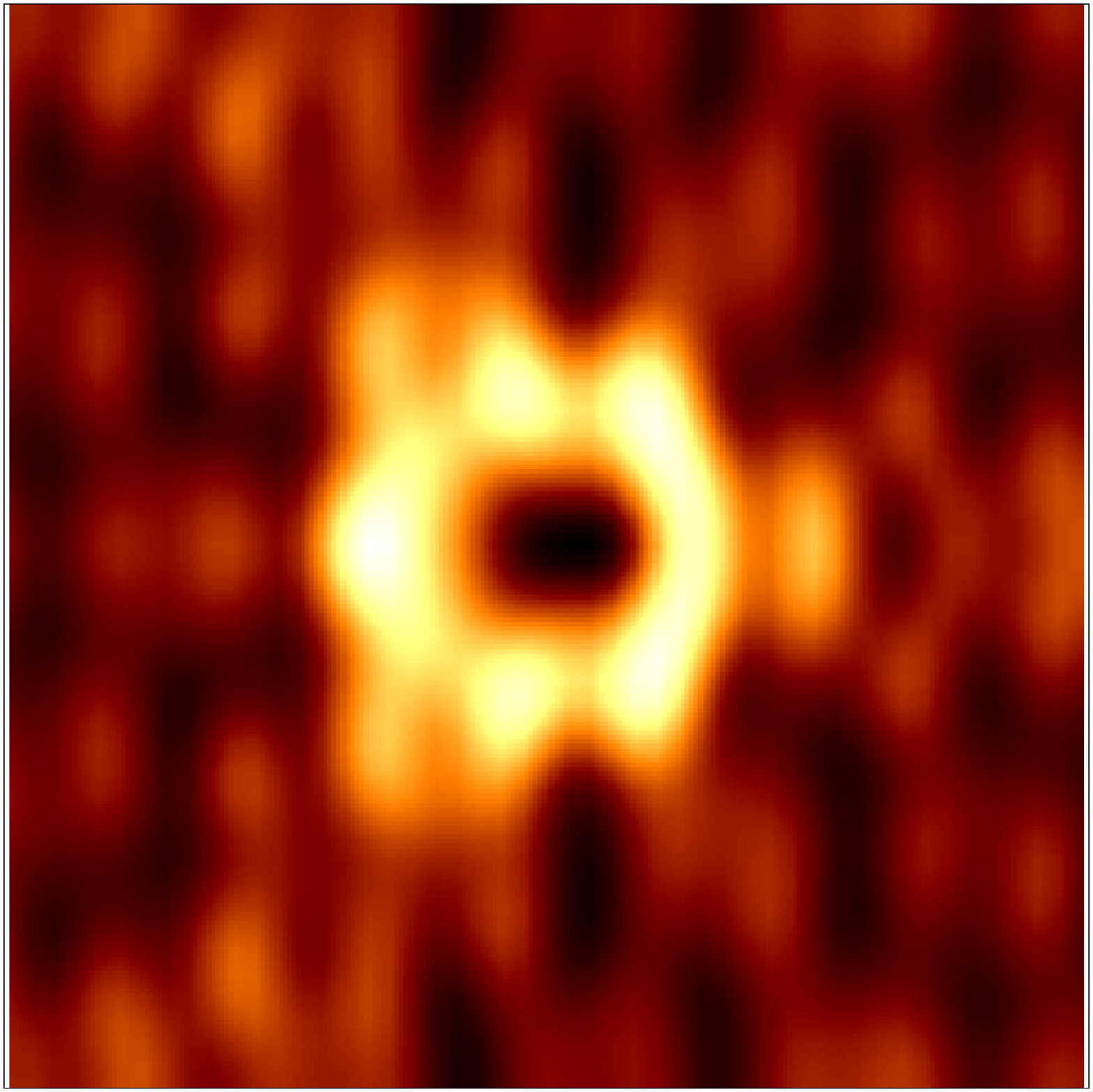} & \includegraphics[width=0.15\textwidth]{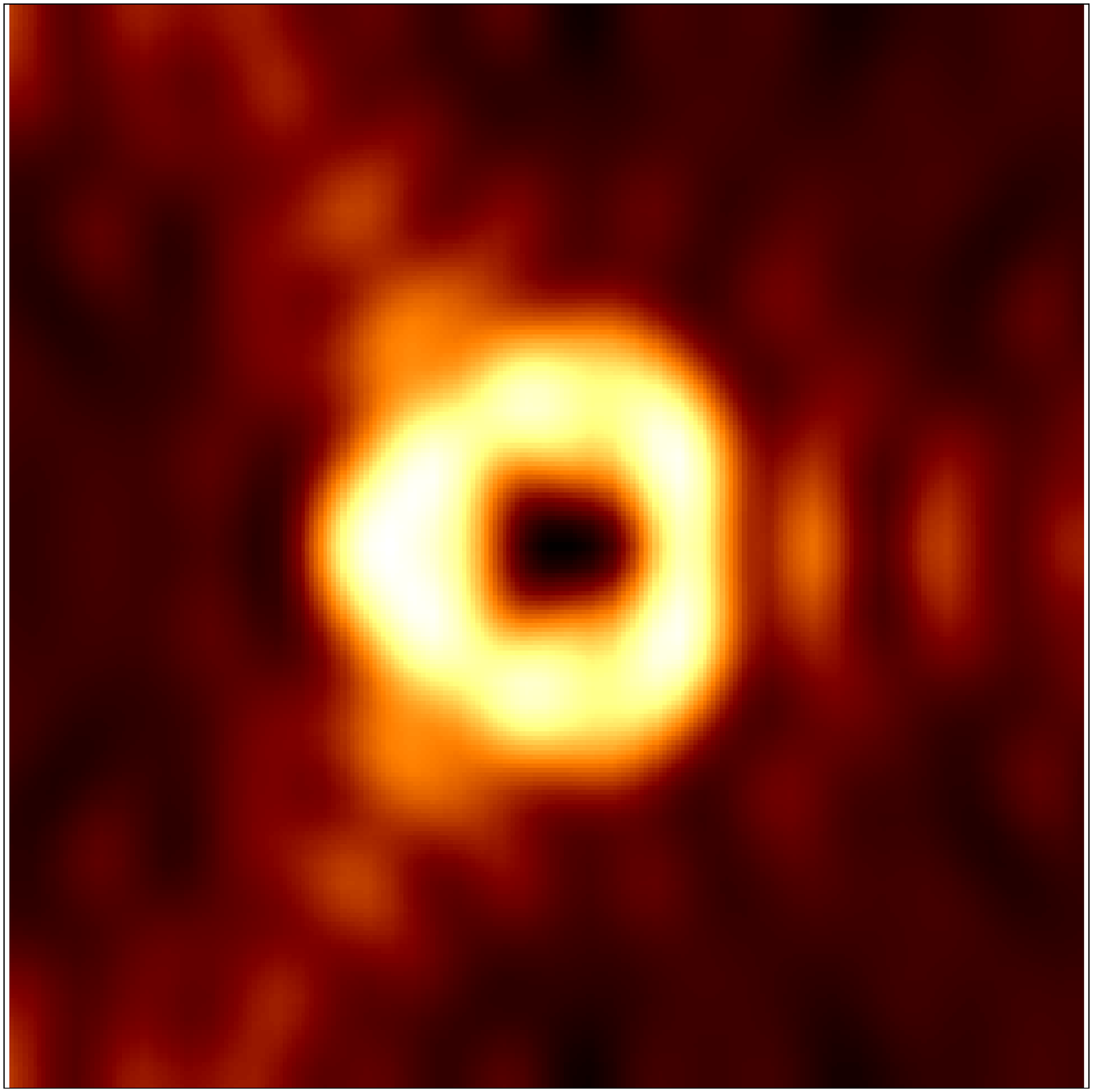} & \includegraphics[width=0.15\textwidth]{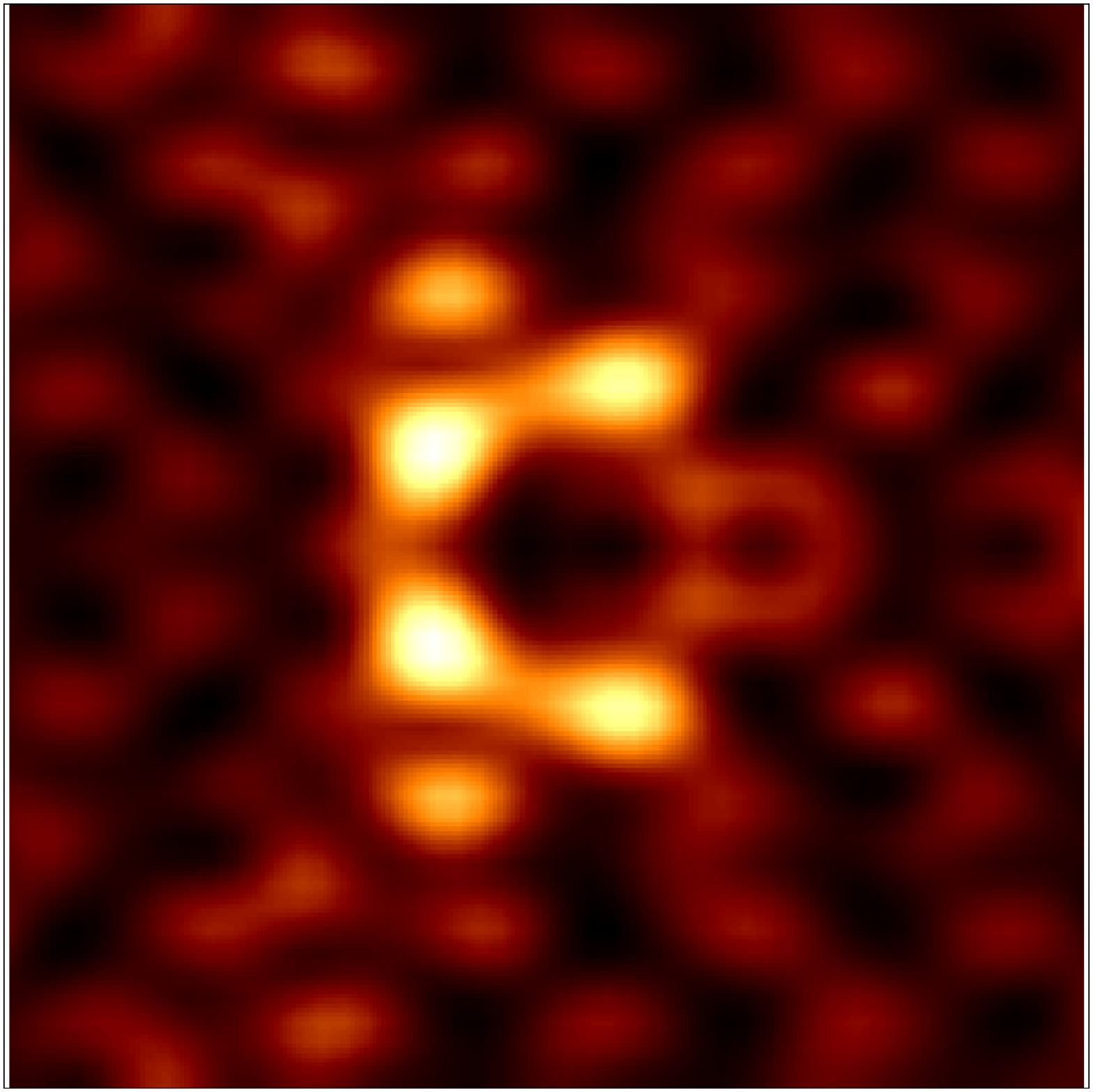} & \includegraphics[width=0.15\textwidth]{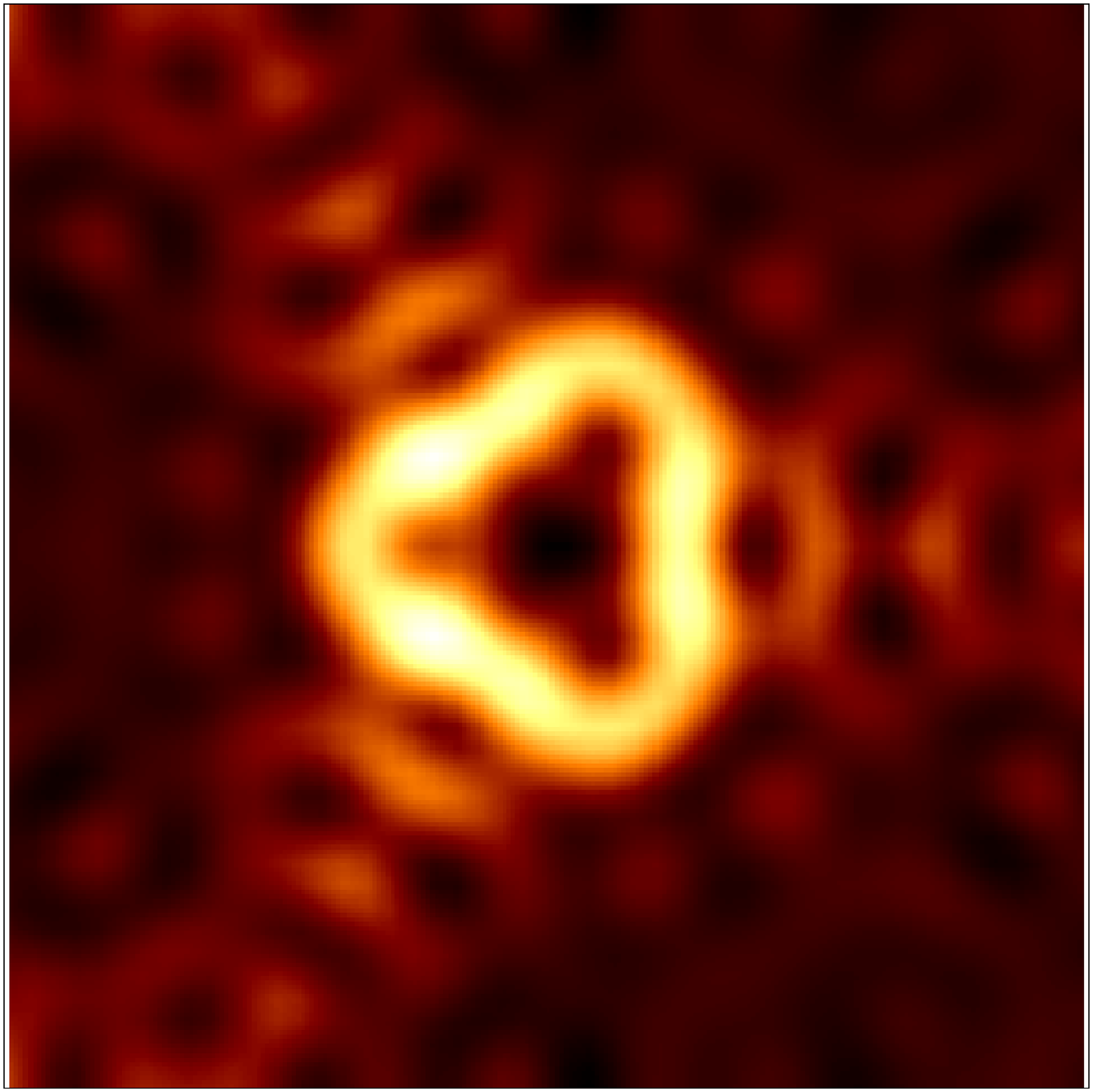} & \includegraphics[width=0.15\textwidth]{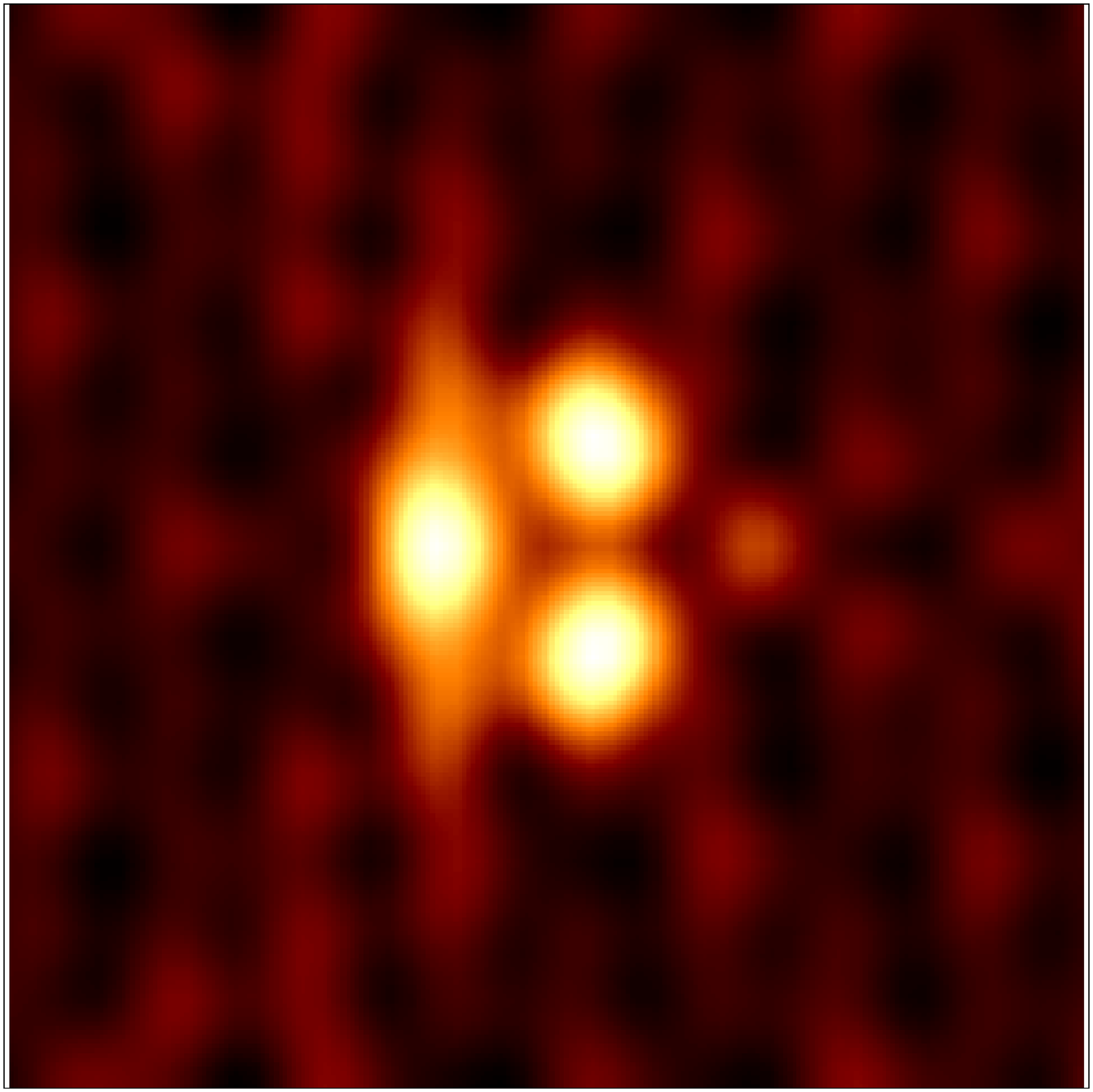} & \includegraphics[width=0.15\textwidth]{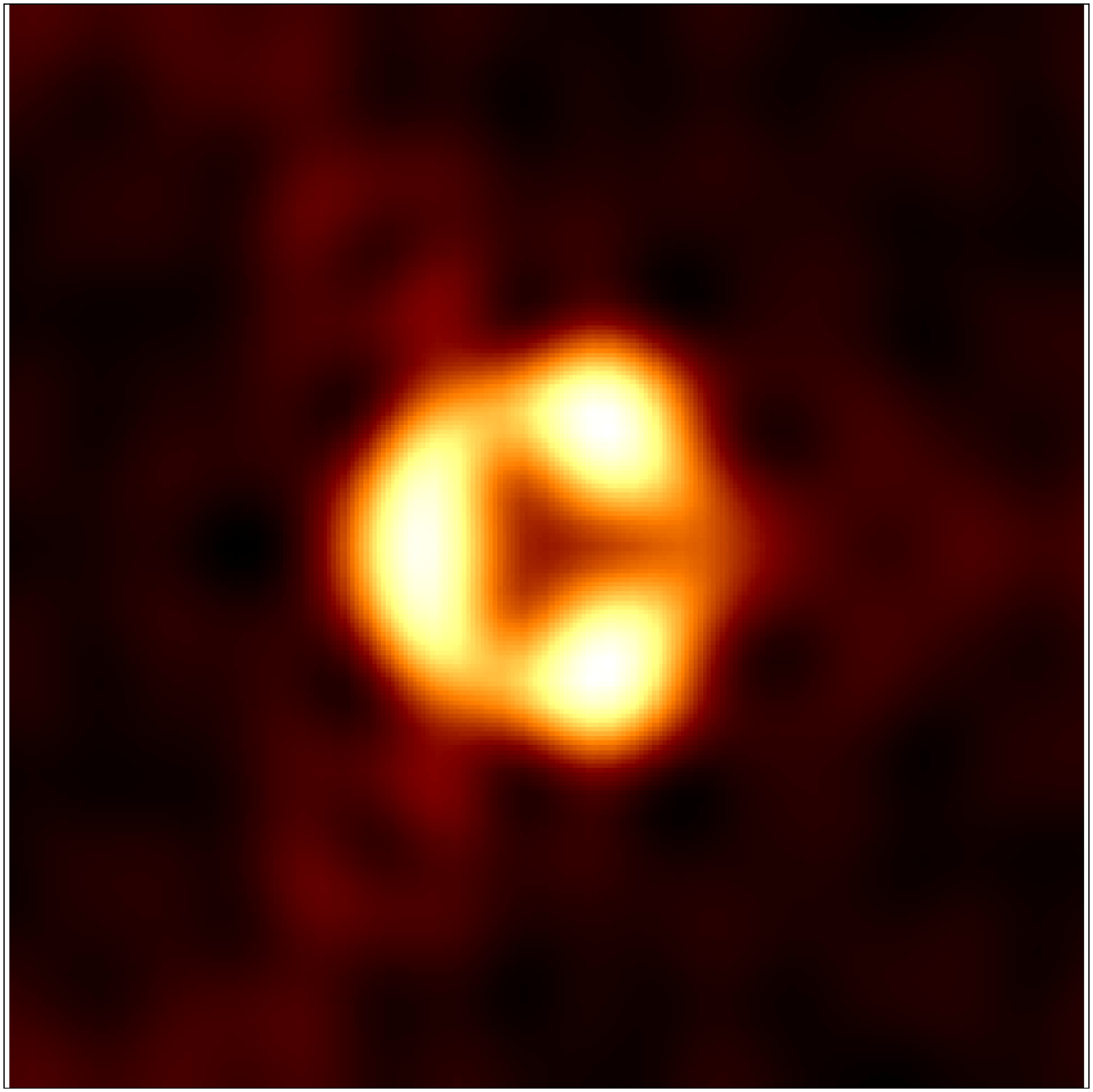}\tabularnewline
\hline 
\end{tabular}
\par\end{centering}
\protect\caption{Constant height STM images of N-doped graphene at 4 $\textrm{\AA}$ tip-sample distance and $\pm 0.4$ V bias.
Comparison between Bardeen's model and revised Chen's method with two different choices of the
$C_{\nu\beta}$ weighting coefficients in Eq.\ (\ref{eq:M2_orig}) (see section \ref{sub:Coeff} (ii) and (iii) for details) for
three tungsten tip models: $\mathrm{W_{blunt}}$, $\mathrm{W_{sharp}}$ and $\mathrm{W_{C-apex}}$. Room temperature
was assumed corresponding to the STM experiments in Ref.\ \citep{Telychko-Graphene-N}. The N defect is located in the
middle of the images. \label{fig:Comparison-Graphene}}
\end{figure}

Fig.\ \ref{fig:Comparison-Graphene} and Table \ref{tab:Corr-Fig2} show qualitative and quantitative comparisons
of the revised Chen's method with Bardeen's method using three different tungsten tip models,
which have also been used in previous studies of STM imaging of HOPG \citep{Teobaldi2012,Mandi-HOPGstipe}.
Moreover, we compare two different choices of the $C_{\nu\beta}$ weighting coefficients in Eq.\ (\ref{eq:M2_orig}) of the
revised Chen's method, see section \ref{sub:Coeff} (ii) and (iii), and good qualitative agreement is obtained.

We find that using these methods the current dip above the N atom is always present in the STM images
in Fig.\ \ref{fig:Comparison-Graphene}. The degree of agreement between
Bardeen's and revised Chen's method, reported as correlation coefficients between corresponding STM images in
Table \ref{tab:Corr-Fig2}, depends on the actual tip geometry and electronic structure, hence the bias voltage.
Let us recall that we expand the tip wavefunctions/density of states around the tip apex atom and calculate the $C_{\nu\beta}$
coefficients in the Wigner-Seitz sphere of the tip apex atom in the revised Chen's method. The accuracy of such an expansion
depends strongly on the neighboring sub-apex atoms' electronic structure and on the tip-sample distance.
For example, for sharp tips the contribution of sub-apex atoms to the tunneling current is more important than for blunt tips
\citep{Mandi-HOPGstipe}. On the other hand, the larger the tip-sample distance the better the agreement of STM
images between the two methods. The reason is that with increasing tip-sample separation the effect of the local tip geometry
decreases. At larger tip-sample distances we find that the current dip above the N atom vanishes and a rounded triangular pattern
is obtained leaving the three nearest neighbor C atoms visible, similarly to the Tersoff-Hamann results in
Fig.\ \ref{fig:Comparison-Graphene-TH}.

\begin{table}[h]
\begin{centering}
\begin{tabular}{|c|c|c|c|c|c|c|}
\hline
Correlation coefficients & \multicolumn{2}{c|}{$\mathrm{W_{blunt}}$ tip} & \multicolumn{2}{c|}{$\mathrm{W_{sharp}}$ tip} & \multicolumn{2}{c|}{$\mathrm{W_{C-apex}}$ tip}\tabularnewline
\cline{2-7}
between STM images in Fig.\ \ref{fig:Comparison-Graphene}& $-0.4$ V & $+0.4$ V & $-0.4$ V & $+0.4$ V & $-0.4$ V & $+0.4$ V\tabularnewline
\hline
Bardeen-Revised Chen ($C_{\nu\beta}$ in Eq.(\ref{eq:coeff}))          & 0.81 & 0.82 & 0.78 & 0.89 & 0.73 & 0.91\tabularnewline
Bardeen-Revised Chen ($C_{\nu\beta}\approx\sqrt{n^{TIP}_{\nu\beta}}$) & 0.71 & 0.81 & 0.62 & 0.79 & 0.74 & 0.92\tabularnewline
\hline
\end{tabular}
\par\end{centering}
\protect\caption{Quantitative comparison between Bardeen's model and revised Chen's method:
calculated correlation coefficients between STM images in Fig.\ \ref{fig:Comparison-Graphene}.
\label{tab:Corr-Fig2} }
\end{table}

Overall, we find good agreement between Bardeen's and revised Chen's method in Fig.\ \ref{fig:Comparison-Graphene}
and Table \ref{tab:Corr-Fig2}. Calculated correlation values are above 0.7 except for the $\mathrm{W_{sharp}}$ tip
at $-0.4$ V bias and $C_{\nu\beta}\approx\sqrt{n^{TIP}_{\nu\beta}}$. We find better correlation for
$C_{\nu\beta}$ in Eq.(\ref{eq:coeff}) than for the $C_{\nu\beta}\approx\sqrt{n^{TIP}_{\nu\beta}}$ approximation used in the
revised Chen's method, with the exception of the $\mathrm{W_{C-apex}}$ tip. Moreover, systematically better correlation between
Bardeen's and revised Chen's method is found for $+0.4$ V than for $-0.4$ V bias voltage.
Note that the large difference between STM images of the $\mathrm{W_{sharp}}$ tip with different bias polarities can be
explained by the asymmetric electronic structure of the tip apex around the Fermi level \citep{Teobaldi2012,Mandi-HOPGstipe}.

\begin{figure}[h]
\begin{centering}
\begin{tabular}{|cc|cccccc|}
\cline{3-8}
\multicolumn{2}{c|}{} & \multicolumn{2}{c|}{$\mathrm{W_{blunt}}$ tip} & \multicolumn{2}{c|}{$\mathrm{W_{sharp}}$ tip} & \multicolumn{2}{c|}{$\mathrm{W_{C-apex}}$ tip}\tabularnewline
\cline{3-8}
\multicolumn{2}{c|}{} & \multicolumn{1}{c|}{$-0.4$ V} & \multicolumn{1}{c|}{$+0.4$ V} & \multicolumn{1}{c|}{$-0.4$ V} & \multicolumn{1}{c|}{$+0.4$ V} & \multicolumn{1}{c|}{$-0.4$ V} & $+0.4$ V\tabularnewline
\hline
\begin{turn}{90}
Bond lengths
\end{turn} &
\begin{turn}{90}
$-10$\%
\end{turn} & \includegraphics[width=0.15\textwidth]{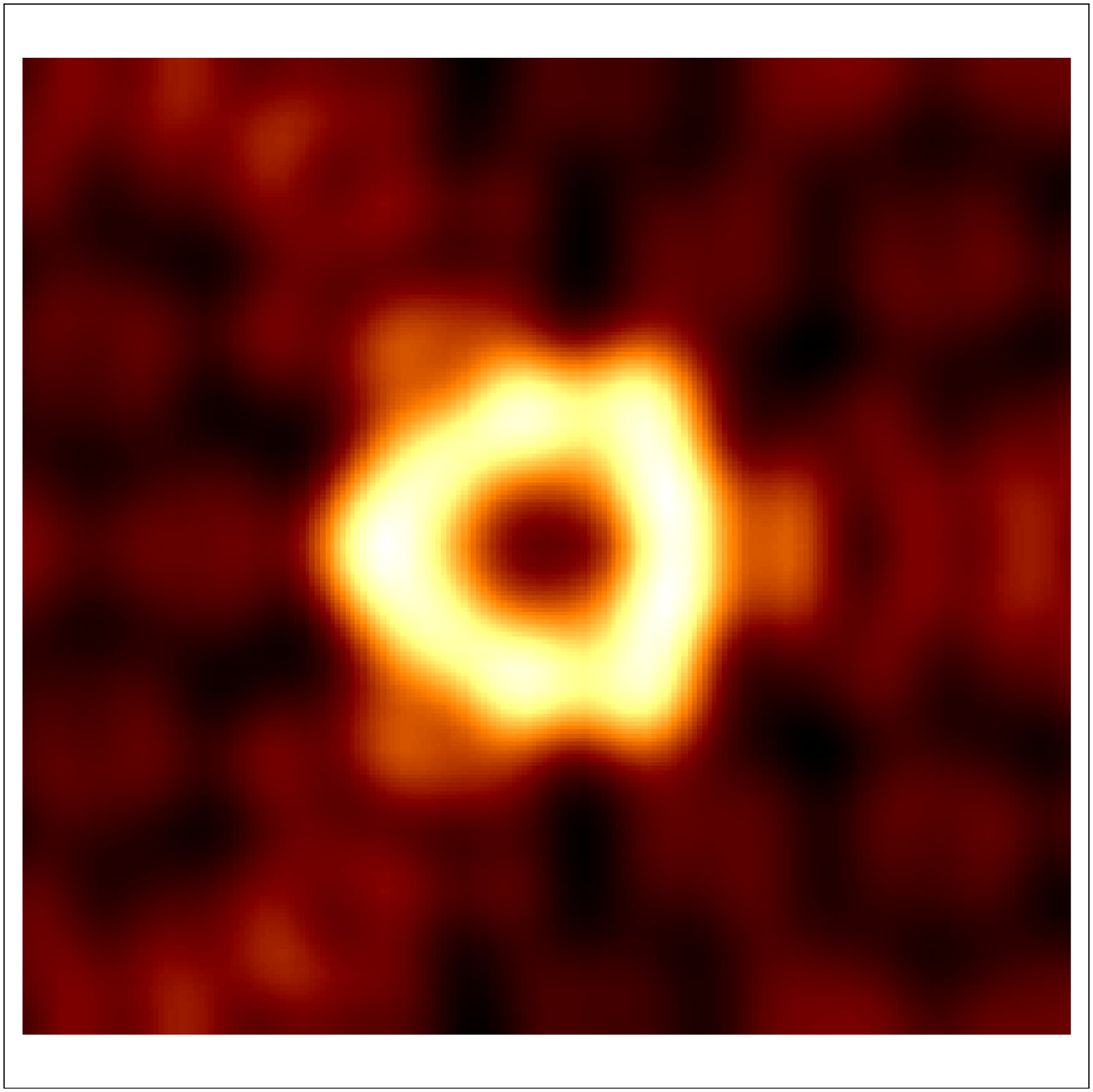} & \includegraphics[width=0.15\textwidth]{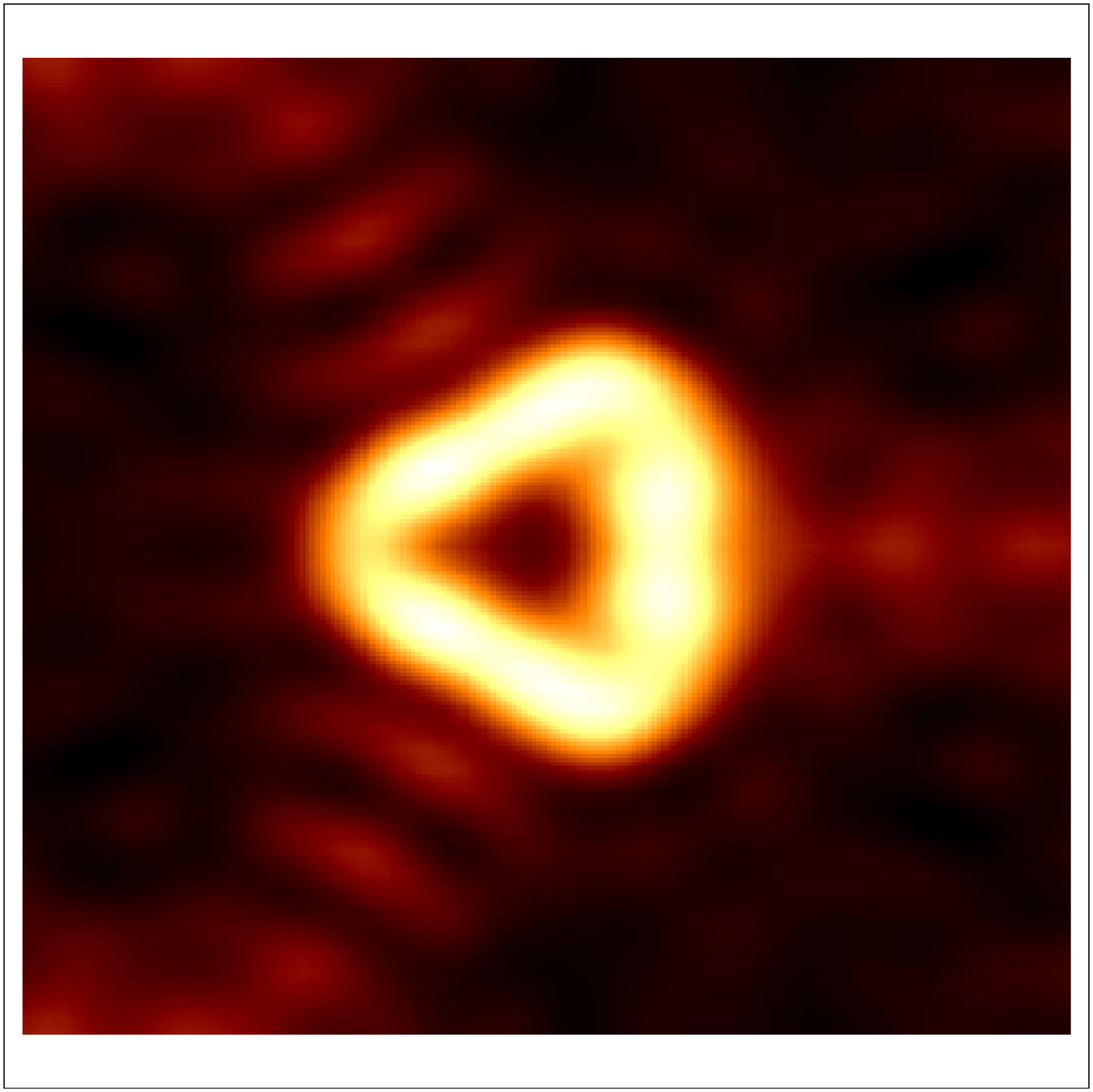} & \includegraphics[width=0.15\textwidth]{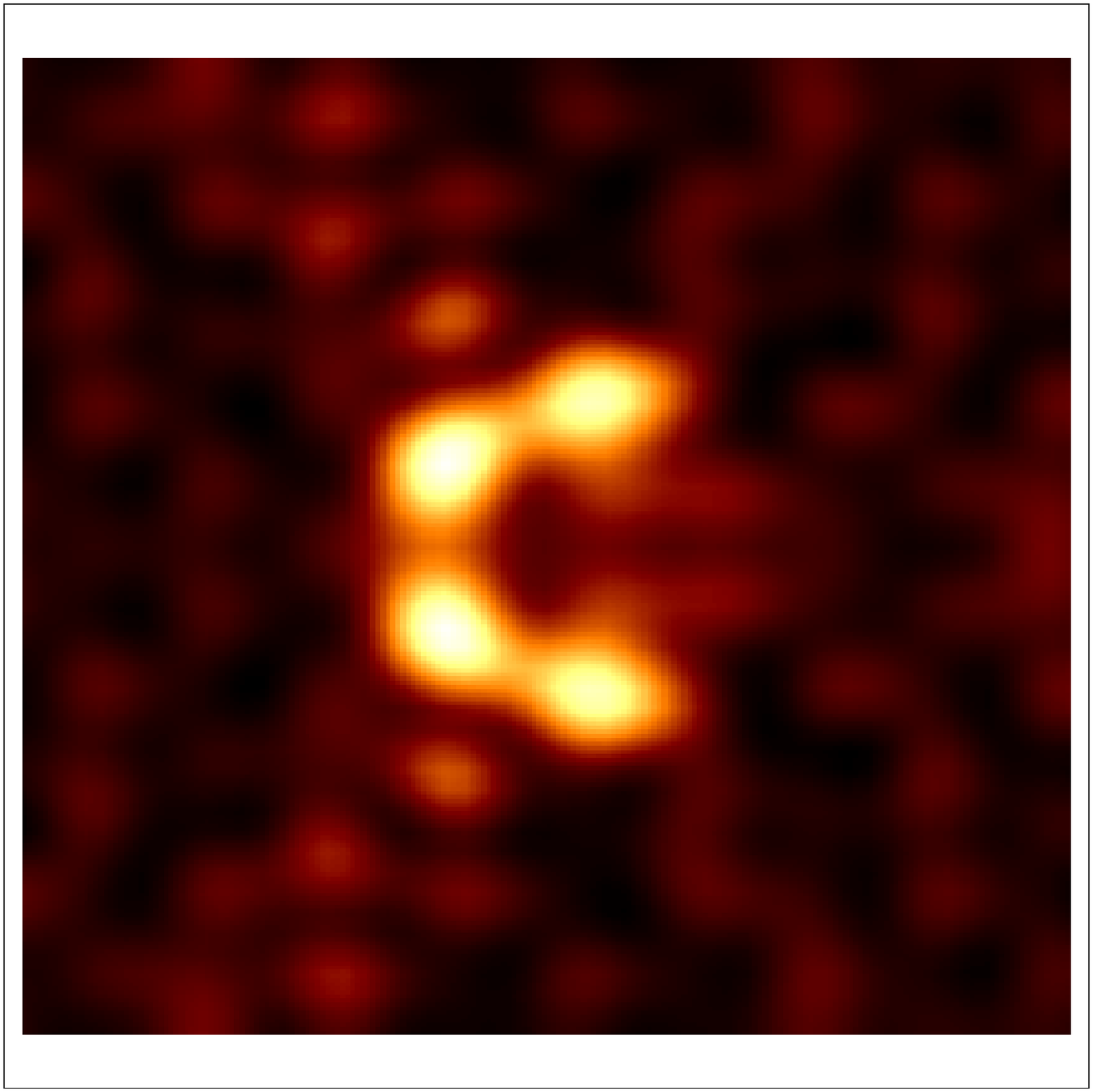} & \includegraphics[width=0.15\textwidth]{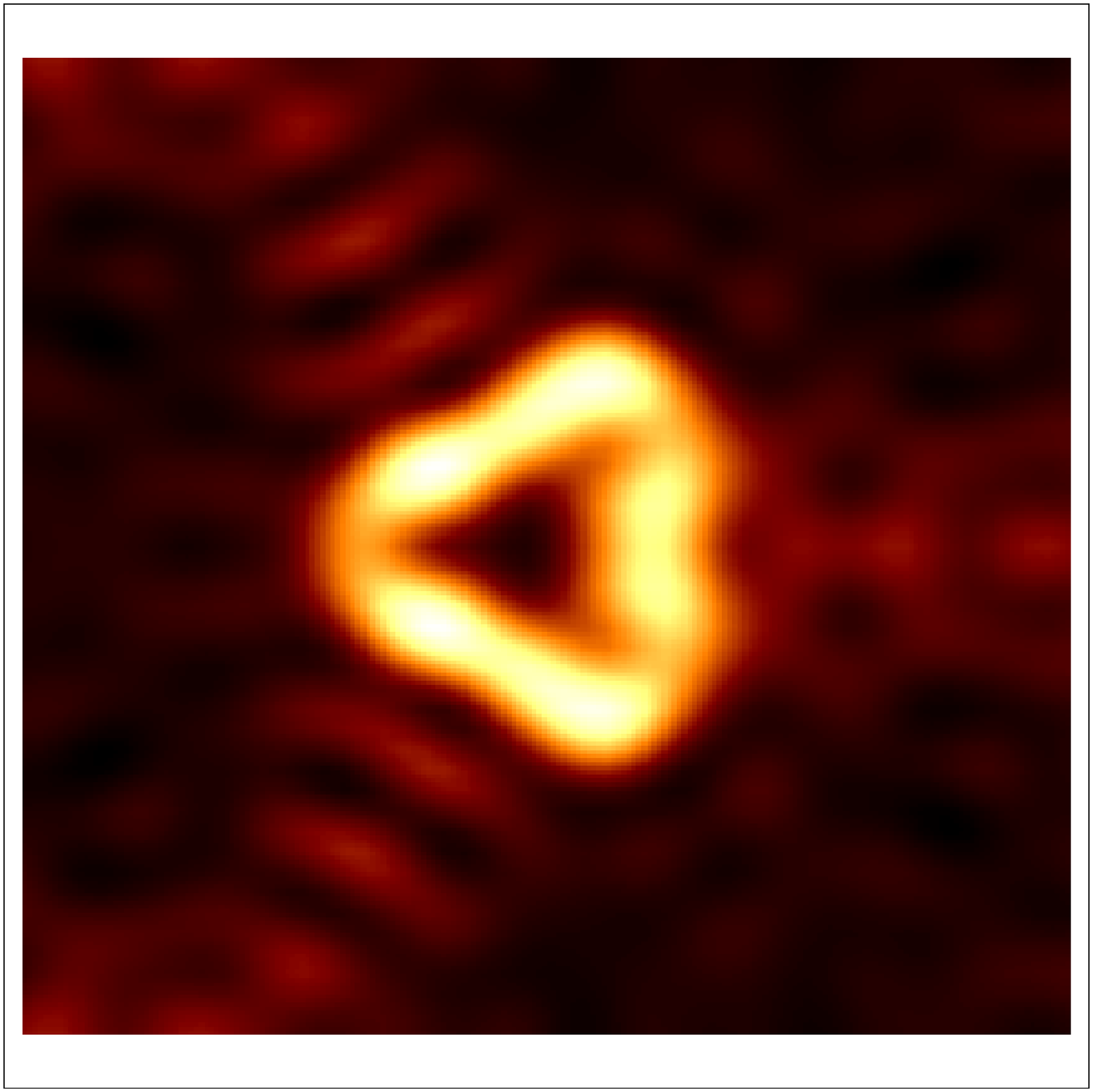} & \includegraphics[width=0.15\textwidth]{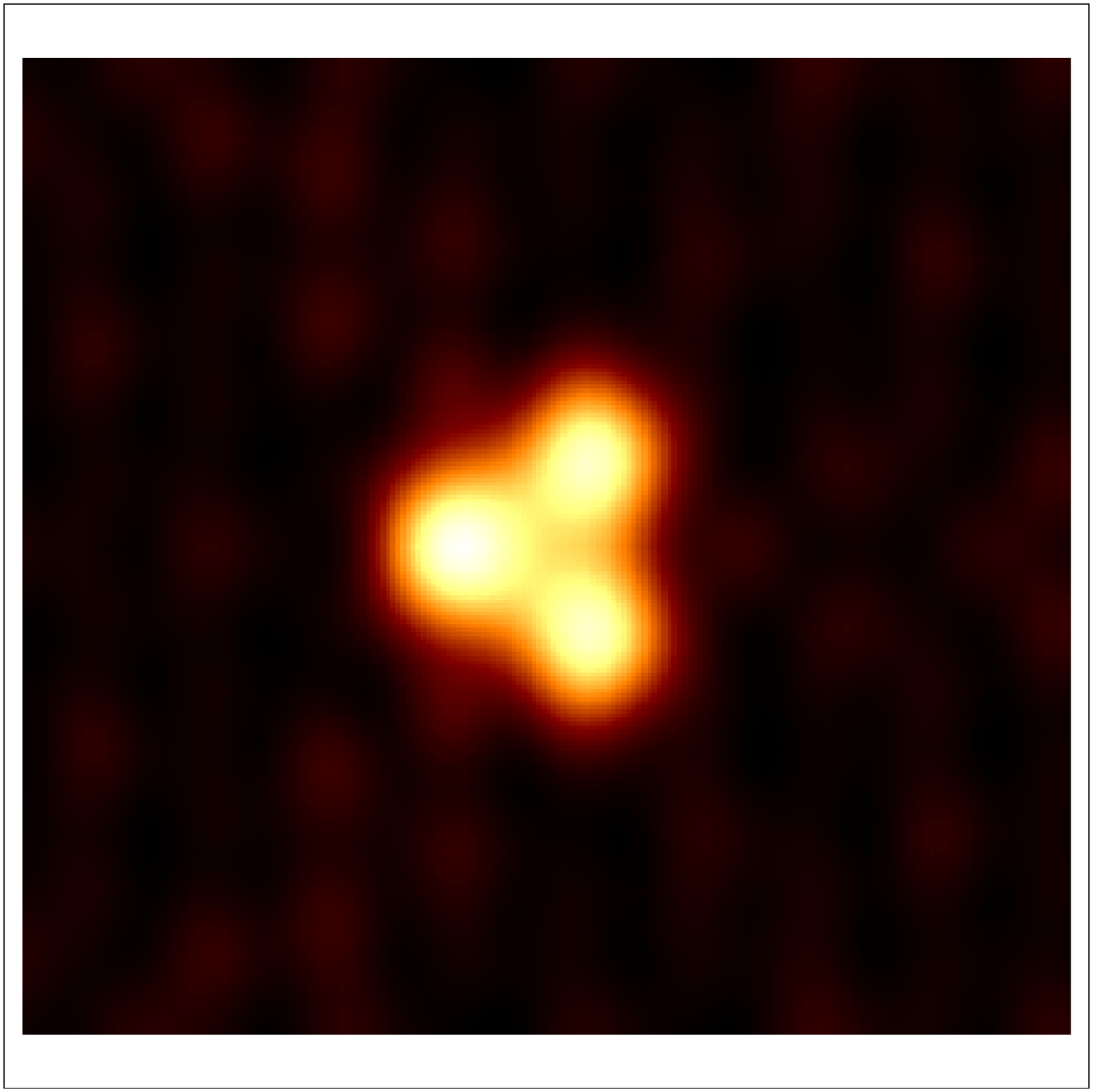} & \includegraphics[width=0.15\textwidth]{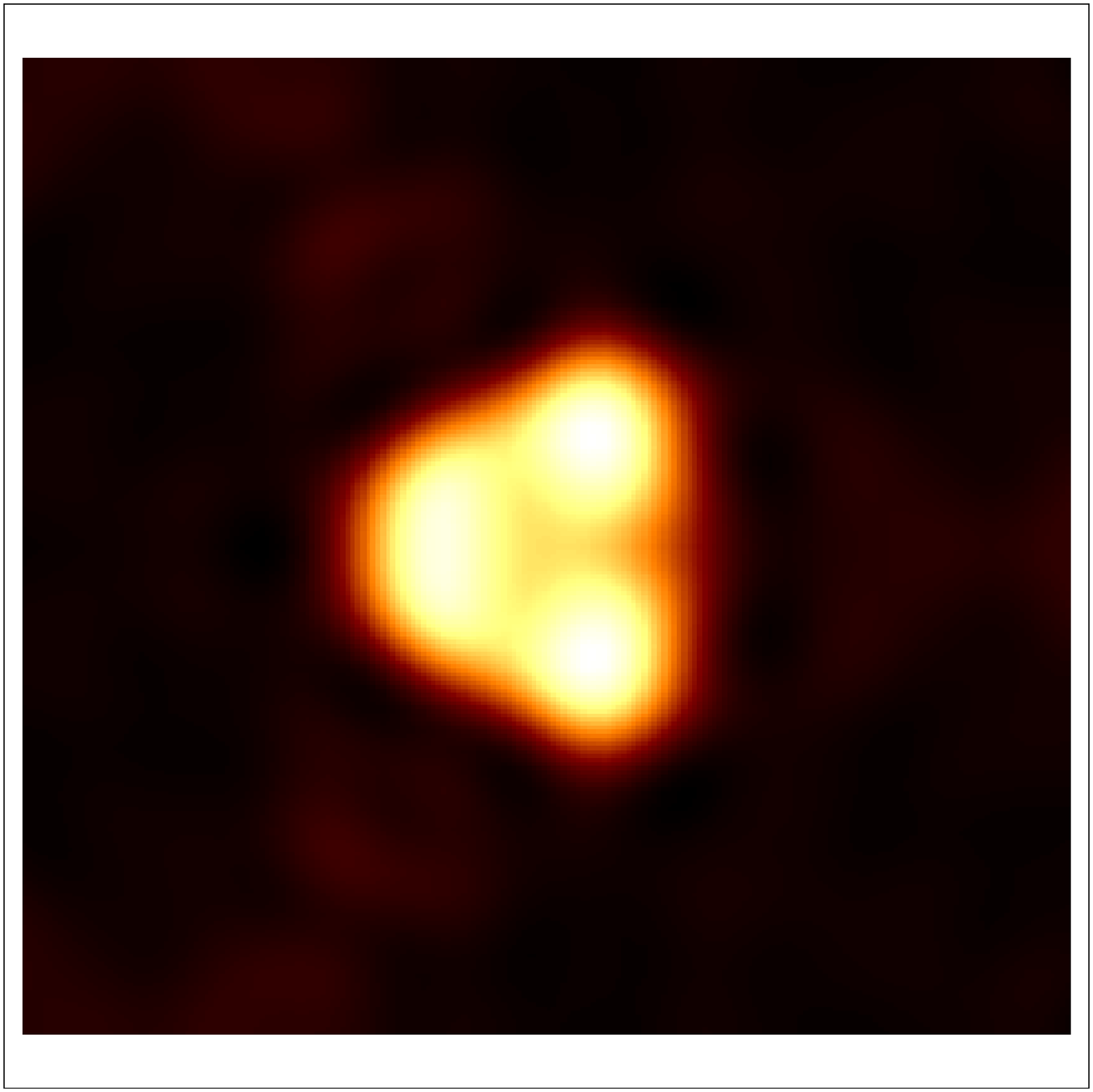}\tabularnewline
\cline{1-2}
\begin{turn}{90}
Ground state
\end{turn} &
\begin{turn}{90}
structure
\end{turn} & \includegraphics[width=0.15\textwidth]{chen_blunt_-0.4V_58.eps} & \includegraphics[width=0.15\textwidth]{chen_blunt_+0.4V_58.eps} & \includegraphics[width=0.15\textwidth]{chen_sharp_-0.4V_58.eps} & \includegraphics[width=0.15\textwidth]{chen_sharp_+0.4V_58.eps} & \includegraphics[width=0.15\textwidth]{chen_Capex_-0.4V_58.eps} & \includegraphics[width=0.15\textwidth]{chen_Capex_+0.4V_58.eps}\tabularnewline
\cline{1-2}
\begin{turn}{90}
Bond lengths
\end{turn} &
\begin{turn}{90}
$+10$\%
\end{turn} & \includegraphics[width=0.15\textwidth]{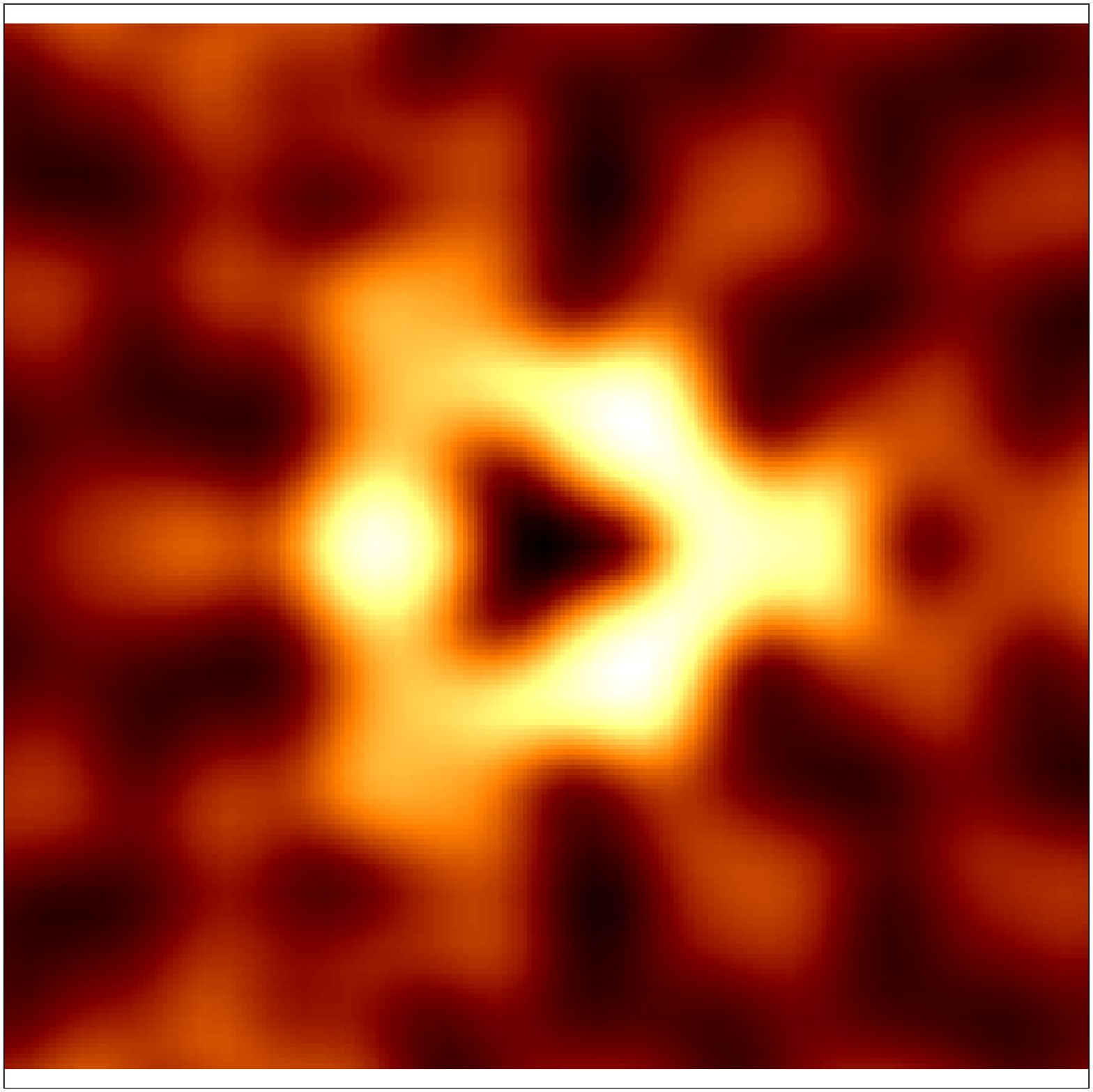} & \includegraphics[width=0.15\textwidth]{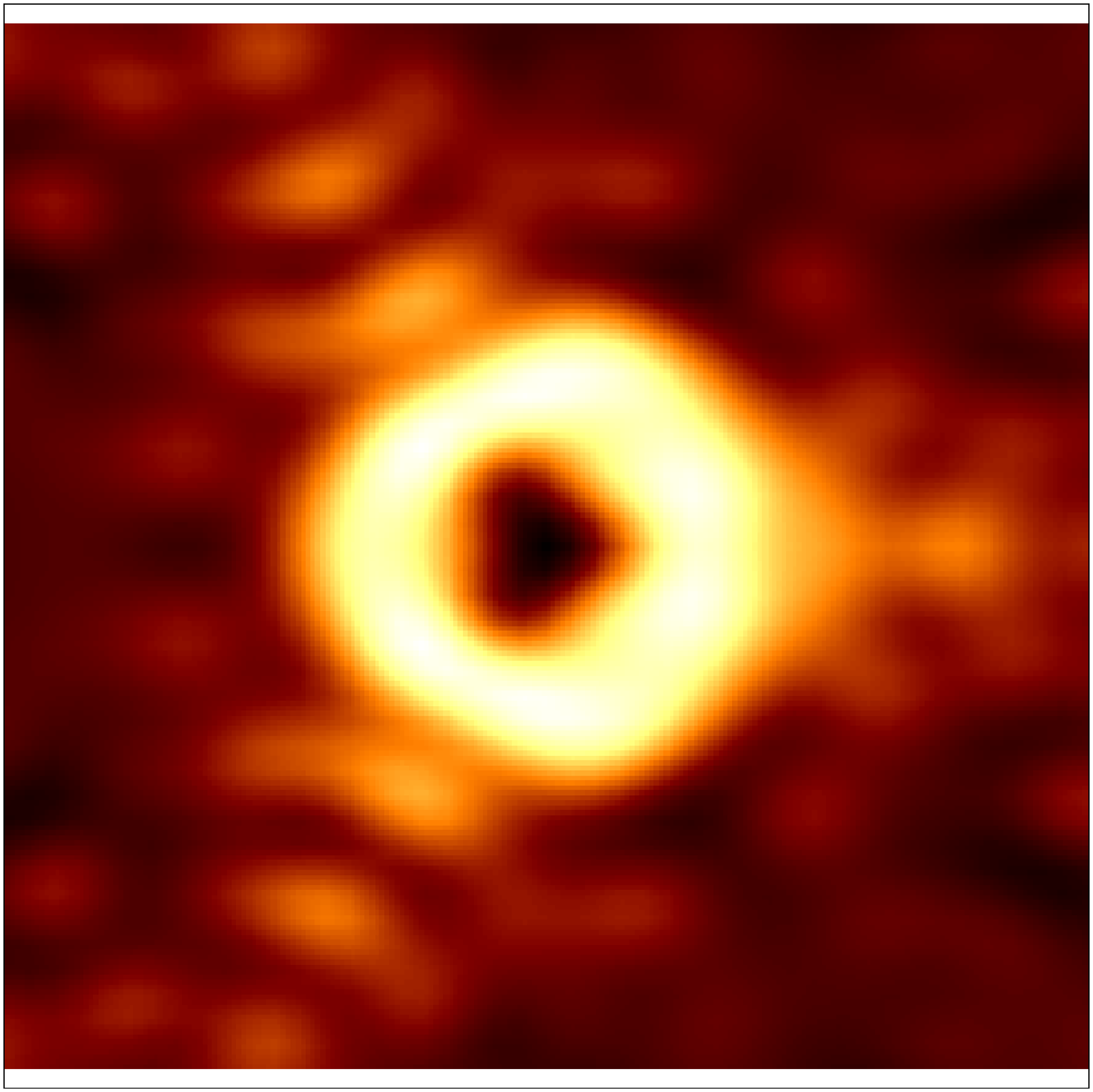} & \includegraphics[width=0.15\textwidth]{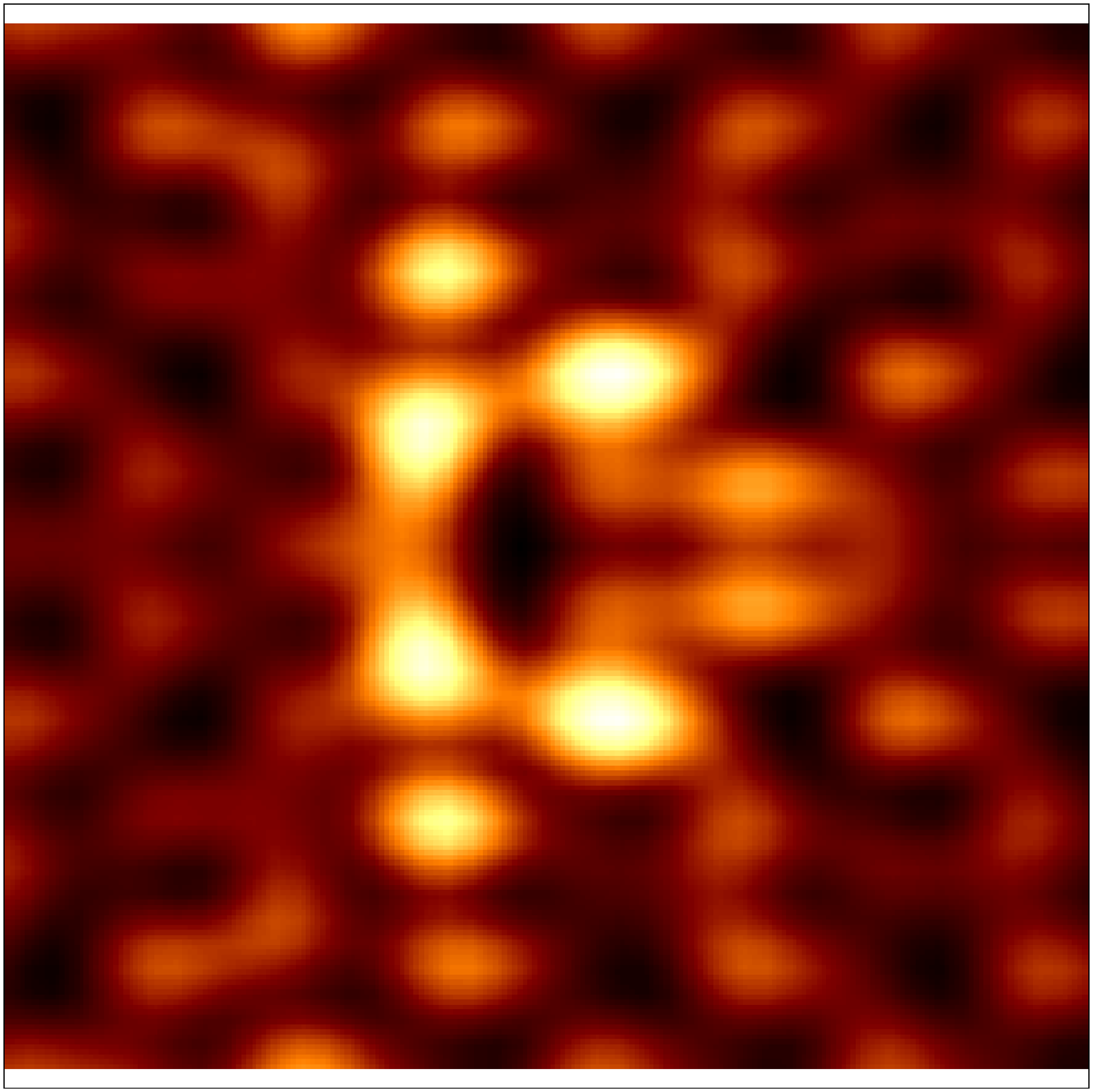} & \includegraphics[width=0.15\textwidth]{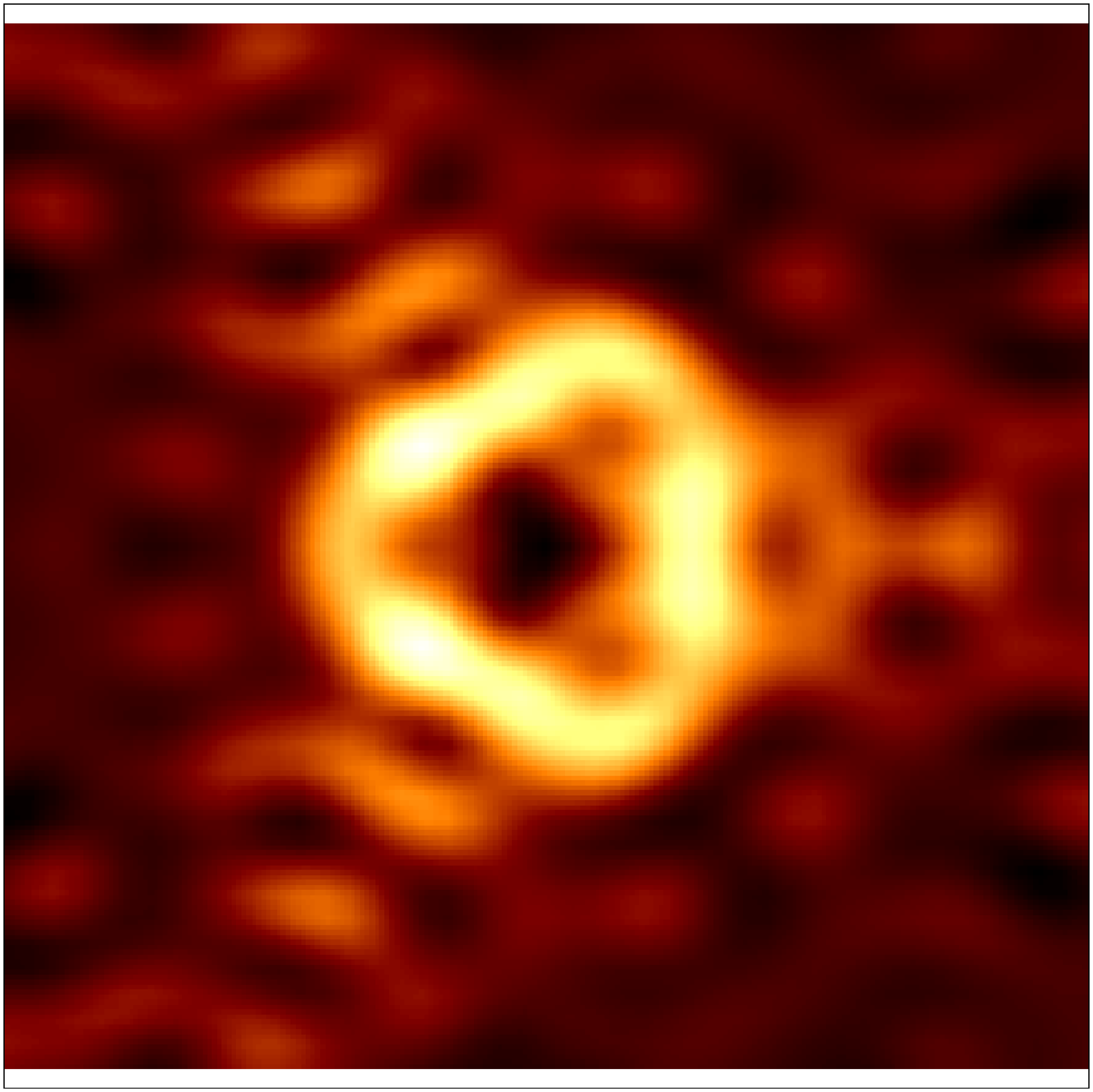} & \includegraphics[width=0.15\textwidth]{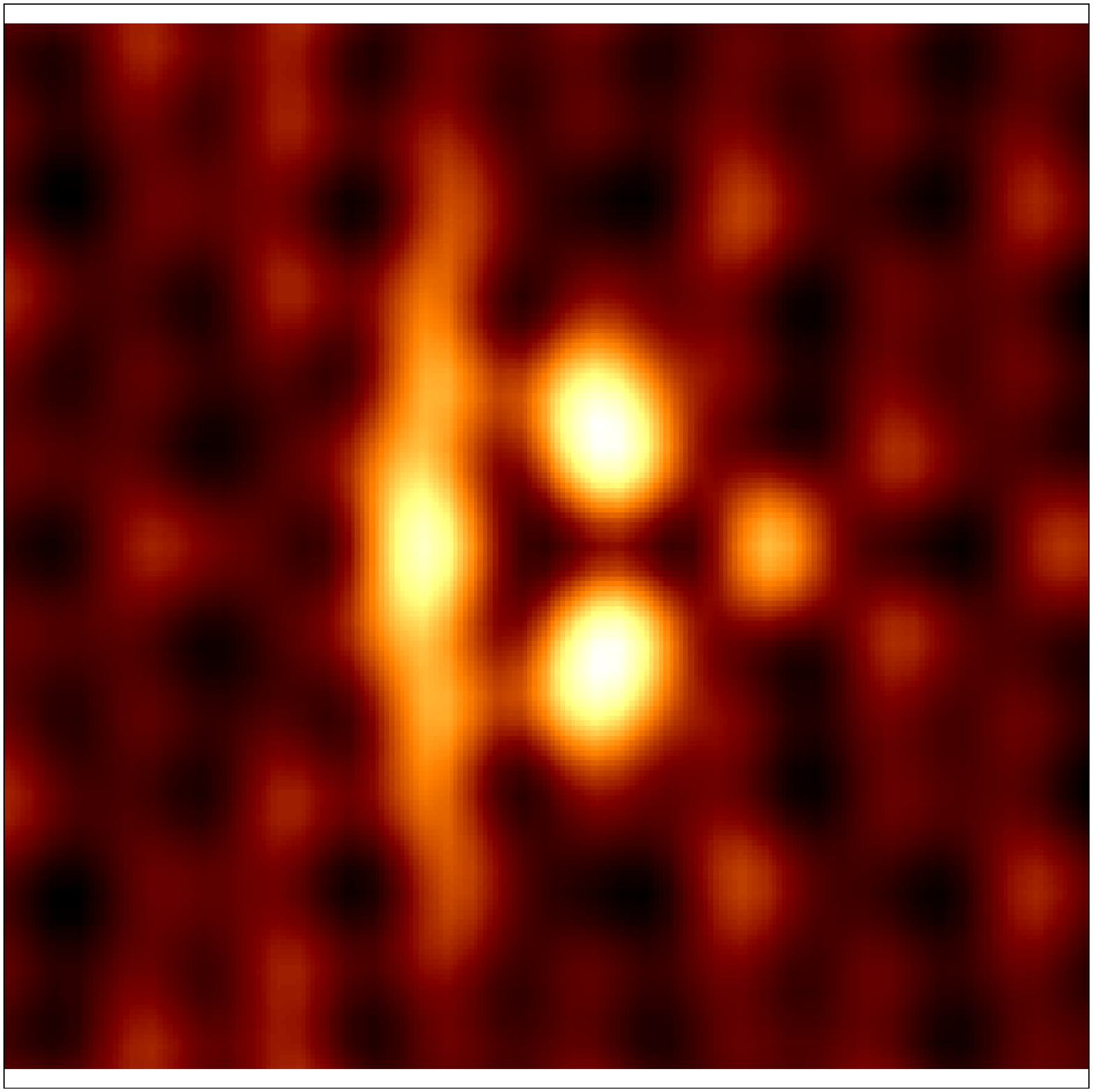} & \includegraphics[width=0.15\textwidth]{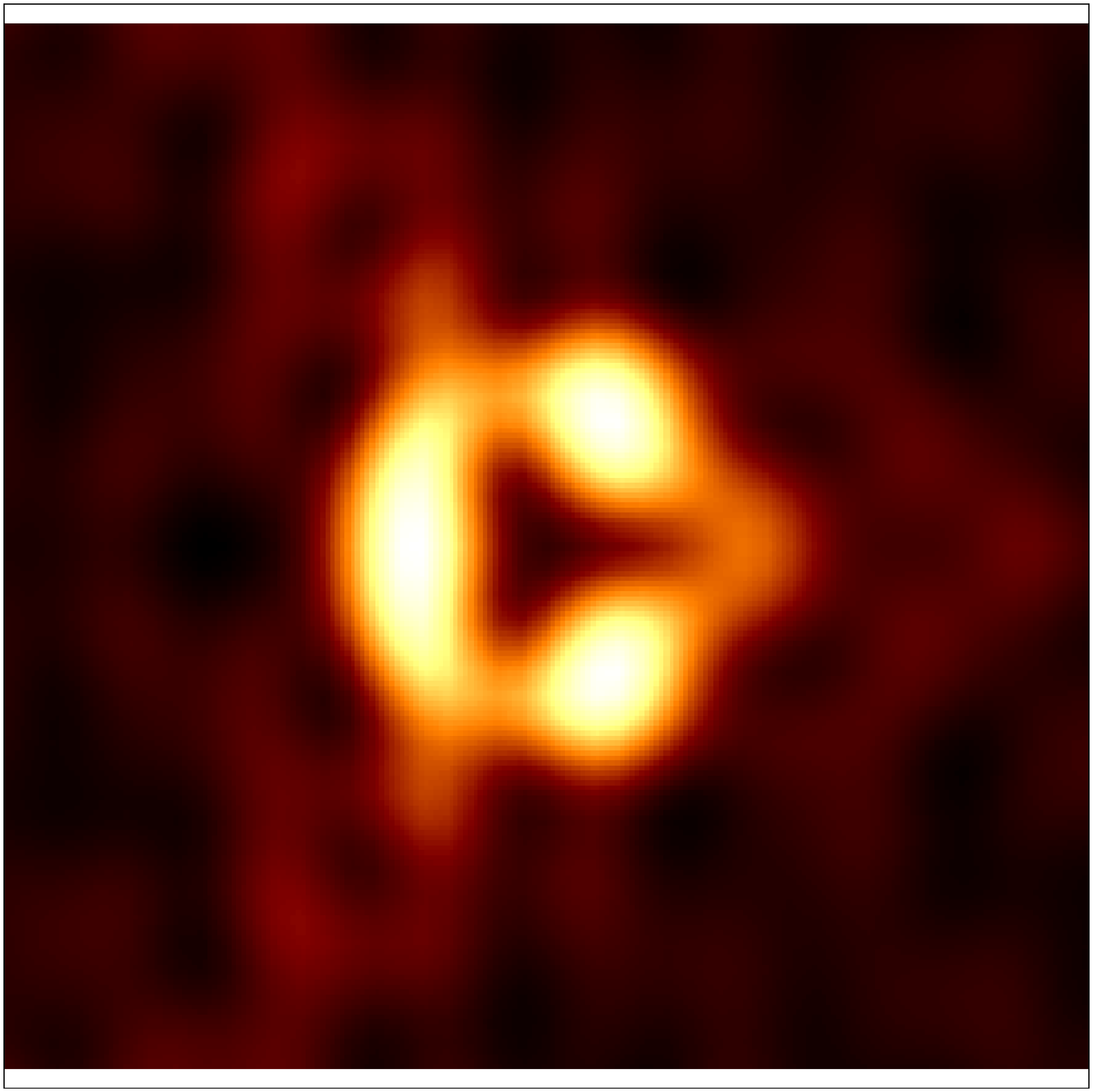}\tabularnewline
\hline
\end{tabular}
\par\end{centering}
\protect\caption{Effect of strain on constant height STM images of N-doped graphene at 4 $\textrm{\AA}$
tip-sample distance and $\pm 0.4$ V bias using the revised Chen's method with three tungsten tip models: $\mathrm{W_{blunt}}$,
$\mathrm{W_{sharp}}$ and $\mathrm{W_{C-apex}}$. The ground state N-doped graphene geometry obtained by DFT and two other
structures with bond lengths varied by $\pm 10$\% relative to the ground state are compared.\label{fig:Bond-Graphene}}
\end{figure}

It is interesting to investigate the effect of strain on the obtained STM contrast.
Fig.\ \ref{fig:Bond-Graphene} shows a comparison between STM images calculated for three different N-doped graphene geometries
with varying bond lengths by $\pm 10$\% relative to the ground state structure, which has been obtained by DFT calculation with
C-N and C-C bond lengths of 1.42 \AA. Generally, we observe that the main features of the STM contrast do not change with the
applied strain. This is quantitatively confirmed by correlation coefficients being above 0.93 for each tip and bias combination
calculated between images within each column of Fig.\ \ref{fig:Bond-Graphene}. We find a tendency of spatially extended
brighter features in the STM images upon elongation of the bonds.

\begin{figure}[h]
\begin{centering}
\begin{tabular}{|c|c|c|c|c|c|}
\hline 
$\varphi_{0}=0\text{\textdegree}$ & $\varphi_{0}=30\text{\textdegree}$ & $\varphi_{0}=60\text{\textdegree}$ & $\varphi_{0}=90\text{\textdegree}$ & $\varphi_{0}=120\text{\textdegree}$ & $\varphi_{0}=150\text{\textdegree}$\tabularnewline
\hline 
\hline 
\includegraphics[width=0.15\columnwidth]{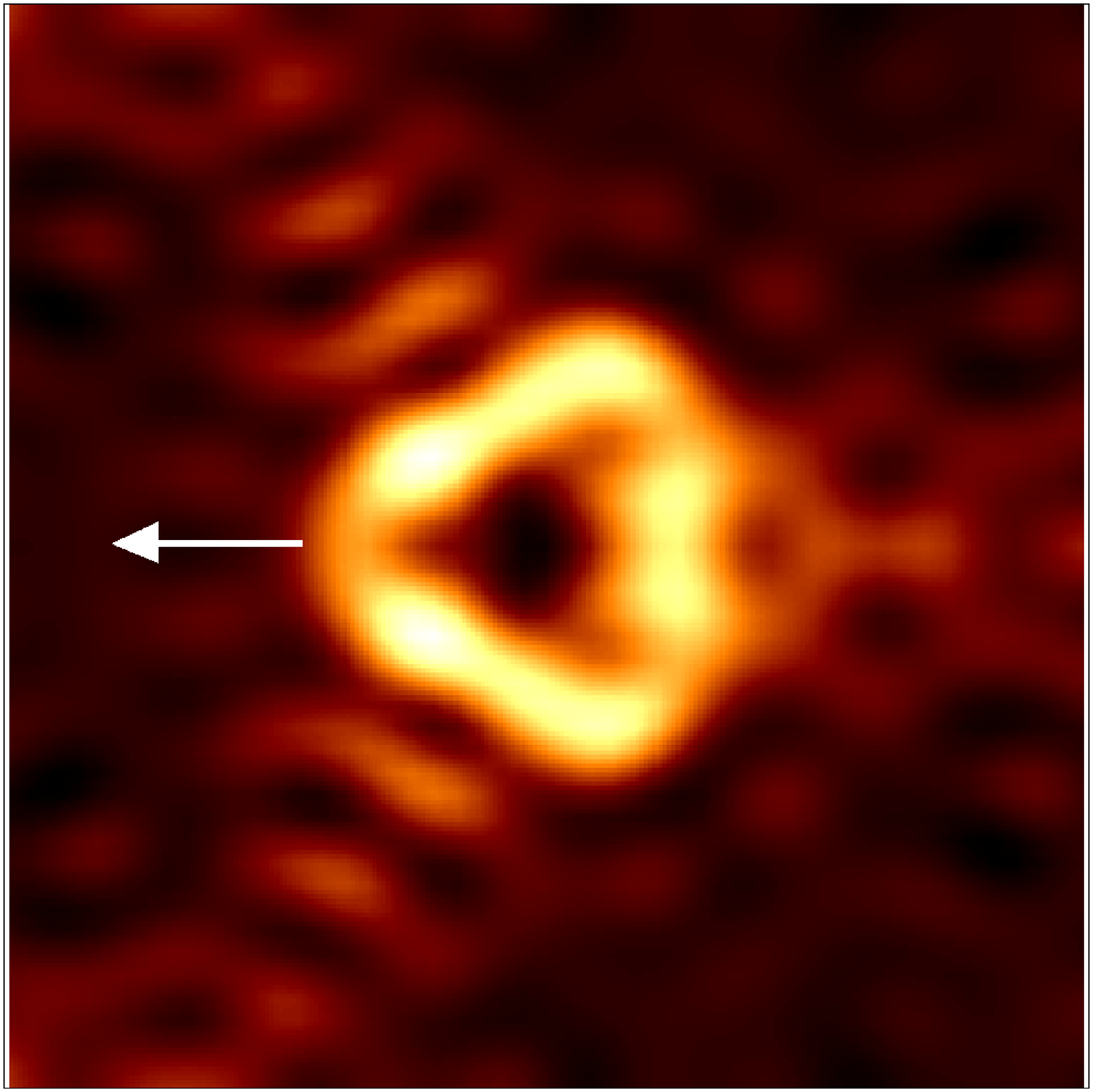} & \includegraphics[width=0.15\columnwidth]{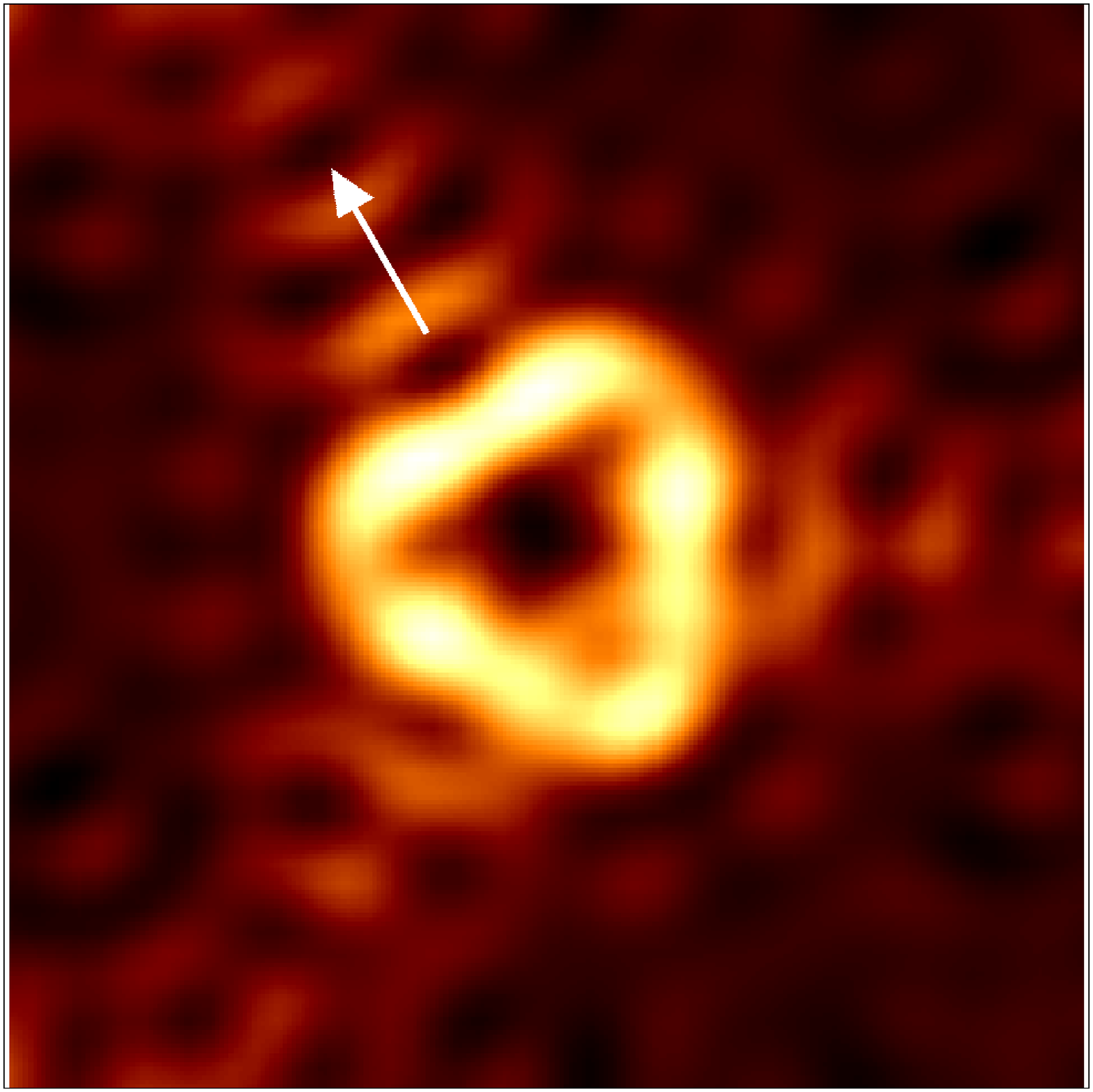} & \includegraphics[width=0.15\columnwidth]{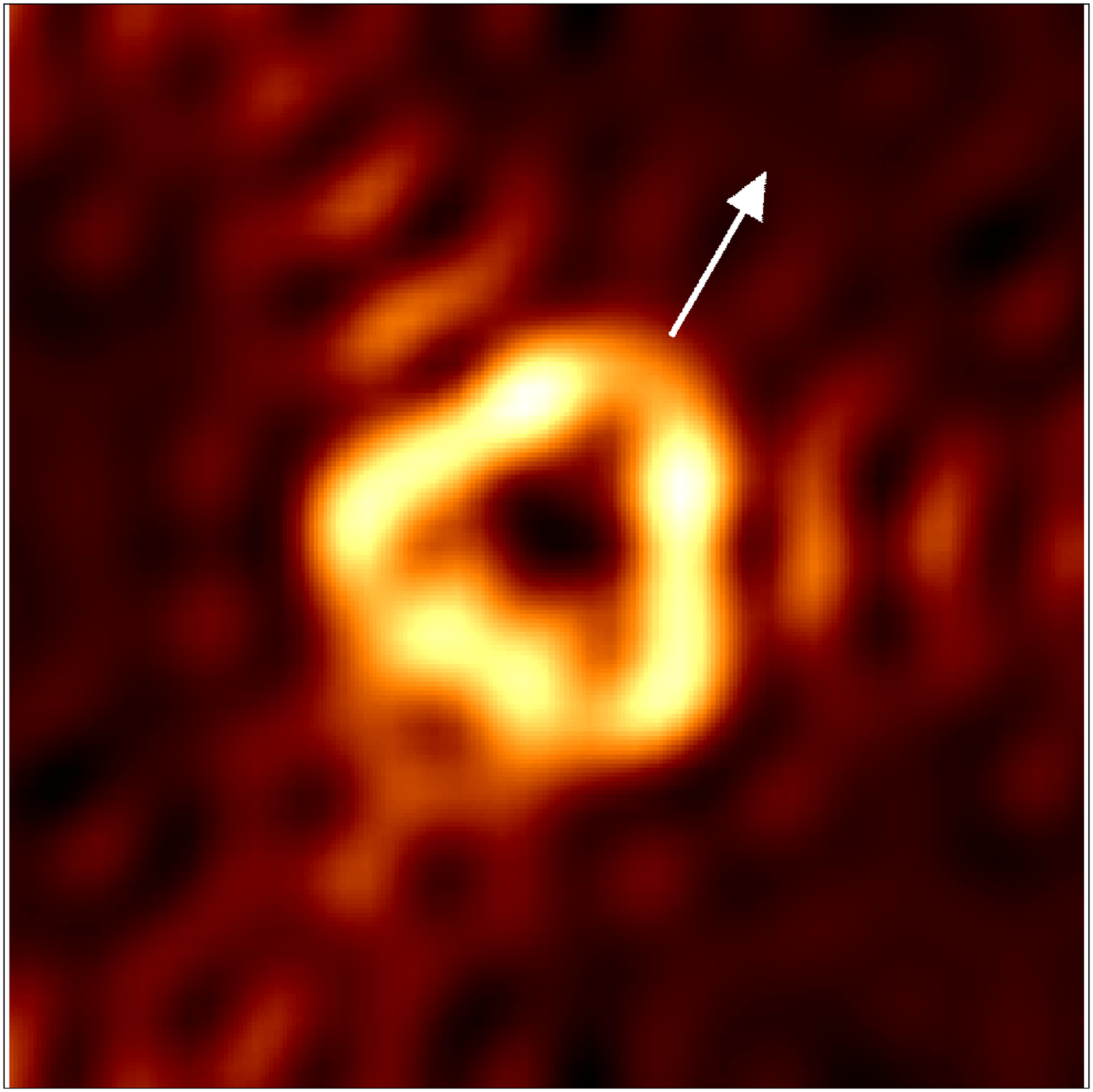} & \includegraphics[width=0.15\columnwidth]{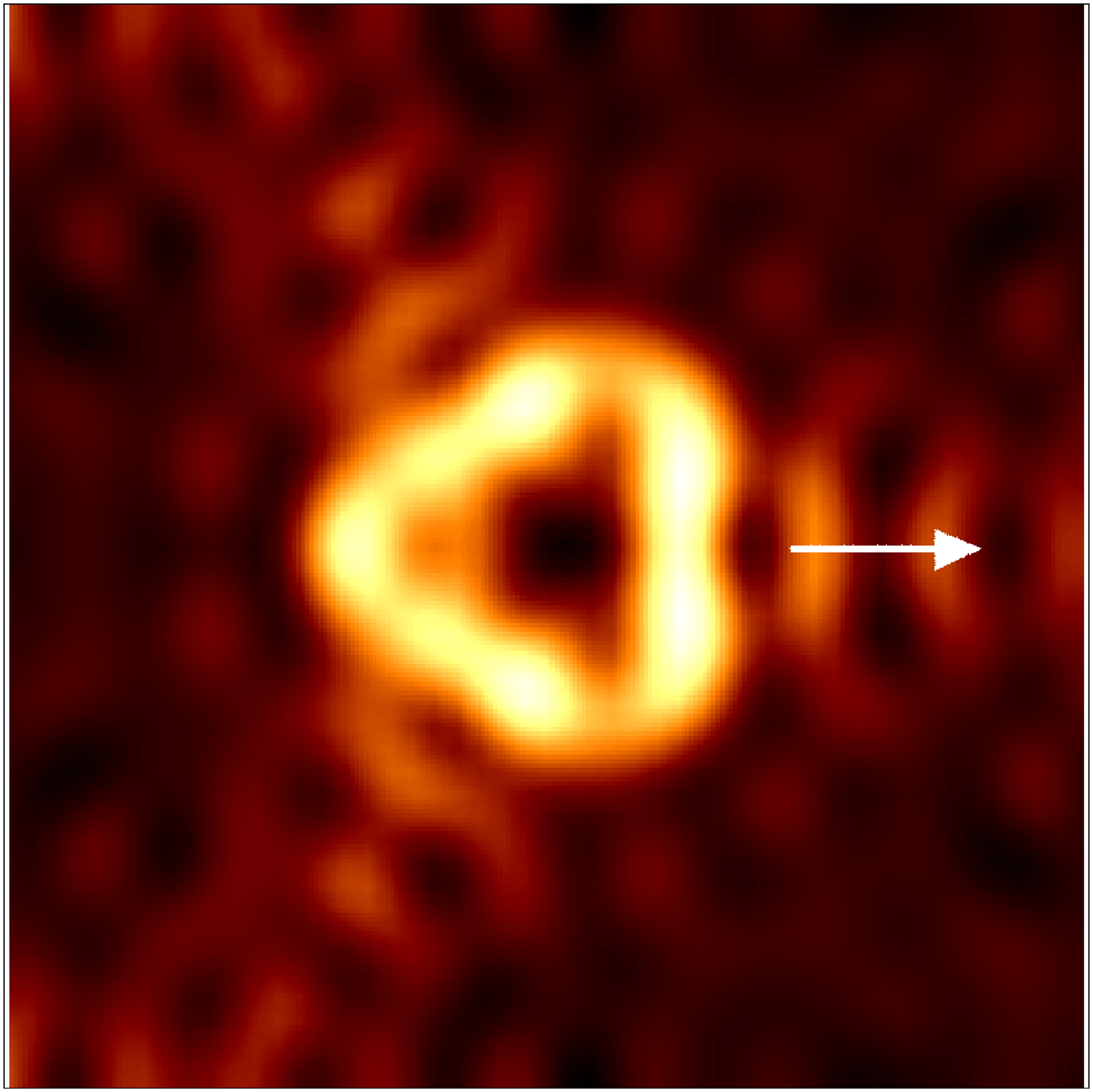} & \includegraphics[width=0.15\columnwidth]{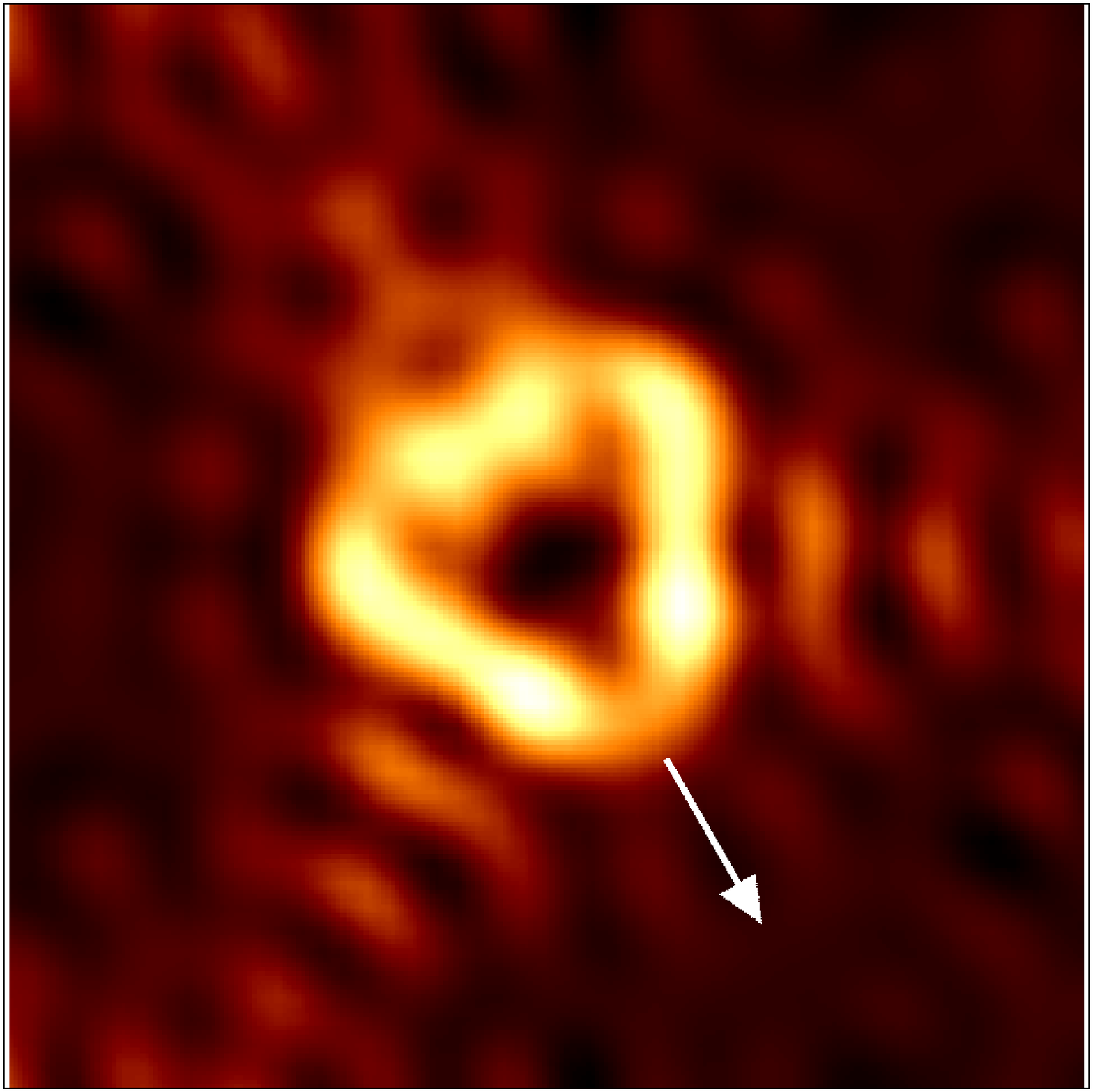} & \includegraphics[width=0.15\columnwidth]{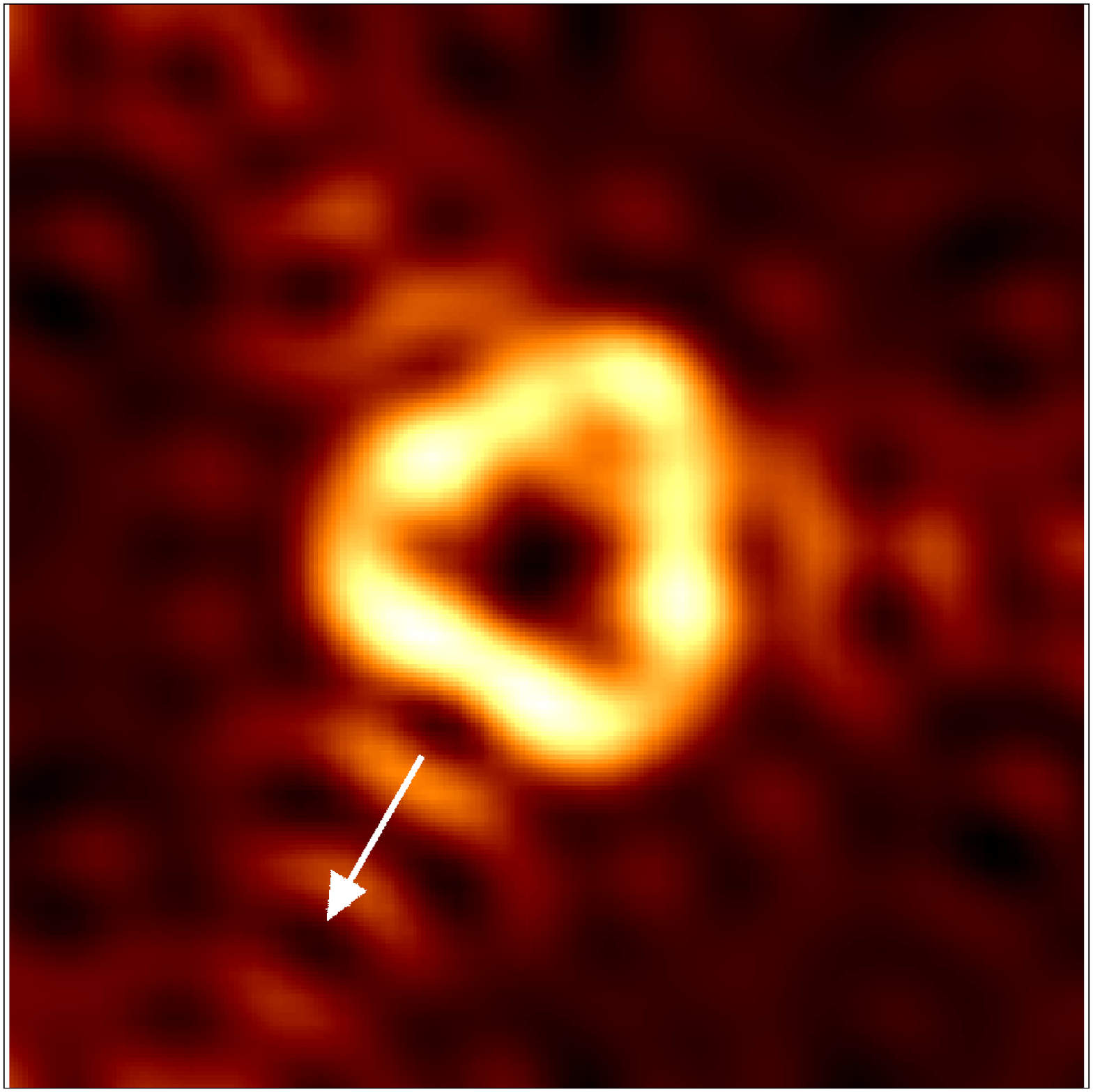}\tabularnewline
\hline 
\end{tabular}
\par\end{centering}
\protect\caption{Tip rotation effect on the constant height STM images of N-doped graphene calculated with $\mathrm{W_{sharp}}$
tip at 4 $\textrm{\AA}$ tip-sample distance and $+0.4$ V bias voltage. The rotation axis of the tip is perpendicular to the
surface ($\vartheta_0=\psi_0=0\text{\textdegree}$). The orientations of the brightest features are indicated by white arrows in
each STM image.
\label{fig:rotation-Graphene}}
\end{figure}

In the following we investigate the effect of tip rotations on the STM images. Constant height STM images have been calculated
above the N-doped graphene surface with the tungsten tip models at 4 $\textrm{\AA}$ tip-sample distance. We considered tip
rotations around the axis perpendicular to the sample surface. This corresponds to $\vartheta_0=0\text{\textdegree}$, and in this
case rotations with respect to $\varphi_0$ and $\psi_0$ are equivalent, thus, we fixed $\psi_0=0\text{\textdegree}$ and rotated
the tip by $\varphi_0$ in 10\textdegree{} steps. Since a more asymmetric tip is expected to have a larger tip rotational effect,
we present results obtained by the $\mathrm{W_{sharp}}$ tip at $+0.4$ V bias voltage. Selected STM images are shown in
Fig.\ \ref{fig:rotation-Graphene}.

Due to the $C_{2v}$ symmetry of the tip, the same image is obtained for $\varphi_0=180\text{\textdegree}$ as for
$\varphi_0=0\text{\textdegree}$. We find that the current dip above the N atom is always present independently of the degree of
tip rotation by $\varphi_0$, but the intensity of the current above the surrounding C atoms changes with the tip rotation. There
are certain directions denoted by white arrows in Fig.\ \ref{fig:rotation-Graphene}, where the brightest features occur that
correspond to the largest currents above or close to nearest neighbor C atoms. These indicated directions rotate twice faster
than the tip rotation by $\varphi_0$ itself. The finding that such kind of tip rotations, where the $z$ axis of the tip is not
tilted with respect to the $z$ axis of the sample ($\vartheta_0=0\text{\textdegree}$), affect the secondary features of the STM
image is in agreement with previous results using the 3D-WKB method \citep{Mandi-ArbTipOriW110,Mandi-HOPGstipe}.

\begin{figure}[h]
\begin{centering}
\includegraphics[width=0.5\columnwidth]{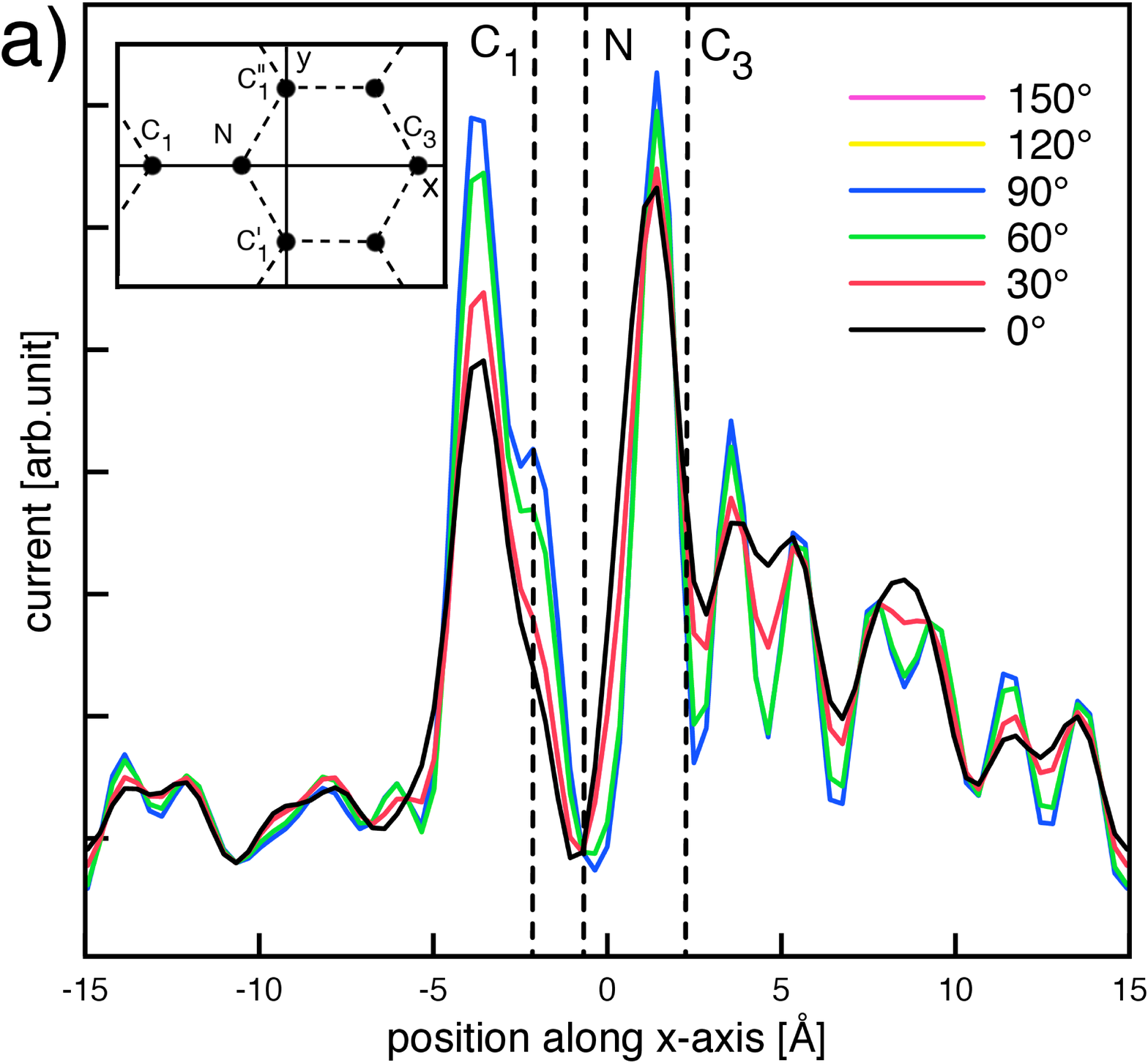}\includegraphics[width=0.5\columnwidth]{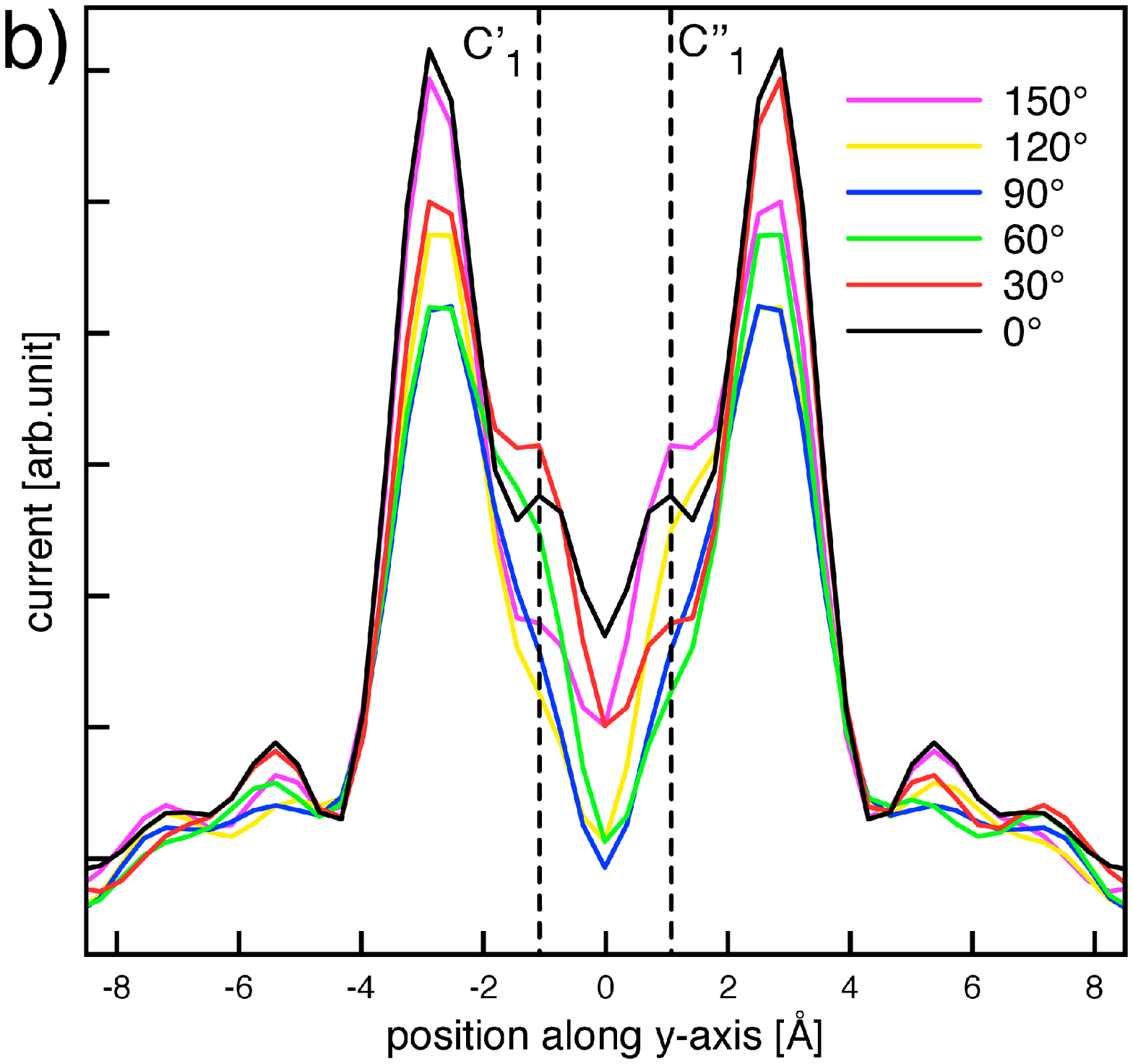} 
\par\end{centering}
\protect\caption{\label{fig:Current-profiles}Current profiles along $x$ and $y$ directions of the N-doped graphene surface
(see inset) as a function of tip rotation with respect to $\varphi_0$. ($\vartheta_0=\psi_0=0\text{\textdegree}$)}
\end{figure}

In Fig.\ \ref{fig:Current-profiles} we extracted line sections of the constant height STM images presented in
Fig.\ \ref{fig:rotation-Graphene}. The line along the $x$ direction contains the N atom and its nearest (C$_1$) and
third nearest neighbor (C$_3$) carbon atoms. The line along the $y$ direction contains the other two nearest neighbor
carbon atoms (C$'_1$ and C$''_1$), see the inset of Fig.\ \ref{fig:Current-profiles}a). The symmetries of the sample and
the tip are reflected in these line sections as well. We find indeed that the current value above the N atom is insensitive to
the tip rotation, and it is almost the smallest current value in the entire scan area. We can also see that the brightest
features, i.e., the largest current values of the STM images are actually not located above the carbon atoms, but rather above
the hollow positions of the honeycomb lattice.

\subsection{Mn$_2$H on Ag(111) surface\label{sub:MnH-on-Ag(111)}}

\begin{figure}[h]
\begin{centering}
\includegraphics[width=5cm]{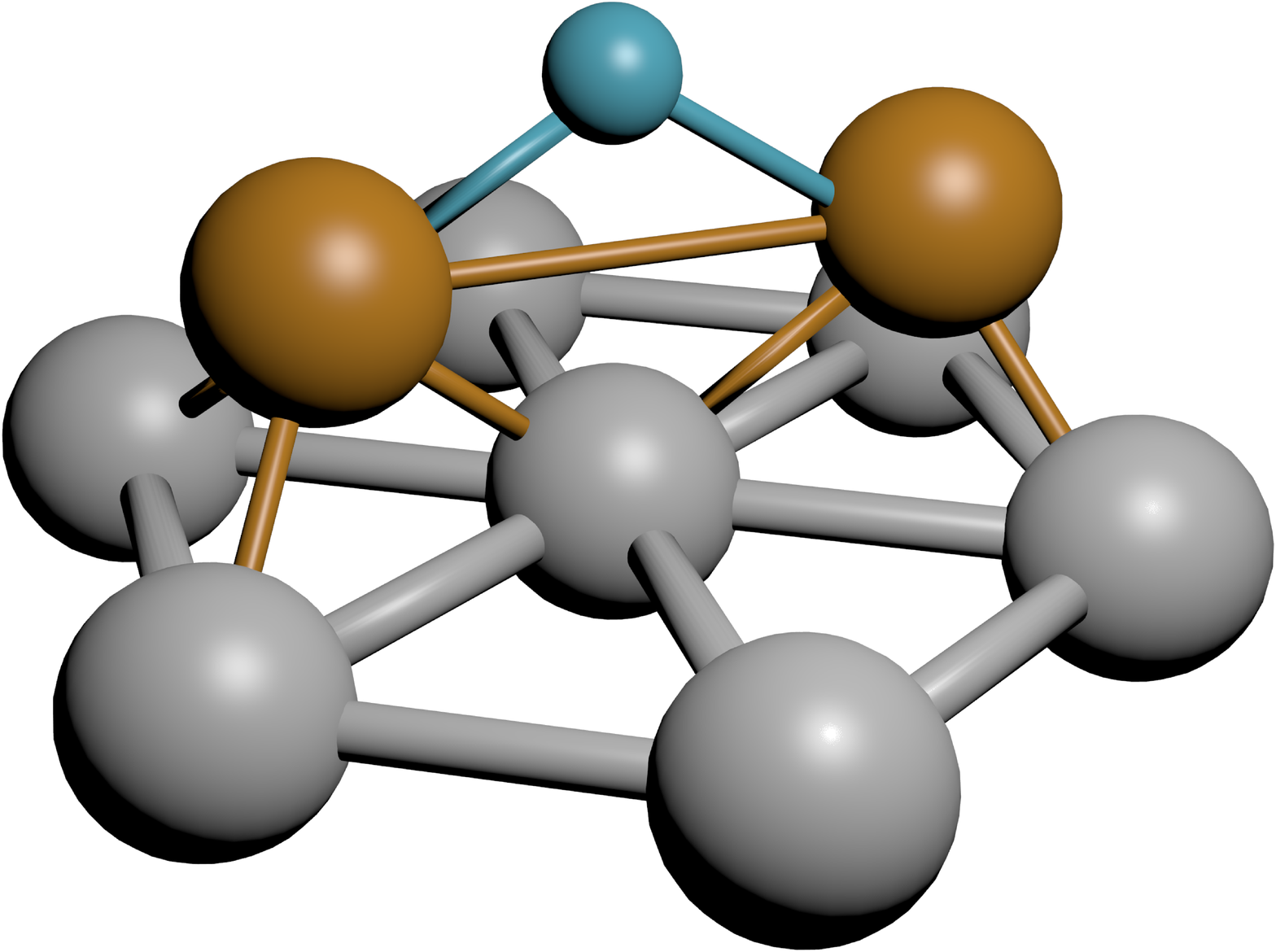}\includegraphics[width=5cm]{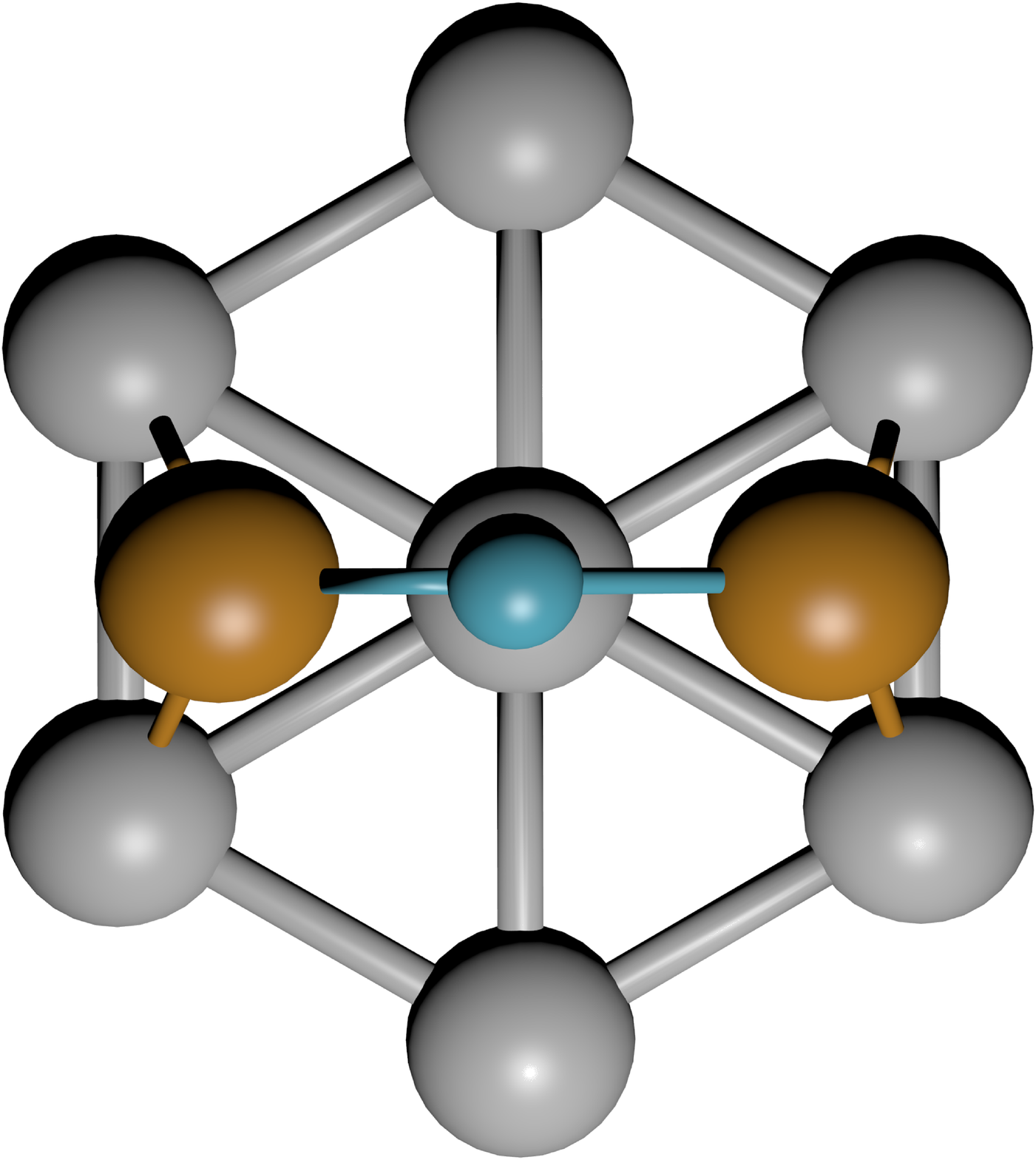}
\par\end{centering}
\protect\caption{\label{fig:AgMn2H}Calculated relaxed geometry of antiferromagnetic Mn$_2$H on the Ag(111) surface.
Data is taken from Ref.\ \citep{Sachse-AgMn2H}.}
\end{figure}

Sachse {\it{et al.}} found that Mn$_2$H on the Ag(111) surface can produce STM images with single or double features
depending on the magnetic coupling between Mn atoms \citep{Sachse-AgMn2H}. Double features have been obtained at positive bias
employing the Tersoff-Hamann method for an antiferromagnetic Mn-Mn coupling, which corresponds to the energetically favored
ground state. The calculated relaxed geometry of antiferromagnetic Mn$_2$H on the Ag(111) surface is shown in
Fig.\ \ref{fig:AgMn2H}. We consider this system and perform an investigation of its STM imaging depending on three employed
tunneling models: Bardeen, revised Chen and Tersoff-Hamann. Using the decomposition of the tunneling current according to
Eq.\ (\ref{eq:M2_interference}) in the revised Chen's method, we are able to identify the physical origin of the observed dip
above Mn$_2$H.

\begin{figure}[h]
\begin{centering}
\begin{tabular}{|c|cccc|}
\cline{2-5} 
\multicolumn{1}{c|}{} & \multicolumn{2}{c|}{Ag(001) tip} & \multicolumn{2}{c|}{Ag(111) tip}\tabularnewline
\cline{2-5} 
\multicolumn{1}{c|}{} & \multicolumn{1}{c|}{$-0.1$ V} & \multicolumn{1}{c|}{$+0.1$ V} & \multicolumn{1}{c|}{$-0.1$ V} & $+0.1$ V\tabularnewline
\hline 
\begin{turn}{90}
Bardeen
\end{turn} & \includegraphics[angle=90,width=0.2\columnwidth]{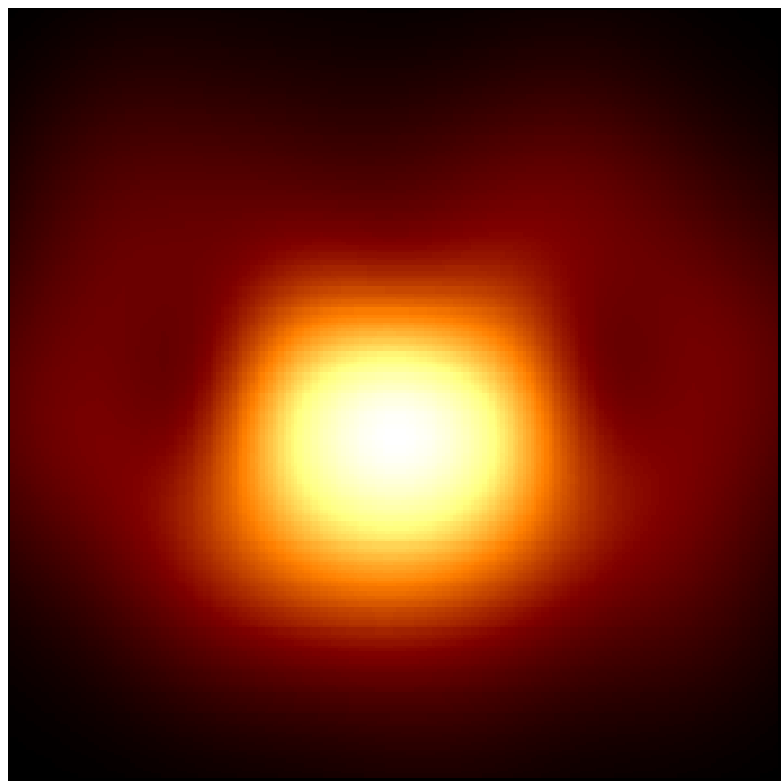} & \includegraphics[angle=90,width=0.2\textwidth]{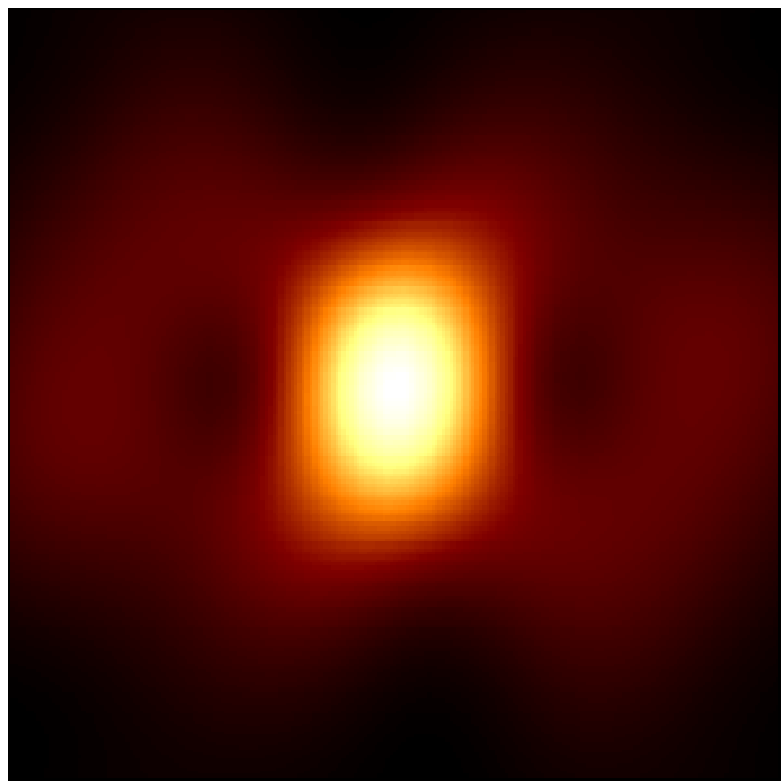} & \includegraphics[angle=90,width=0.2\textwidth]{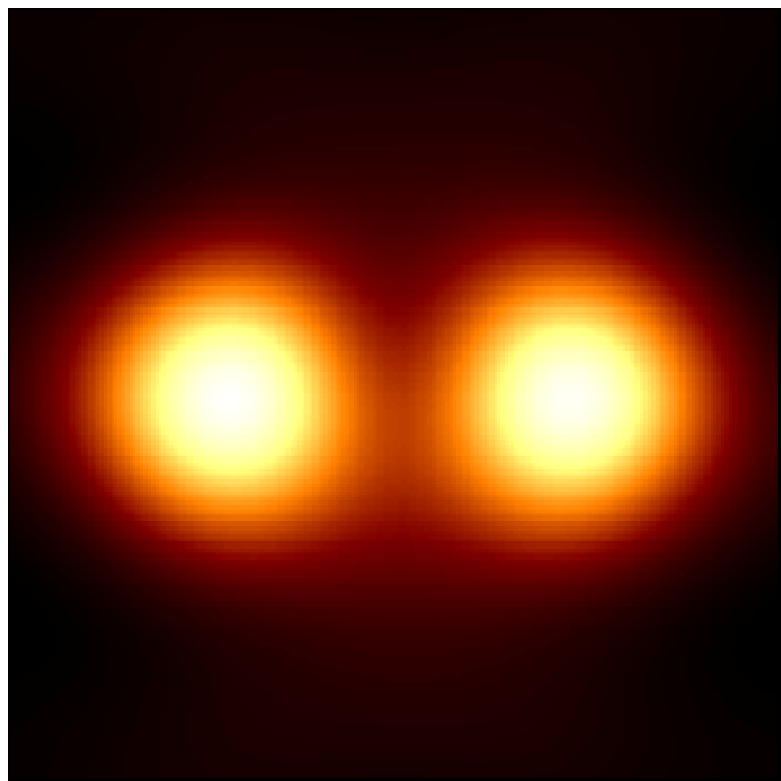} & \includegraphics[angle=90,width=0.2\textwidth]{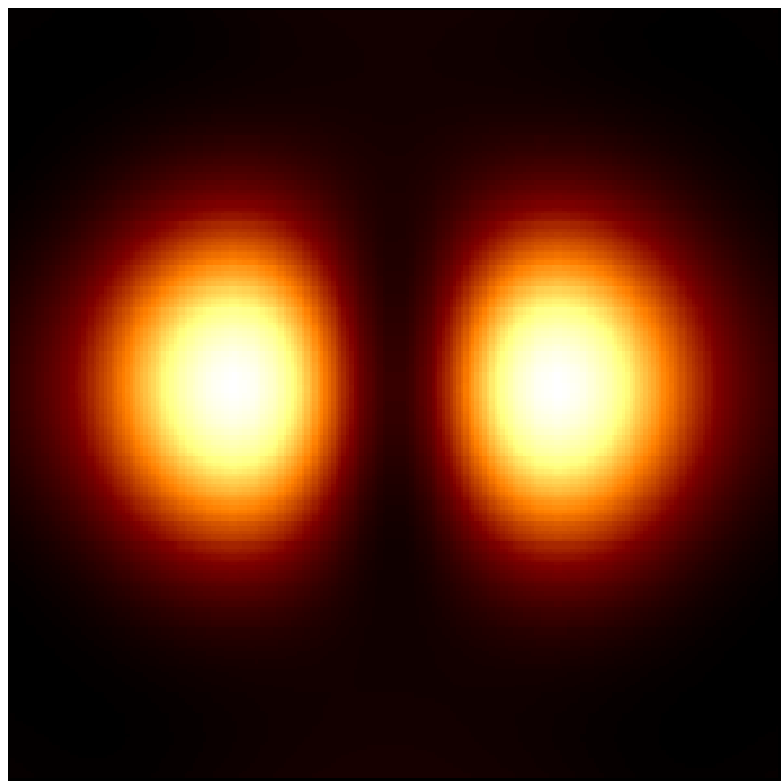}\tabularnewline
\cline{1-1} 
\begin{turn}{90}
Revised Chen
\end{turn} & \includegraphics[angle=90,width=0.2\textwidth]{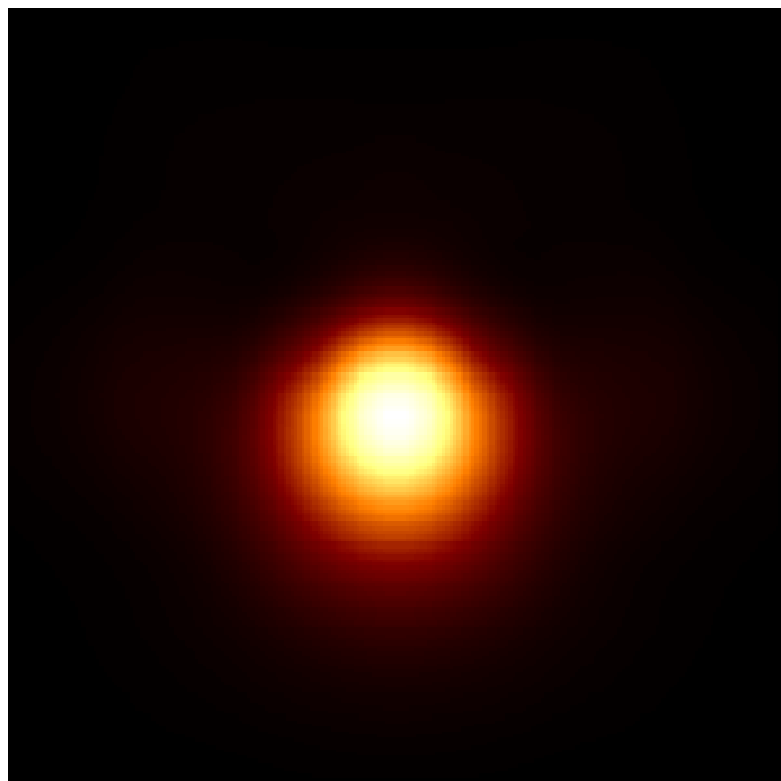} & \includegraphics[angle=90,width=0.2\textwidth]{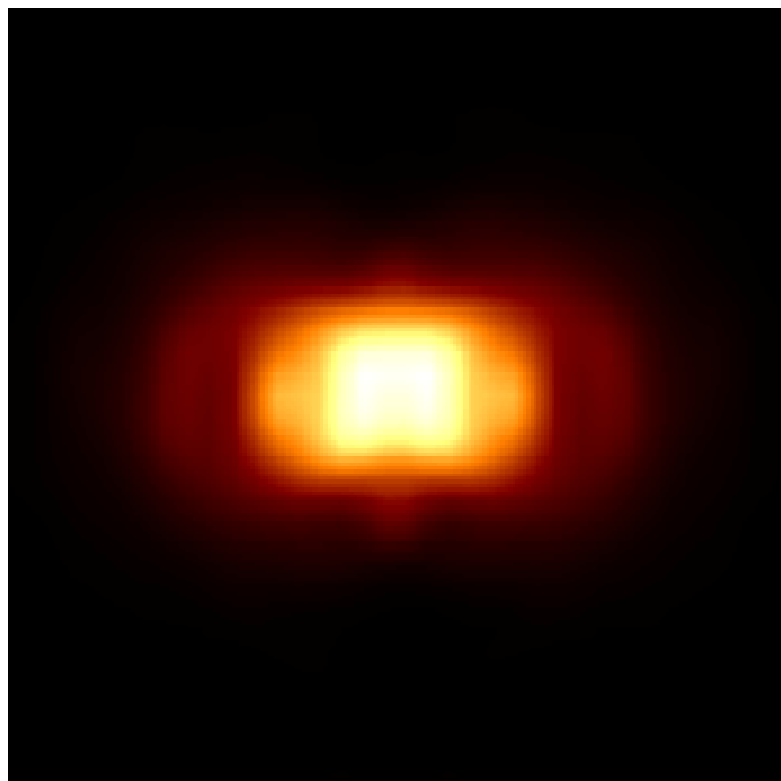} & \includegraphics[angle=90,width=0.2\textwidth]{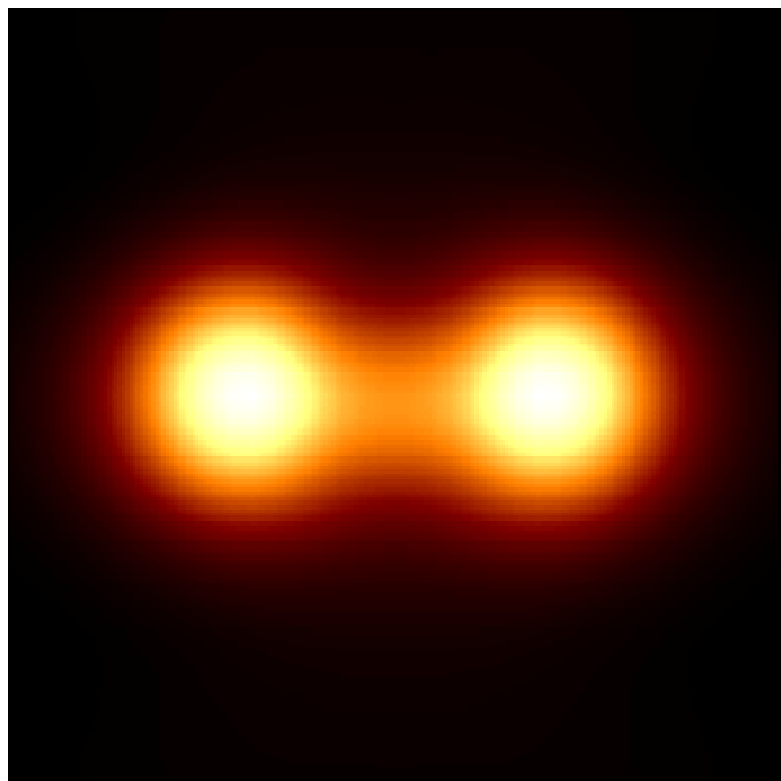} & \includegraphics[angle=90,width=0.2\textwidth]{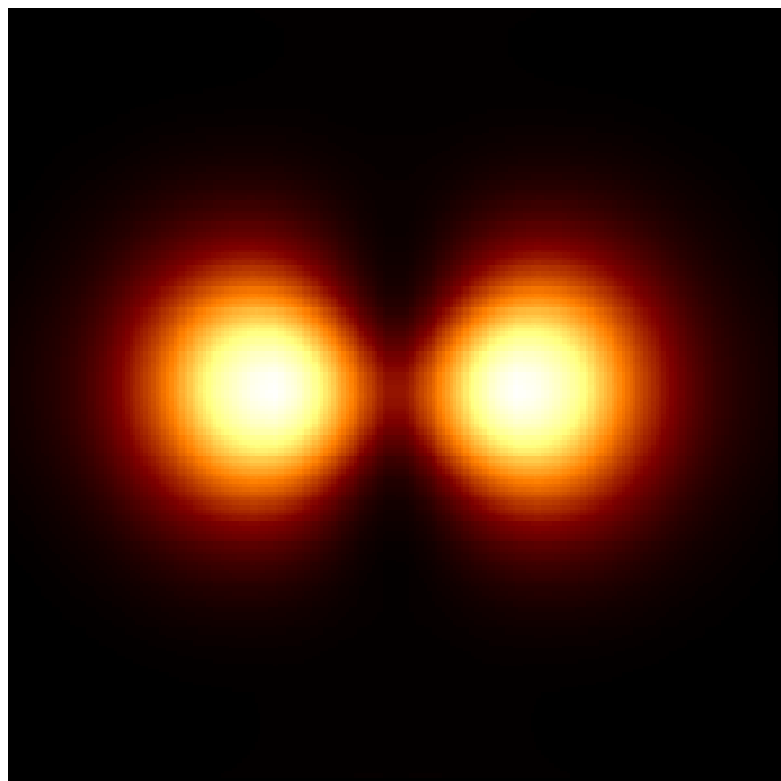}\tabularnewline
\cline{1-1} 
\begin{turn}{90}
Tersoff-Hamann
\end{turn} & \includegraphics[angle=90,width=0.2\textwidth]{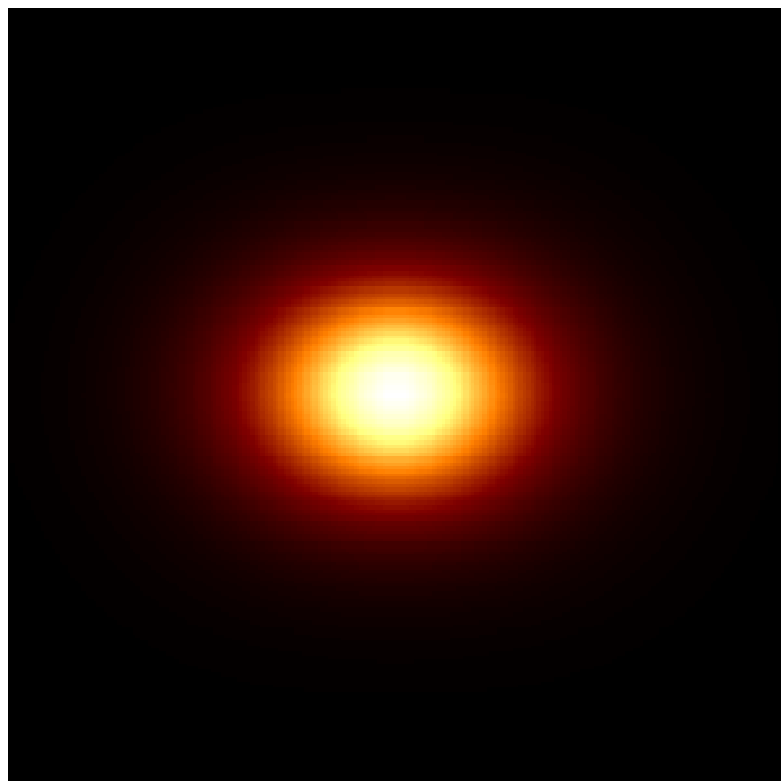} & \includegraphics[angle=90,width=0.2\textwidth]{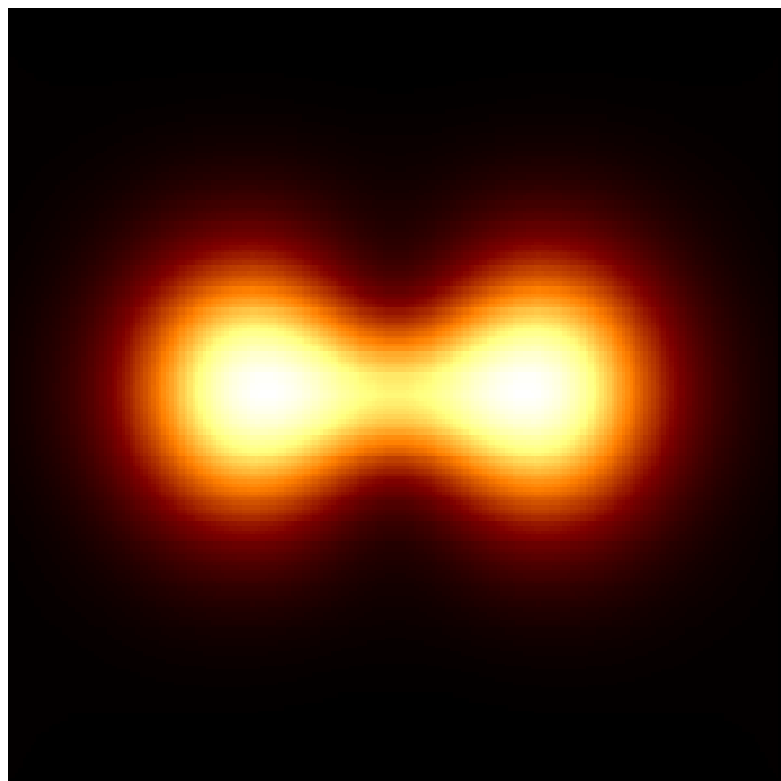} & \includegraphics[angle=90,width=0.2\textwidth]{plot_tersoff_-0.1V.eps} & \includegraphics[angle=90,width=0.2\textwidth]{plot_tersoff_+0.1V.eps}\tabularnewline
\hline 
\end{tabular}
\par\end{centering}
\caption{Constant height STM images of antiferromagnetic Mn$_2$H on the Ag(111) surface simulated at 5 $\textrm{\AA}$
Ag surface-tip distance and $\pm 0.1$ V bias with three different methods (Bardeen, revised Chen, Tersoff-Hamann) and
two blunt tip models: Ag(001) and Ag(111). A temperature of 7 K was assumed corresponding to the STM experiments in
Ref.\ \citep{Sachse-AgMn2H}. Note that results of the Tersoff-Hamann model are shown for comparison reasons
only, and no explicit tip electronic structure is considered there. \label{fig:Comparing-AgMn2H}}
\end{figure}

The calculated constant height STM images at small bias voltages ($\pm 0.1$ V) using two silver blunt tip models
[Ag(001) and Ag(111)] are shown in Fig.\ \ref{fig:Comparing-AgMn2H}. Correlation coefficients between STM images
obtained by Bardeen's and revised Chen's method are reported in Table \ref{tab:Corr-Fig7}.
First of all, we find excellent quantitative agreement between the STM images obtained by Bardeen's and
revised Chen's method for the Ag(111) tip and good agreement for the Ag(001) tip. Recalling that the revised
Chen's method is 25 times faster than Bardeen's method in practical STM calculations, this clearly indicates that
our proposed model is a very promising tool for STM simulations in the future. Moreover, the results in
Fig.\ \ref{fig:Comparing-AgMn2H} show that the geometry and electronic structure of the tip have
a considerable effect on the STM imaging of Mn$_2$H/Ag(111): Ag(001) tip provides single protrusion and Ag(111) tip provides
double features of the STM images at both bias polarities using both Bardeen's and revised Chen's methods. However, the
Tersoff-Hamann model provides qualitative agreement with these at selected bias voltages only: At $-0.1$ V a single protrusion is
obtained, while at $+0.1$ V a double feature is visible. The comparison of the Tersoff-Hamann results with those obtained by the
revised Chen's method suggests a contradiction with the general assumption of Ag tips being of $s$ orbital character
\citep{Sachse-AgMn2H}. In order to understand the components of the tunneling current above the H atom, the decomposition
according to Eq.\ (\ref{eq:M2_interference}) in the revised Chen's method is employed.

\begin{table}[h]
\begin{centering}
\begin{tabular}{|c|c|c|c|c|}
\hline
Correlation coefficients & \multicolumn{2}{c|}{Ag(001) tip} & \multicolumn{2}{c|}{Ag(111) tip}\tabularnewline
\cline{2-5}
between STM images in Fig.\ \ref{fig:Comparing-AgMn2H} & $-0.1$ V & $+0.1$ V & $-0.1$ V & $+0.1$ V\tabularnewline
\hline
Bardeen-Revised Chen ($C_{\nu\beta}$ in Eq.(\ref{eq:coeff})) & 0.72 & 0.88 & 0.97 & 0.95\tabularnewline
\hline
\end{tabular}
\par\end{centering}
\protect\caption{Quantitative comparison between Bardeen's model and revised Chen's method:
calculated correlation coefficients between STM images in Fig.\ \ref{fig:Comparing-AgMn2H}.
\label{tab:Corr-Fig7} }
\end{table}

\begin{figure}[h]
\begin{centering}
\includegraphics[width=0.5\columnwidth]{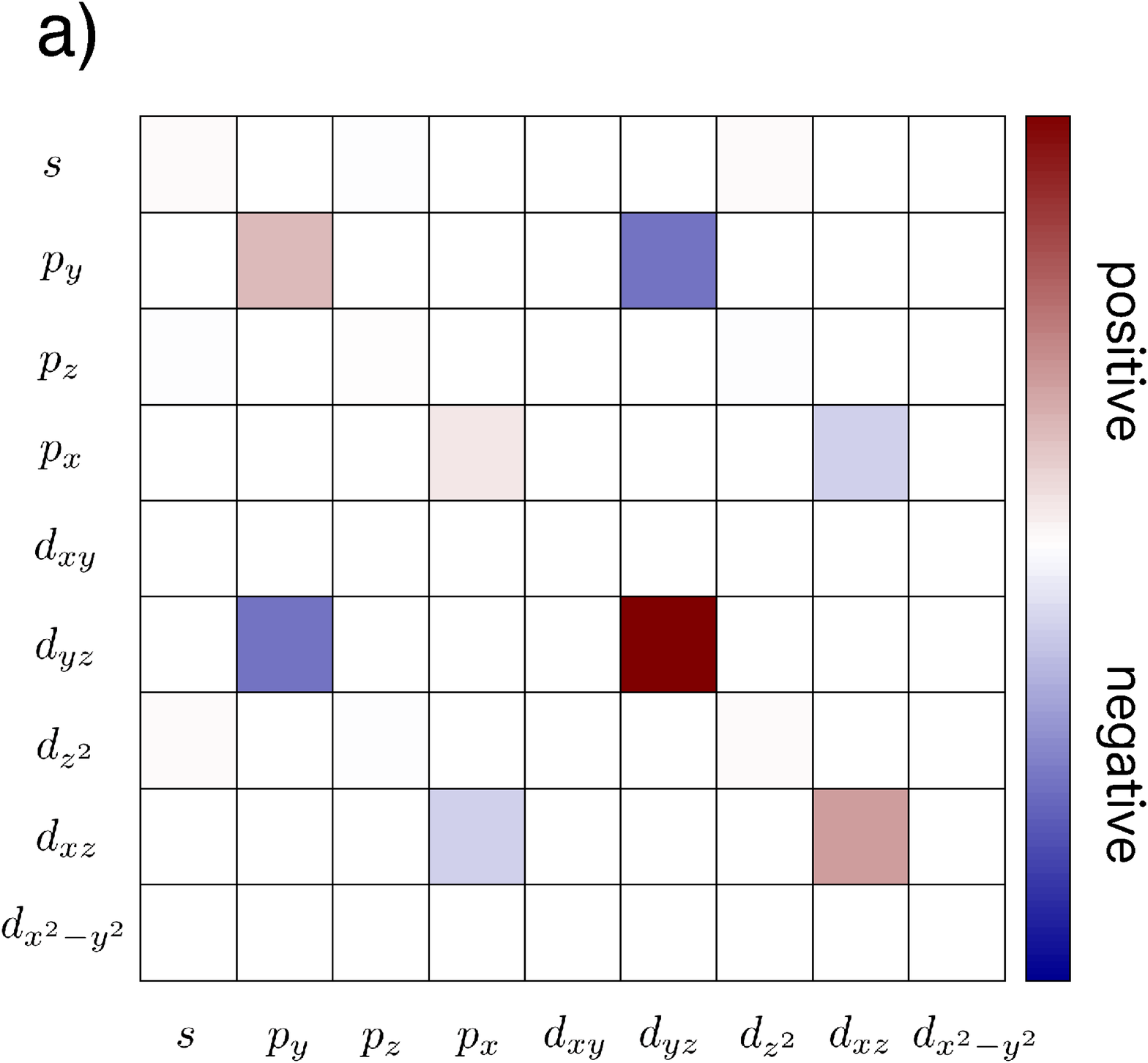}\includegraphics[width=0.5\textwidth]{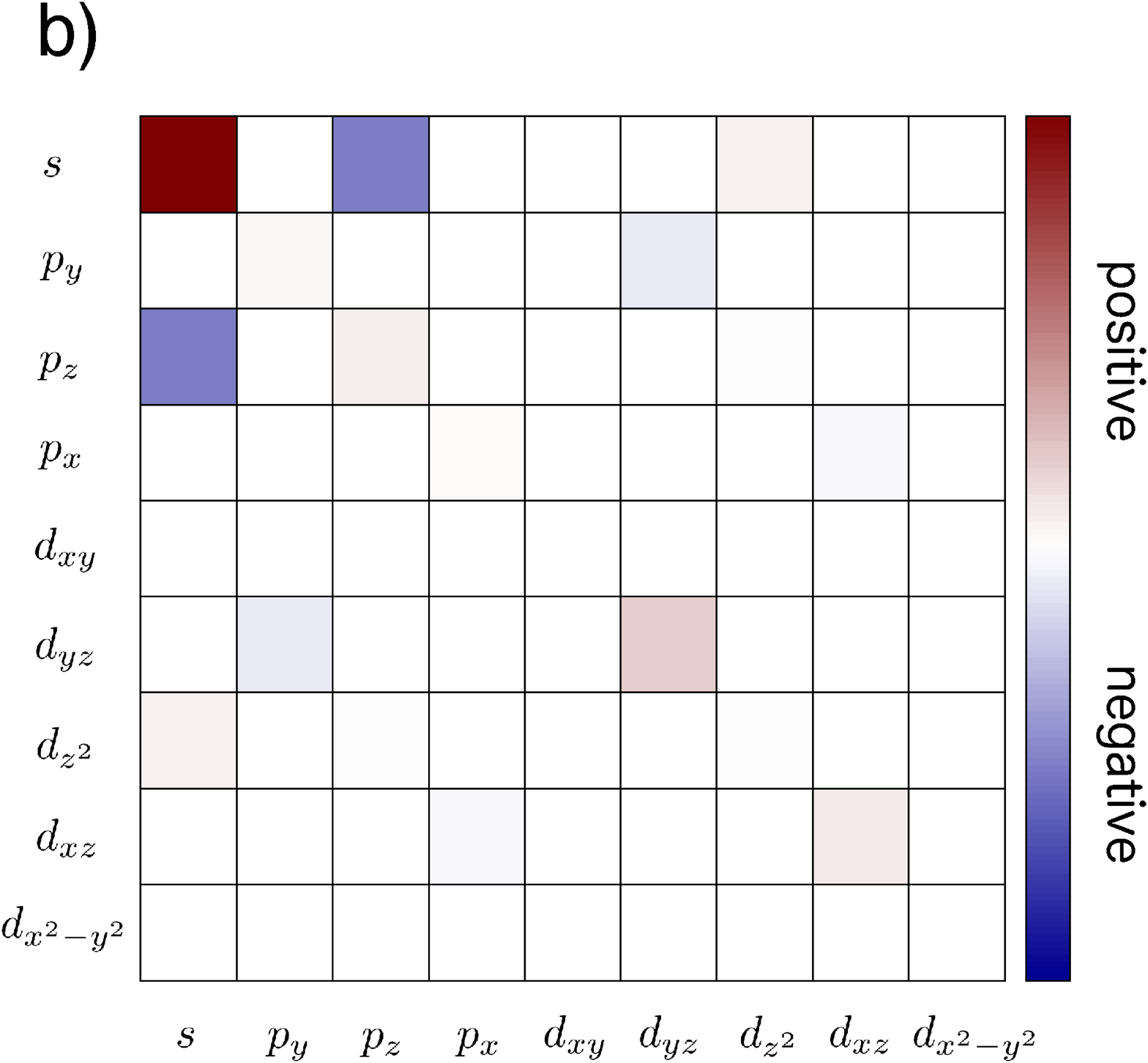}
\par\end{centering}
\protect\caption{\label{fig:Decomposition}Decomposition of the tunneling current 1.83 $\textrm{\AA}$ above the H atom in
Mn$_2$H/Ag(111) (corresponding to Fig.\ \ref{fig:Comparing-AgMn2H}) using Eq.\ (\ref{eq:M2_interference}).
Diagonal: direct (positive) contributions, off-diagonal: interference (positive or negative) contributions to the current.
a) Ag(001) tip at bias voltage $V=+0.1$ V, b) Ag(111) tip at bias voltage $V=-0.1$ V.}
\end{figure}

Fig.\ \ref{fig:Decomposition} shows the results of the current decomposition according to tip orbital characters.
Interestingly, we find that the Ag(001) tip does not behave as an $s$-type tip at $+0.1$ V bias
[see Fig.\ \ref{fig:Decomposition}a)]. The major contributions are from the $p_{x}$, $p_{y}$, $d_{xz}$ and $d_{yz}$ tip
orbitals, and there are destructive interferences arising from $p_{x}-d_{xz}$ and $p_{y}-d_{yz}$ tip orbitals. This
explains the qualitative disagreement between the Tersoff-Hamann and revised Chen's results for the Ag(001) tip at $+0.1$ V
bias. On the other hand, using the Ag(111) tip at $-0.1$ V bias, the major contribution is clearly from the tip's $s$ orbital
[see Fig.\ \ref{fig:Decomposition}b)]. Apart from that there is a strong $s-p_z$ destructive tip interference that is missing in
the Tersoff-Hamann model, causing the observed qualitative difference in the STM images for the Ag(111) tip at $-0.1$ V bias.
Moreover, we find similar current decomposition characteristics for the Ag(001) tip at $-0.1$ V bias and for the Ag(111) tip at
$+0.1$ V bias as Fig.\ \ref{fig:Decomposition}b) shows, with a dominating $s$ orbital contribution from the tip. For these
tip and bias voltage combinations a good qualitative agreement of the STM images between revised Chen's and Tersoff-Hamann results
is obtained. Our findings suggest that although the quality of the STM contrast (single or double feature) is mainly determined by
the electronic states of the sample surface that can be captured by employing the Tersoff-Hamann model,
the tip electronic structure and in the present case an $s-p_z$ destructive tip interference can cause a contrast change.

\begin{figure}[h]
\begin{centering}
\begin{tabular}{|c|cccc|}
\cline{2-5}
\multicolumn{1}{c|}{} & \multicolumn{2}{c|}{Ag(001) tip} & \multicolumn{2}{c|}{Ag(111) tip}\tabularnewline
\cline{2-5}
\multicolumn{1}{c|}{} & \multicolumn{1}{c|}{$-0.1$ V} & \multicolumn{1}{c|}{$+0.1$ V} & \multicolumn{1}{c|}{$-0.1$ V} & $+0.1$ V\tabularnewline
\hline
\begin{turn}{90}
$T=$ 7 K
\end{turn} & \includegraphics[angle=90,width=0.2\textwidth]{plot_chen_Ag001_-0.1V.eps} & \includegraphics[angle=90,width=0.2\textwidth]{plot_chen_Ag001_+0.1V.eps} & \includegraphics[angle=90,width=0.2\textwidth]{plot_chen_Ag111_-0.1V.eps} & \includegraphics[angle=90,width=0.2\textwidth]{plot_chen_Ag111_+0.1V.eps}\tabularnewline
\cline{1-1}
\begin{turn}{90}
$T=$ 70 K
\end{turn} & \includegraphics[angle=90,width=0.2\textwidth]{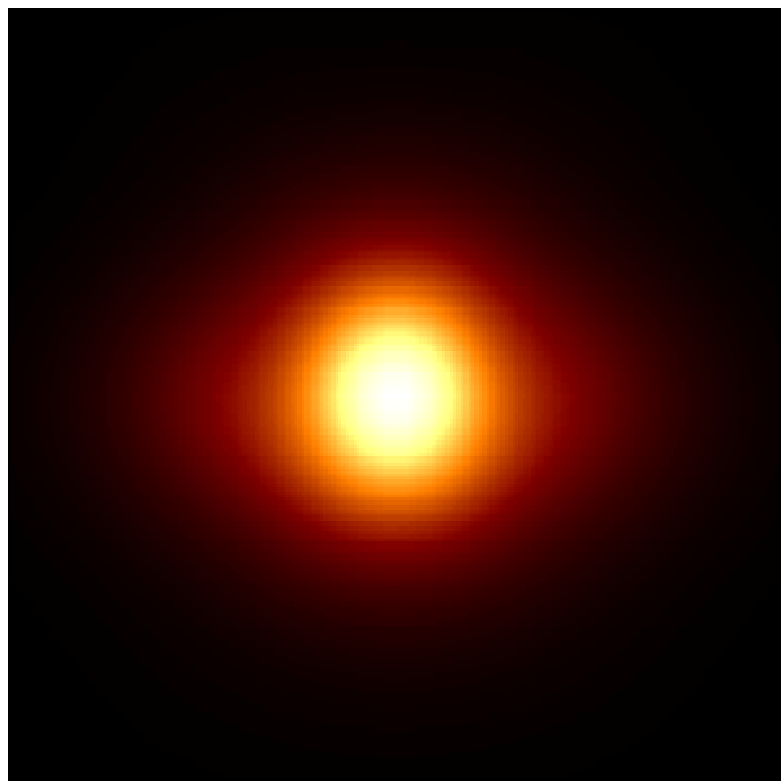} & \includegraphics[angle=90,width=0.2\textwidth]{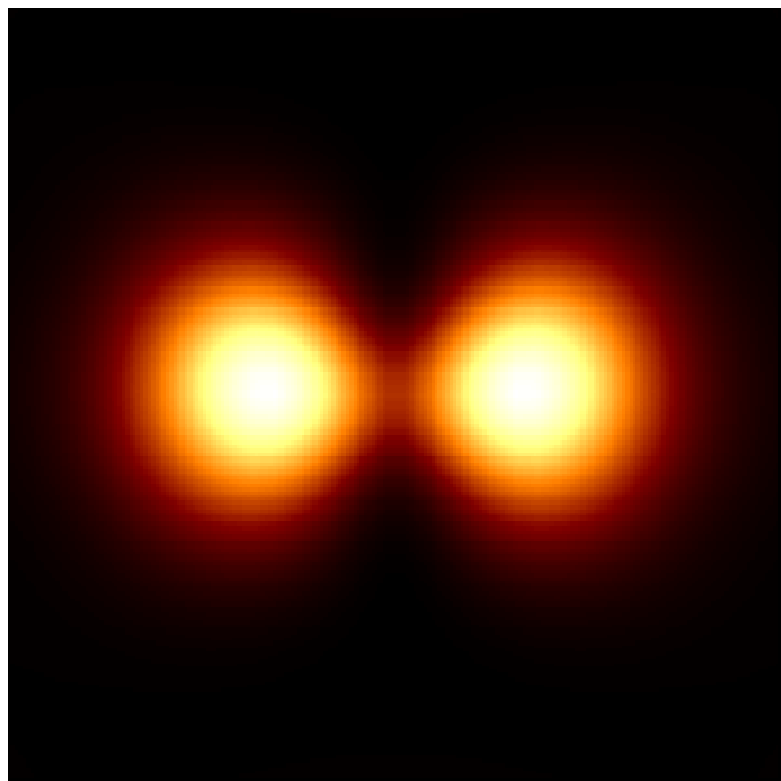} & \includegraphics[angle=90,width=0.2\textwidth]{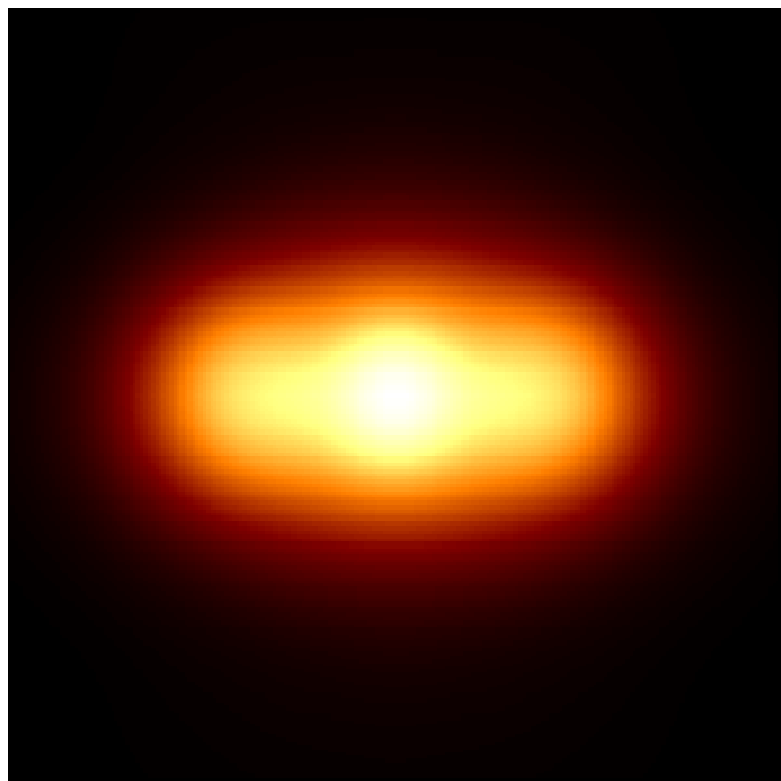} & \includegraphics[angle=90,width=0.2\textwidth]{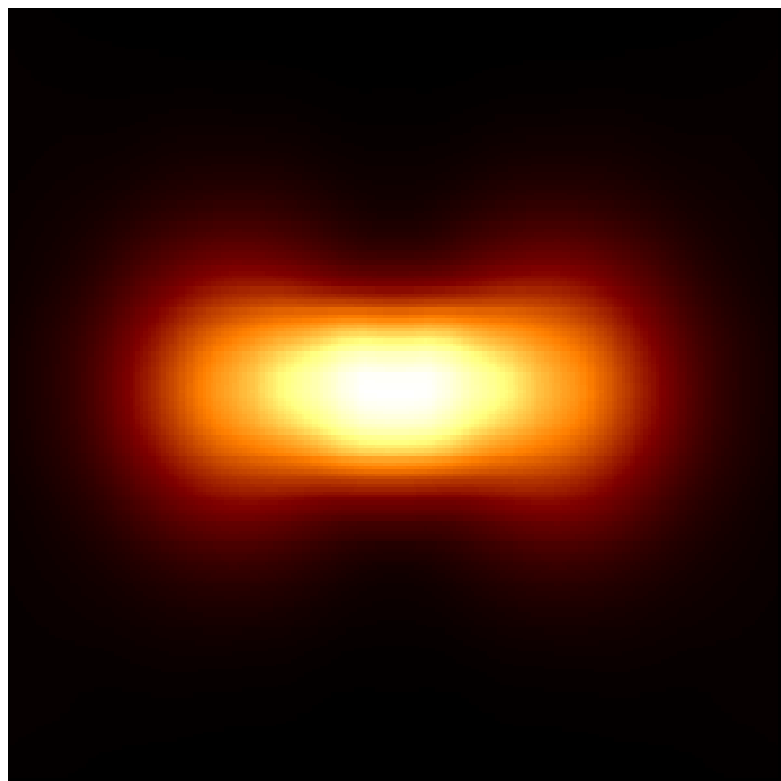}\tabularnewline
\hline
\end{tabular}
\par\end{centering}
\caption{Effect of temperature on constant height STM images of antiferromagnetic Mn$_2$H on the Ag(111)
surface simulated at 5 $\textrm{\AA}$ Ag surface-tip distance and $\pm 0.1$ V bias using the revised Chen's method with
two blunt tip models: Ag(001) and Ag(111). Temperatures of $T=$ 7 K and 70 K are compared.\label{fig:Temperature-AgMn2H}}
\end{figure}

It is important to highlight the effect of temperature on the obtained STM contrast.
Temperature enters into the tunneling model exactly in the same fashion as in the well established Bardeen's model, i.e.,
in two ways: (i) the thermal broadening of the electron states, see Eq.(\ref{eq:Delta}),
and (ii) the energy window for calculating the tunneling channels, where nonzero temperature results in an extension of the energy
window due to the Fermi distribution, see Eq.(\ref{eq:Bardeen tunneling}).
Fig.\ \ref{fig:Temperature-AgMn2H} shows a comparison between STM images calculated at $T=$ 7 K and 70 K, the former corresponding
to the temperature used in the experiments of Ref.\ \citep{Sachse-AgMn2H}. We find a diverse behavior of the STM contrast
at the higher temperature depending on the tip and bias voltage. The contrast (single protrusion) is preserved for the Ag(001) tip
at $-0.1$ V only. The other three images show different contrasts at the two temperatures. Upon increasing the temperature,
for the Ag(001) tip at $+0.1$ V the single protrusion contrast changes to double features, while for the Ag(111) tip at both bias
voltages the double protrusion contrast changes to an elongated single feature with the maximal current above the H atom of
Mn$_2$H. This diversity of simulated STM contrasts points to the importance of the correct choice of temperature in
STM simulations if a meaningful explanation of given experimental STM data is desired.

\section{Conclusions\label{sec:Conclusions}}

We revised Chen's derivative rule for electron tunneling for the purpose of computationally efficient STM simulations based on
first principles electronic structure data. The revised Chen's model includes the electronic structure and arbitrary spatial
orientation of the tip by taking appropriate weighting coefficients of tunneling matrix elements of different tip orbital
characters. Interference of tip orbitals in the STM junction is included in the model by construction. We demonstrated the
reliability of the model by applying it to two functionalized surfaces of recent interest where quantum interference effects play
an important role in the STM imaging process: N-doped graphene and an antiferromagnetic Mn$_2$H complex on the Ag(111) surface.
We found that the revised Chen's model is 25 times faster than the Bardeen method concerning computational time, while
maintaining good agreement. Our results show that the electronic structure of the tip has a considerable effect
on STM images, and the Tersoff-Hamann model does not always provide sufficient results in view of quantum interference effects.
For both studied surfaces we highlighted the importance of interference between $s$ and $p_z$ tip orbitals that can cause a
significant contrast change in the STM images.
Moreover, our findings show that stretched bonds have a minor effect on the main features of the STM contrast, and temperature is
an important factor to be taken into account in STM simulations if aiming at accuracy in comparison with experiments.
Our method implemented in the BSKAN code, thus, provides a fast and reliable tool for calculating STM images based on
Chen's derivative rule taking into account the electronic structure and local geometry of the tip apex.

\section*{Acknowledgments}

The authors thank P.\ Mutombo and P.\ Jel\'inek for insightful discussions concerning N-doped graphene and
T.\ Sachse and W.\ A.\ Hofer for useful discussions and electronic structure data of Mn$_2$H/Ag(111).
Financial support of the Hungarian State E\"otv\"os Fellowship, Hungarian Scientific Research Fund project
OTKA PD83353, the Bolyai Research Grant of the Hungarian Academy of Sciences, and the New Sz\'echenyi Plan of Hungary
(Project ID: T\'AMOP-4.2.2.B-10/1--2010-0009) is gratefully acknowledged.
Usage of the computing facilities of the Wigner Research Centre for Physics and the BME HPC Cluster is kindly acknowledged.

\end{document}